\documentclass[12pt,twoside]{mitthesis}
\usepackage{lgrind}
\usepackage{cmap}
\usepackage[T1]{fontenc}
\usepackage{graphicx}
\usepackage[usenames, dvipsnames]{color}
\usepackage{changepage}
\usepackage{amssymb}
\usepackage{multirow}
\usepackage{algorithm}
\usepackage{mathtools}
\usepackage{lipsum}
\usepackage{float}
\usepackage{varwidth}
\usepackage[noend]{algpseudocode}
\usepackage{xr}
\usepackage{tablefootnote}
\usepackage{url}
\usepackage[utf8x]{inputenc}
\DeclareUnicodeCharacter{"2061}{}
\begin{document}

\title{Educational Content Linking for Enhancing Learning Need Remediation in MOOCs}

\author{Shang-Wen Li}

\setcounter{savepage}{\thepage}
\begin{abstractpage}
Since its introduction in 2011, there have been over 4000 MOOCs on various subjects on the Web, serving over 35 million learners. MOOCs have shown the ability to democratize knowledge dissemination and bring the best education in the world to every learner. However, the disparate distances between participants, the size of the learner population, and the heterogeneity of the learners' backgrounds make it extremely difficult for instructors to interact with the learners in a timely manner, which adversely affects learning experience. To address the challenges, in this thesis, we propose a framework: educational content linking. By linking and organizing pieces of learning content scattered in various course materials into an easily accessible structure, we hypothesize that this framework can provide learners guidance and improve content navigation. Since most instruction and knowledge acquisition in MOOCs takes place when learners are surveying course materials, better content navigation may help learners find supporting information to resolve their confusion and thus improve learning outcome and experience. To support our conjecture, we present end-to-end studies to investigate our framework around two research questions: 1) can manually generated linking improve learning? 2) can learning content be generated with machine learning methods? For studying the first question, we built an interface that present learning materials and visualize the linking among them simultaneously. We found the interface enables users to search for desired course materials more efficiently, and retain more concepts more readily. For the second question, we propose an automatic content linking algorithm based on conditional random fields. We demonstrate that automatically generated linking can still lead to better learning, although the magnitude of the improvement over the unlinked interface is smaller.
\end{abstractpage}

\cleardoublepage


\pagestyle{plain}

\chapter{Introduction}

Since 2011, a revolution called MOOCs (Massive Open Online Courses) began in university education \cite{1, 2}. Today, a mere five years after the first MOOC was launched, over 4000 MOOCs, from science and engineering to humanities and law, have been offered on the Web and have served over 35 million learners using platforms such as Coursera, edX, Udacity, and FutureLearn \cite{3, 4, 5, 6, 7}. These MOOCs have been created by over 500 top institutions in the world and taught by top instructors. In addition, MOOCs allow free enrollment and enable learners around the globe to take courses without the need for physical presence. Thus, MOOCs have the potential to transcend time and space, democratize knowledge dissemination, and bring opportunities to learners in every corner of the world.

MOOCs inspire a new model to deliver quality education. In conventional residential education, we have classes of much smaller sizes. These classes are taught in thousands of institutions on the same subject with only slight variation. In contrast, MOOCs adopt a distributed model. This model can accumulate the investment of offering these classes in institutions and instructors, and allow course builders to allocate their time and efforts more efficiently in implementing various state-of-the-art and research-based pedagogies, such as active learning, mastery learning, and cooperative learning, in the course \cite{8, 9, 10}. Thus, MOOCs provide enormous educational value to learners and instructors. Evidence suggests that well-designed MOOCs alone can lead to high levels of student learning and satisfaction \cite{8}. In addition, we have observed a growing trend of college instructors' adoption in blended classrooms \cite{11}. In blended learning, residential classroom instructors utilize existing MOOC content to save their efforts in course material preparation, and thus they can focus on interacting with students to create a learner-centered environment \cite{12, 13, 14}.

However, the open and free character of MOOCs has also created a set of challenges that are not observed in conventional education, that is, the sheer size of the learner body, and the heterogeneity of their backgrounds \cite{15}. A MOOC typically has thousands to tens of thousands of learners with various demographics, course preparedness, learning goals, and motivations. With such size and heterogeneity of a class, conventional one-size-fits-all pedagogy is not sufficient. For example, in the same MOOC, some learners may struggle with elementary concepts due to having insufficient prerequisite background, while another group of learners may already have years of experience in the industry of the relevant area and their learning goal is to update their job skills. Due to the distant nature, learners in MOOCs usually rely on self-regulated learning to resolve their own learning needs. For instance, a learner who is confused about a topic in the lecture video may choose to pause the video, turn to textbook or discussion forum for more understandable description, and return to the video when this learner feels crystal clear about the underlying topic. In this way, different learners may take various learning paths and learning materials for their diverse learning needs. Nonetheless, because of the unfamiliarity of learners with the course subject as well as the amount of learning content in a MOOC, it is usually cumbersome for learners to find suitable content.

To address the challenges, in this thesis we propose a framework: \textit{educational content linking}. This framework allows linking and organizing scattered educational materials in a MOOC, as well as visualizing conceptual relations across these materials. Since the visualization can provide guidance for learners to navigate through materials, we expect this framework can help learners achieve self-regulated learning by allowing them to find appropriate information efficiently.

\section{Motivation}

One-to-one tutoring has been shown to be extremely effective in enhancing learning outcomes \cite{16}. However, this approach is too costly, and thus for many years, educators have dreamt of achieving similar effective size to the one-to-one model with a scalable approach \cite{16}. A significant number of studies have tried to unveil how people learn and why one-to-one tutoring is so successful in improving learning \cite{17, 18}. One of the key factors could be constructive struggle: much research have shown that keeping learners in a state of engagement between boredom and confusion has a substantial positive impact on learning \cite{19, 20, 21}. Outside the laboratory, such strategy has also been commonly applied to keep learners engaged, e.g., by asking students questions, inserting quizzes into lecture videos, or providing instructional scaffolding (Instructors provide sufficient support to learn a concept, while, during the entire learning process, support is taken away gradually to promote learners developing deeper-level knowledge).

Timely responses to confusion play a crucial role in the success of this strategy, and failing to do so can affect learning in the opposite way, such as causing frustration, or making learners stop participation. In a MOOC scenario, the incredibly low instructor-to-learner ratio and the heterogeneous background of learners make responding to learning needs extremely challenging. To address the problem, typically instructors can provide pre-defined hints, optional course materials, or even intelligent tutoring systems (ITS) to serve various needs and confusions. Another alternative is relying on learners to discover answers themselves in the course forums or on the Web. Both approaches are helpful but with several downsides. Providing hints, optional materials or ITS, even with the help from state-of-the-art machine learning methods, means lots of hand-crafting, such as designing banks of responses or individualized pathways for different needs. This approach is neither scalable nor generalizable from course to course. Furthermore, application of this approach in undergraduate-level or graduate-level subjects, which are the focus of MOOCs, is more cumbersome, since concepts in such a level of subjects are much more complicated. In contrast, the alternative way is much more scalable. The model of learnersourcing \cite{22} can potentially generate responses to diverse learning needs at scale. However, due to the amount of generated responses and needs, matching between the two is challenging. For example, although ideally a learner can find answers for every confusion in the MOOC's forum or from the Web, looking for useful contents from the huge database can be troublesome. The searching is more difficult for beginning learners since they may have trouble in describing their needs.

\section{Educational content linking}

Therefore in this thesis we propose a third way, \textit{educational content linking}. In this framework learning contents scattered in different types of course materials, such as lecture videos, slides, discussions forums, or quizzes, are linked based on their conceptual relation. A tree is then built based on the linking and presented to learners along with the content. Visualizing the relation provides guidance for learners to navigate through the content. Thus, we surmise that learners can find appropriate content for their various learning needs with much less effort, and tailor the learning path to suit their background. Furthermore, this framework has two extra upsides. First, since we do not limit this framework to any types of materials, \textit{educational content linking} can work with both approaches described in the previous section seamlessly. Second, since a conceptual relation is the only property that has to be inferred, this framework is simple enough to be realized with state-of-the-art machine learning and human language technologies (HLTs). The simplicity of the framework also means that this model can potentially work well in general cases rather than certain constrained environments.

\begin{figure}[t]
\includegraphics[width=9cm]{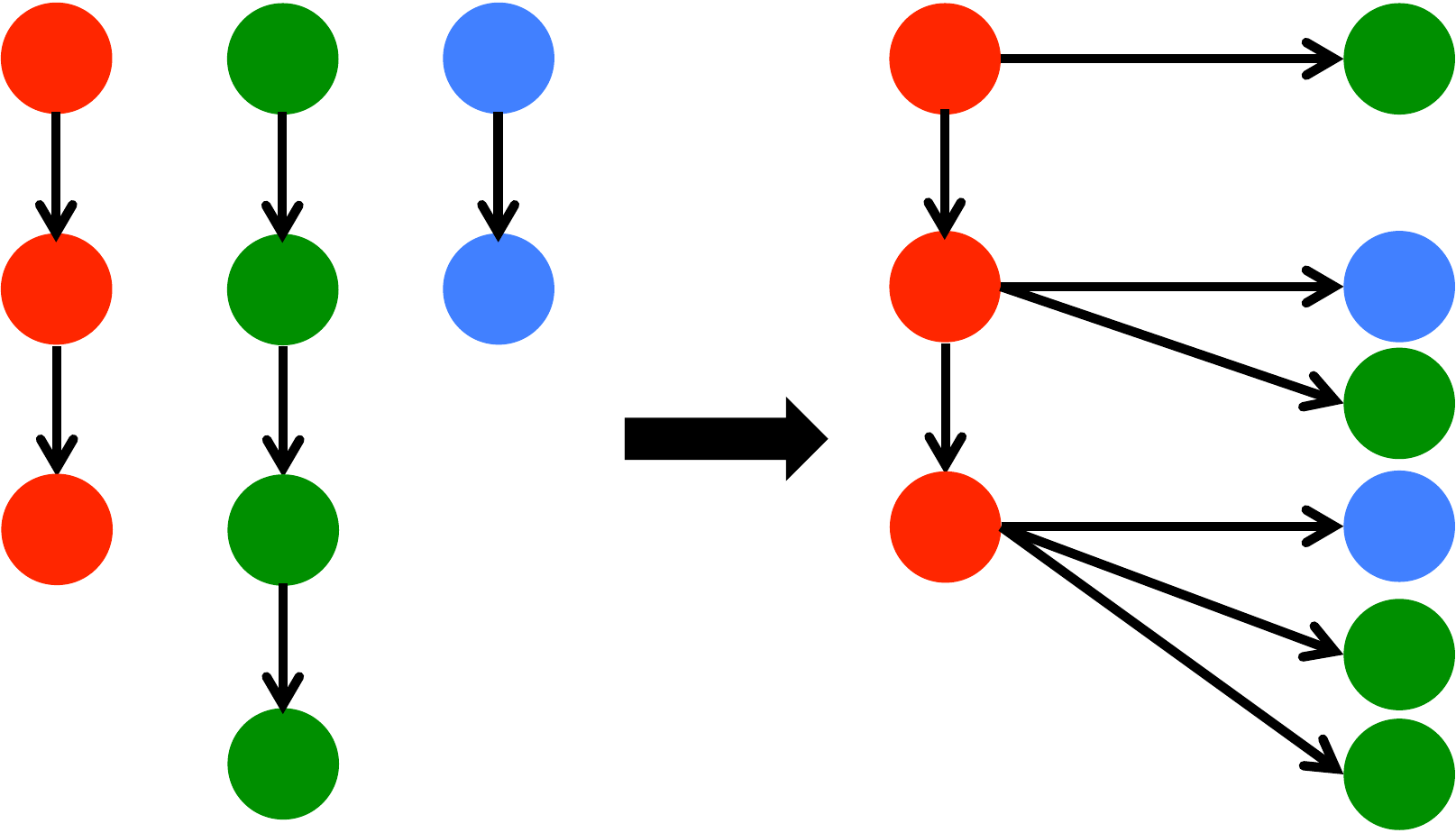}
\centering
\caption{Schematics of the transformation of several independent course materials to a linked structure. Each color illustrates one type of education material. Note that here linking refers to conceptual relation across materials.}
\label{fig:linking_diagram}
\end{figure}

In Fig.~\ref{fig:linking_diagram} we give an example to illustrate how \textit{educational content linking} works by comparing course materials presented in the traditional way to our proposed "linked" way. In the figure, different types of materials are represented in different colors. Content in each type of materials is segmented into smaller units, called learning objects in this thesis, and represented as nodes here. In this framework, the only requirement for learning objects is that an object should convey concepts in a self-contained way so that learners can understand. Thus, in realization of the framework an object can be any reasonable unit such as a textbook section, a discussion thread, or a vignette of video. 

The left-hand side of Fig.~\ref{fig:linking_diagram} illustrates how materials are presented in MOOCs conventionally. Objects are aligned in sequence based on syllabus, table of contents, or user-created time. Various types of materials are made available to learners as disjoint entities. In this scenario, a student interested in a specific concept cannot easily look up relevant information from various materials, e.g., from lectures or slides to sections of the textbook or discussions. In addition, the amount of user-generated content, such as discussions, is usually too large to be accessed efficiently if only organized chronologically.

In contrast, in \textit{educational content linking}, the courseware is linked across material types and presented as a tree, which is illustrated on the right-hand side of Fig.~\ref{fig:linking_diagram}. In this tree, one type of course materials is specified as the trunk, as shown in red nodes in the figure. This type of materials is utilized to extract the syllabus represented by the trunk. The rest of the materials are employed to build leaves of the tree. Each leaf, as represented in blue and green nodes in the figure, corresponds to a learning object that is related to an object from the trunk material. In this framework, conceptual relations among learning objects are visualized in addition to the original sequential presentation of materials. Thus, we expect that learners can better compare content from varied materials and identify useful information for them more efficiently. 

\begin{figure}[t]
\includegraphics[width=14cm]{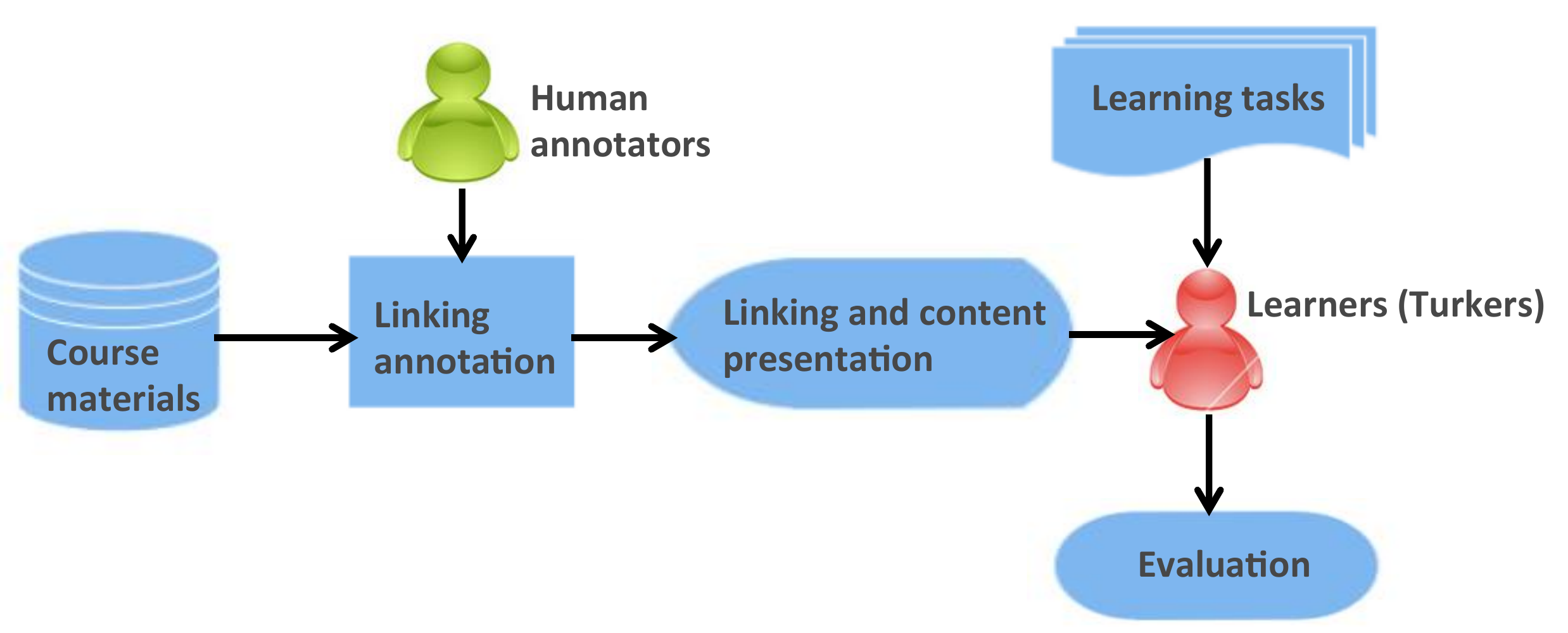}
\centering
\caption{The experimental pipeline for approaching the question: if we are able to link course materials using human annotators, would it help learners?}
\label{fig:rq1}
\end{figure}

\begin{figure}[t]
\includegraphics[width=14cm]{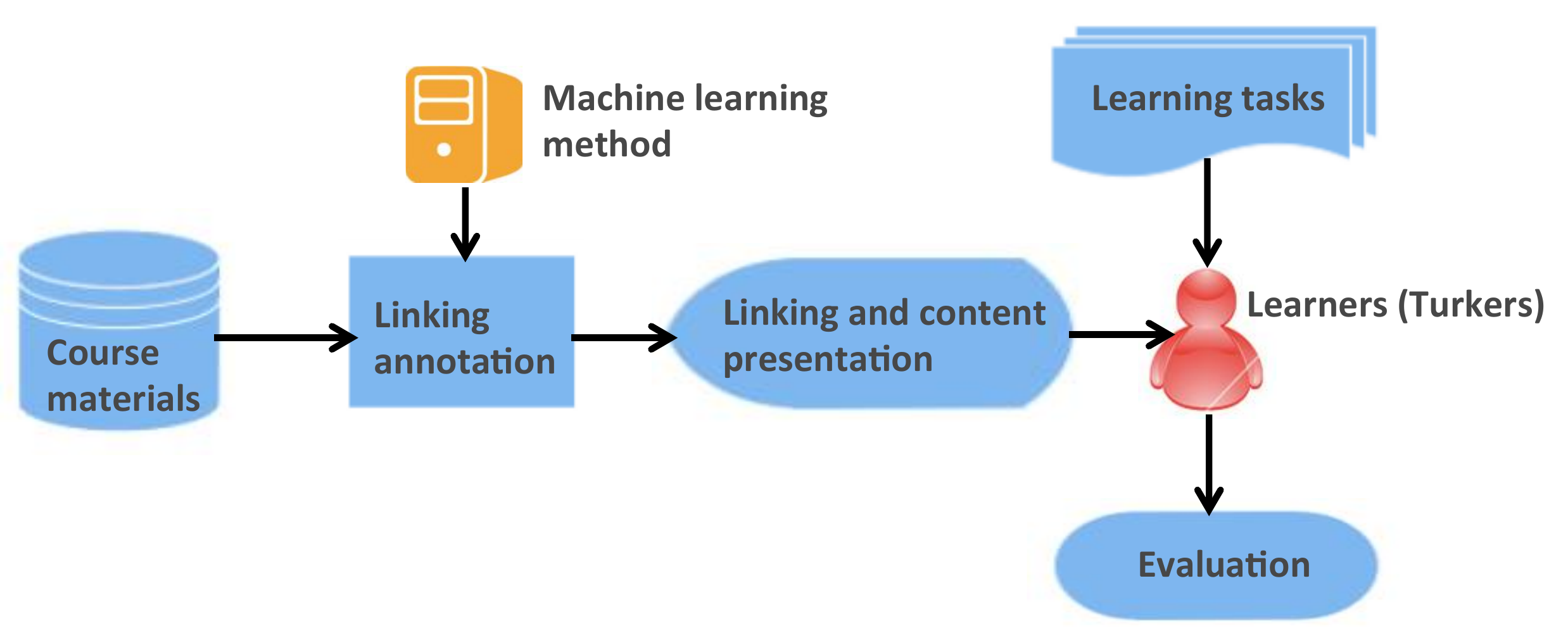}
\centering
\caption{The experimental pipeline for approaching question: can the courseware be linked at scale with machine learning methods?}
\label{fig:rq2}
\end{figure}

The goal of this thesis is therefore to prove our hypothesis: \textit{educational content linking} can help learners find desired information at scale. We focus our investigation on two research questions: 1), if we are able to link course materials using human annotators, would it help learners?, and 2), can the courseware be linked at scale with machine learning methods? Figs.~\ref{fig:rq1} and \ref{fig:rq2} outline the steps we take in this thesis to approach the two questions.

Fig.~\ref{fig:rq1} shows how we proceed in question 1. In the investigation, we first choose subjects of MOOCs to focus on and collect corresponding course materials. Human annotators are then recruited to label conceptual relations among learning objects in these materials. After that, we design an interface to present material content along with linking to users. With the interface, user studies are conducted to observe how learners accomplish learning tasks with different strategies of material presentation. We recruit workers from Amazon Mechanical Turk (Turkers) \cite{23} to realize the study on a massive scale with reasonable cost. Task results are analyzed to explore how linking can affect learning.

As for the second question, our investigation is summarized in Fig.~\ref{fig:rq2}. We adopt a similar pipeline to approach this question, except that we replace human annotators with a machine learning algorithm to label the linking. With the automation, implementation of \textit{educational content linking} can be scalable. 

\section{Contributions}
The primary contributions of this thesis can be summarized as follows:
\begin{itemize}
  \item Proposed a framework of courseware presentation that allows learners to navigate much more easily. We have found that learners, especially novices, can find desired information faster without sacrificing accuracy, and can retain concepts more readily with our proposed approach. This framework can also be easily integrated with different pedagogies to further improve learning.
  \item Developed an end-to-end study with Turkers to explore effects of proposed framework on learning. The pipeline is a practical solution to investigate various pedagogies on a massive scale.
  \item Proposed a method based on machine learning and human language technologies, or HLTs, to discover linking automatically. We showed that realizing \textit{educational content linking} can be scalable at least for STEM (Science, Technology, Engineering and Math) courses. Results suggested that learners still benefit from linking labeled automatically, with a slightly smaller improvement than with the handcrafted system.
\end{itemize}

\section{Thesis overview}
The remainder of this thesis is organized as follows:
\begin{itemize}
  \item Chapter 2 lays the groundwork for \textit{educational content linking} by covering related research in education and HLTs. It also provides descriptions about the MOOCs and course materials used in this thesis.
  \item Chapter 3 describes the detail on how we approach the first research question: can linking help learning? We discuss the annotation of linking, the implementation of an interface which presents course content and conceptual relation, the conducting of user study, and the results.
  \item Chapter 4 presents an automatic linking method based on machine learning and HLTs. By analyzing how linking labeled with this method affects learning, this chapter investigates the second research question: can linking be done at scale?
  \item Chapter 5 reviews the experiments and contributions of this thesis, and proposes directions for future research.
\end{itemize}

\chapter{Background}

This chapter gives background concepts of the three main building blocks in this thesis: \textbf{learning science} to motivate the entire framework as well as supervise the system design and learning interface implementation; \textbf{crowdsourcing and learnersourcing} to recruit participants and effort at scale; \textbf{machine learning and understanding} to fuel the automation of the system. We review related literature and offer background material in these four domains. Additionally, a description of course materials used in experiments throughout this thesis will also be provided.

\section{Learning science}
For many years education practitioners and researchers endeavor to discover better ways of learning from a variety of aspects \cite{18, 29, 30}. Researchers try to unveil the mechanism of learning, knowledge acquisition, and long-term memory establishment from cognitive science and psychology; practitioners design theory-grounded and evidence-based approaches in their classes to improve student performance. The mental state of learners and its effect on learning is one of the most discussed topics. Constructive struggle shows positive impact on learning performance by keeping learners in a mental state of boredom and confusion alternatively \cite{19, 20, 21}. Jean Piaget proposed a theory describing how cognitive disequilibrium, such as confusion, can drive a human to develop new knowledge schema or rebuild an existing one, i.e., motivate the process of learning \cite{55}. However, without being properly resolved in time, confusion can lead to frustration or even dropping out \cite{21, 37}. Active learning emphasizes the engaging of learners in discussion, high-order thinking, problem solving, or peer teaching \cite{31}; it has demonstrated a positive impact on learning outcomes by increasing enthusiasm in students and maintaining their interest in the course \cite{32, 33}. Cognitive load theory suggests that a complicated learning environment can overwhelm limited working memory of a human, cause distraction or frustration, and is detrimental to learning outcomes \cite{34, 35, 36}. Although these theories seem contradictory at first glance, all of them imply the importance of balancing between challenging learners with confusion and easing their load with a proper response. 

\subsection{Tutoring at scale}
Due to the delicacy of the learning mechanism, one-to-one tutoring, which is the model where learners can receive maximum attention from teachers, has set a benchmark in education that is hard to match \cite{16}. However, a one-to-one model is cost prohibitive. In order to provide quality education to each and every learner, the idea of intelligent tutoring systems (ITSs) has been proposed \cite{17, 38, 39}. ITS is a computer system that provides immediate feedback or hints to students based on their current learning states. For example, when asked to write a piece of code solving "square root of a number x" with guess-and-check algorithm, in an ITS students can first choose strategy "start with a guess, g". After implementing corresponding code, students can choose following strategy such as "check whether g times g equals x", "claim g as the answer", or "make a new guess". The system gives feedback such as congratulating learners, asking to try again, or providing hints either on each step or waiting until the students have submitted solutions. The example here shows an ITS applied in a problem-solving task. Actually, the framework can be implemented for different tasks to assist students in different learning stages \cite{40, 41}. Since the tutoring is based on a computer, ITS can help many more learners at the same time.

Although ITS has been shown to be effective on improving learning outcomes at much larger scale than one-to-one tutoring \cite{17}, authoring such a system takes a lot of effort \cite{38}. A great tutor is made of an abundance of knowledge derived from years of his/her experience in teaching. Thus, codifying the knowledge and designing instruction strategies in the system, e.g., deciding what exactly the feedback to the learner is, is an extremely complex task. While solutions involving automated methods such as machine learning exist for some components in the system \cite{42, 43, 106}, the state-of-the-art artificial intelligence techniques are insufficient in solving the entire problem. Thus, it is usually done by handcrafted rules to codify instruction strategies. Because of the effort that has to be taken, authoring an ITS from scratch is still expensive.

Because of the demand for human input, a peer-to-peer model is proposed for scalability. Learnersourcing demonstrates how learners can collectively contribute to improving learning material and interfaces for future learners, and engage in a meaningful learning activity simultaneously \cite{22}. Mitros and Sun presented a similar framework that allows a community of students and instructors to jointly create and polish tutoring resources around a shared skeleton \cite{46}. Glassman and others demonstrated that learners can work collaboratively, generating rich problem solving hints and strategies \cite{45}, as well as designing complex assessment questions \cite{44}. By automatically ranking submissions of a coding problem based on stylistic mastery from novice to experts, AutoStyle can provide students the "just a little better" submissions from others to improve their coding style incrementally \cite{47}. The model of peer grading is another frequently applied strategy to offer learners feedback at scale with minimal instructor input \cite{48, 49}. In addition to receiving knowledge and feedback passively, learners can also take the initiative and seek help from communities in a course forum \cite{1} or even a question-answering (Q\&A) site such as Stack Overflow \cite{50}. In this peer-to-peer model, tutoring resources are created by employing the wisdom from a massive learner body, and thus the required efforts from instructors or experts are greatly reduced. Furthermore, the opportunity of reflecting on others' confusions and preventing the curse of knowledge are the other two pluses \cite{45}. The former allows learners to revisit and rethink their understanding, and the latter bridges the gap between learners and instructors, who sometimes cannot put themselves in students' shoes \cite{51}.

However, content created with this model usually suffers from information overload and chaos \cite{56}. For example, a discussion board in a MOOC may have thousands of users and thousands of simultaneous threads, with great response time and quality. But for a learner who is three days behind in the course schedule, it is already impossible to follow existing discussions \cite{57}. The peer-created materials are overwhelming and cause confusion. Inspired by the requirement of helping students receive suitable responses from an exploding amount of learning content, researchers have begun to explore a scalable means for organizing peer-generated content. Asking peers to tag content they generated is one frequently adopted strategy \cite{54}, but sometimes criticized for the lack of accuracy and consistency \cite{58}. Wise et al. introduce an automated algorithm to identify forum posts that are related to course topics \cite{56}. The detection of structure in discussion threads with natural language understanding is also investigated \cite{52, 53, 142, 143, 144}.

This thesis proposes a framework of responding to learners' confusion with well-organized learning content. In this framework, linking among content is discovered automatically and visualized when learners seek help. We aim to help learners resolve confusion by providing guidance for content navigation. Content generated by instructors and peers are both used, which illustrates the generalizability of our method. A user study is also explored to provide evidence of benefit in learning.

\subsection{Course navigation}
In this thesis, we introduce a method of automatically organizing learning content as well as the resolution of learners' confusion with guidance for navigating content. The importance of guided instruction in teaching is discussed in detail by Kirschner et al. from aspects of human cognitive structure and the expert-novice difference \cite{59}. Furthermore, due to the distant nature of MOOCs and online learning, learners usually depend on self-regulated learning to resolve their own learning needs, and whether the self-regulated learning can be achieved is highly correlated to the efficiency of finding desired learning materials. Hence, there is a rich thread of research in providing guidance for navigating online learning content.

Kim demonstrates how to extract structure from learning videos with learners' collective video interaction and annotation data \cite{22}. With the possibility of non-linear navigation of videos, which is empowered by the extracted structure, learners reported a better learning experience. The LinkedUp project aims at linking open education resources through the use of Uniform Resource Identifier (URI) and Resource Description Framework (RDF) to improve access of content \cite{60}. In Adaptive Educational Hypermedia, materials are organized using a concept map diagram \cite{61, 62}. Study navigator supports the simultaneous access to multiple textbook sections, one for the current concept to learn, and the rest for background knowledge \cite{63}. The alignment between textbook and lecture videos \cite{64}, and the restructuring of encyclopedic resources \cite{65} are also proposed for better navigation. This thesis offers an end-to-end study in content organization and navigation, from the idea of linking and the algorithmic method, to the visualization of relationships and user study.

\section{Crowdsourcing}
In the previous section we discussed the peer-to-peer model of tutoring. This model is actually an application of crowdsourcing. A typical crowdsourcing system relies on crowd workers recruited from the Web (e.g., workers on Amazon Mechanical Turk \cite{23}) to provide human computation for complicated parts (usually the parts that cannot be easily solved with a computer) in the system. By taking advantage of the large-scaled online community, huge problems can be divided and solved at much lower costs.

Wikipedia is one of the most compelling examples of crowdsourcing. This project of recording all human knowledge in the form of an online encyclopedia solicits contributions from anyone with an Internet connection. Since its launch in 2001, its repository now accumulates over 5.2 million articles with comprehensive topic coverage \cite{67}. Games with a purpose (GWAP) proposes the idea of embedding work into games \cite{68}. Researchers disguise a computation problem as an online game. While people play the game, they are actually serving as processors in a giant distributed system and solving the problem without consciously doing so. The ESP game is one of the earliest successes in GWAP \cite{69}. In the ESP game, two players are shown the same picture and they have to independently label the picture with words. Players can earn scores when their labels are matched and the goal of the game is to maximize earned scores within a fixed period of time. The real computation problem behind the game is labeling images with natural language, and players generated annotations for almost 300,000 images in its first four-month period of deployment. Other examples of crowdsourcing projects include translation (e.g., MIT OpenCourseWare [\texttt{http://ocw.mit.edu/courses/translated-courses/}], or talks in TED conferences [\texttt{http://www.ted.com/participate/translate}]), helping scientific discovery (e.g., Foldit [\texttt{https://fold.it/}]), or public health (e.g., Food Source Information [\texttt{http://fsi.colostate.edu/}]). These projects are driven by noble goals (such as the public good) or offering personal benefits (such as fun). These motivations attract a large number of people on the Web and make recruitment of the crowd possible. 

This thesis contributes to this line of work through two crowdsourcing applications: we utilize learning content generated by peers (i.e., course forum) for confusion resolution and recruit online workers as subjects in experiments. In the former application, the incentive for learners to contribute is that their work can not only help their peers, but also themselves and future learners. As for the latter application, workers are partially motivated by the opportunity to learn from MOOC materials. 

\subsection{Micropayment workforces}
However, not every project has a goal that can attract the general public to contribute, and it is usually difficult to design a win-win condition for both researchers and the crowd. A more general approach is to pay the crowd to complete tasks, and there are many online crowdsourcing platforms offering services of matching and payment between task requesters and anonymous online workers. These platforms include Amazon Mechanical Turk (AMT) \cite{23}, CrowdFlower [\texttt{https://www.crowdflower.com/}], and InnoCentive [\texttt{https://www.innocentive.com/}].

These platforms are widely used among data scientists in academia and industry to access the online workers for a variety of tasks. McGraw demonstrates an organic automatic speech recognition systems trained upon spoken utterances collected from the crowd \cite{70}. PlateMate collects object tagging and natural language description for food photos from paid online workers, and allows users to upload photos of their meals and get information about the food intake, composition, and nutrition \cite{71}. Callison-Burch presents a crowdsourcing workflow to evaluate quality of a machine translating system \cite{81}. There is much other research examining the usefulness of these paid crowds for collecting, annotating, enriching, and evaluating data, including collecting spoken caption of images \cite{77}, annotating intention in user-generated content \cite{78}, real-time captioning of spoken content \cite{79}, and user interface evaluation \cite{80}.

In this thesis, we utilize the paid online workers recruited on AMT as experimental subjects in two research domains: natural language data annotation and user study in education research. For the first domain we design workflows in which workers have to understand natural language content in learning objects and label relations among these objects; for the second domain, we design tasks meaningful in learning for workers to complete and measure workers' performance. By providing monetary incentives to the crowd, we are able to complete experiments at a much faster rate.

\subsection{Quality control}
Quality is the most criticized issue of crowdsourcing. Because of the variance in workers' expertise, level of skills, effort, and personal bias, crowdsourcing usually yields noisier results than a conventional paradigm \cite{72}. Furthermore, the geographically disparate nature of crowdsourcing makes it more difficult to communicate the task guideline and keep workers on consistent procedural executions than in a controlled environment such as a laboratory. Hence, there is a rich thread of research in studying how to obtain satisfactory results with crowdsourcing.

According to Allahbakhsh et al., these approaches for quality control can be categorized into two general types, design time and run time \cite{73}. The most common design time approach is filtering workers based on their profile, such as their previous task acceptance rate, their IP address as an estimator of their first language, or performance in a qualification test. The profile filtering is supported in most crowdsourcing platforms. Another design time approach is effective task preparation. This approach investigates how to improve quality through different worker incentives as well as better task description, workflow, and interface. Mason and Watts found intrinsic incentive such as enjoyable tasks has a more positive effect on the result than extrinsic incentive such as monetary rewards \cite{74}. Learnersourcing described above is one of the best examples to offer workers intrinsic incentives \cite{22, 44, 45, 46}. For a better workflow, CrowdForge proposes a framework to divide complex problems into micro-tasks \cite{75}. Since workers on crowdsourcing platforms are more familiar with simple and independent tasks, this dividing strategy has a positive impact on the results. Chen et al. also discuss in detail the importance of clear task description (e.g., the experimental goal, who is eligible, how the result will be reviewed, and the reward strategy) for the quality \cite{76}.

Run time approaches are another type of quality control strategy. The most common way to do so is that experts review the results, and decide which ones are not qualified and should be rejected. This review mechanism is supported in most crowdsourcing platforms today. Another common approach is majority consensus. By introducing redundancy and overlapping in task assignment, majority voting can be employed to decide the real results. Karger et al. introduce a probabilistic approach to model the noisy answers from workers and improve quality \cite{82}. There also exist studies that redesign the workflow to control quality on the fly. Lee and Glass demonstrate a multi-stage speech transcription system \cite{83}. In this system, after each stage of transcription a machine-learning-based low quality detector is trained to filter spammers and provide instantaneous feedback to workers. Many studies have reported that, with proper quality control, crowdsourcing can yield good or near expert-level task results \cite{83, 84, 85}.

This thesis adopts a wide range of quality control approaches to improve reliability of online workers, including majority consensus, expert review, as well as clear task description, workflow, and interface. Moreover, in addition to monetary incentives, our tasks also provide the opportunity to learn from MOOC materials.

\section{Machine learning and human language understanding}

The crowd can solve many complex computational problems at reasonable costs. However, it will be more efficient if we can solve one problem and apply the solution to other similar problems. This is made possible by the recent progress in machine learning \cite{90}. Given data that records computational problems and usually the corresponding solutions provided by a human, research in machine learning explores algorithms or models that can summarize regularities and patterns in the data, and solve relevant but unseen problems with discovered regularities. Thus, with machine learning we can build a model from data annotated by a human (either a trained data scientist or na\"{\i}ve online workers), and apply the model to solve future in-domain problems automatically.

\subsection{Human language technology}
Machine learning is one of the most active research fields in computer science nowadays, and it has extremely diverse applications: stock market prediction \cite{86}, credit card fraud detection \cite{87}, medical diagnosis \cite{88}, and intelligent robotics \cite{89} to name just a few. Among these applications, human language technology (HLT) is one of the domains receiving the most attention.

Human language is pervasive in our daily life, and it is one of the most crucial means for communication and information exchange. Since human language is ubiquitous, there is a rich thread of research concerning HLT, investigating the producing and understanding of human language as well as attempting to improve human-to-human and human-to-machine communication. HLT is an interdisciplinary field that includes natural language and speech processing, computational linguistics, statistics, and psychology. Due to the recent progress in machine learning, there is a trend of applying machine learning techniques to solve HLT problems. For example, Liu investigated machine-learning-based approaches to facilitate the access of rich user-generated content online \cite{91}. Shahaf et al. propose an algorithm to glue pieces of information scattered in various news articles, and create a structured summary for the entire story \cite{92}. Other applications of machine learning in HLT include information retrieval \cite{93}, automatic speech recognition (ASR) \cite{94, 145, 146, 147, 148}, semantic tagging \cite{95}, topic modeling \cite{96}, and automatic question answering \cite{97}.

Since human language is an integral part of education for knowledge transferring, there is also research studying how to improve communication between learners and instructors by understanding the natural language content in learning materials with the aid of machine learning. Glass et al. demonstrated the MIT Lecture Browser. By automatically transcribing speech in lecture videos with ASR techniques, learners can easily browse through the text and identify topics they are interested in more efficiently \cite{98}. On top of the transcribed text, in the FAU Video Lecture Browser, key phrases are also extracted, ranked, and presented along with aligned lecture video. By clicking each key phrase, learners can access corresponding video vignettes for detailed discussion \cite{99}. Without transcribing speech to text beforehand, a method matching spoken search queries to lecture speech directly on audio is also proposed to improve video navigation \cite{100}. Fujii et al. further presented an algorithm to automatically summarize course lectures; thus learners can get the big picture behind each lecture without watching the video from beginning to end \cite{101}.

Beyond the lectures, there also exist studies in understanding of textbook and course forums with HLT and machine learning, since there is an abundance of natural language in these materials. Lin et al. proposed a method to classify genres of discussion threads for improving accessibility of forums \cite{102}. A similar idea is applied to identify questions and potential answers in discussion boards \cite{103}. Li et al. demonstrated how to build a semantic forum that allows semantic search, relational navigation, and recommendation with HLT \cite{107}. An automatic approach to discover relevance among textbook sections was also investigated \cite{63}. Due to the popularity of MOOCs, recent research has begun using HLT in understanding MOOC materials, including intention classification and topic modeling for forum posts \cite{53, 56, 104, 105}, textbook section recommendation for lecture videos \cite{64}, and automated essay scoring \cite{108}.

These research strategies demonstrate that, with HLT, the machine can learn to understand course materials as well as assist the information exchange among teachers, students, and materials. Due to these advantages and the nature of MOOCs (i.e., size of the audience and physical distance among them), HLT can play a crucial role in improving the learning experience and performance in online learning. In this thesis we introduce an HLT-based method to automatically discover relations among various types of MOOC materials, and show its benefit in learning.

\subsection{Conditional random fields (CRF)}
In this thesis we adopt conditional random fields (CRF) to model the relations among learning objects. CRF is an instance of graphical models \cite{109}, which is a graph designed to model the conditional dependence structure among random variables (a random variable is usually used to express the observation in data samples and the hidden classes these samples belong to). The training and inference of CRF is well studied in the machine learning field. Therefore, it is widely used in learning temporal dependence from sequential data, such as speech, text, image, and bioinformatics \cite{95, 110, 112}. Since most course materials can be expressed with sequential structure, we believe the CRF is a perfect match to our problem. In the following we will introduce the mathematical definition, the training, and the inference of CRF.

A general CRF can be defined as follows: given \textit{Y} as the set of unobserved variables, and \textit{X} as the set of observed ones, let G be a factor graph over \textit{X} and \textit{Y}. If, for any \textbf{x}, the conditional probability \textit{p}(\textbf{y}|\textbf{x}) can be factorized according to G, then (\textit{X}, \textit{Y}) is a conditional random field \cite{109}. That is,
\begin{equation} \label{eq:2_1}
p(\textbf{y}|\textbf{x}) = \frac{1}{\textrm{Z}(\textbf{x})}\prod_{a=1}^{\mathrm{A}}\Psi_a(\textbf{y}_a, \textbf{x}_a)
\end{equation}
where \textbf{y} is a vector denoting the assignment to \textit{Y}, \textbf{x} denoting the assignment to \textit{X}, {$\Psi_a$} is the set of factors in G, \textit{a} is the index of factors, and Z(\textbf{x}) is the normalization term.

\begin{equation} \label{eq:2_2}
\textrm{Z}(\textbf{x})=\sum_{\textbf{y}}\prod_{a=1}^{\mathrm{A}}\Psi_a(\textbf{y}_a,\textbf{x}_a).
\end{equation}

\begin{figure}[t]
\includegraphics[width=14cm]{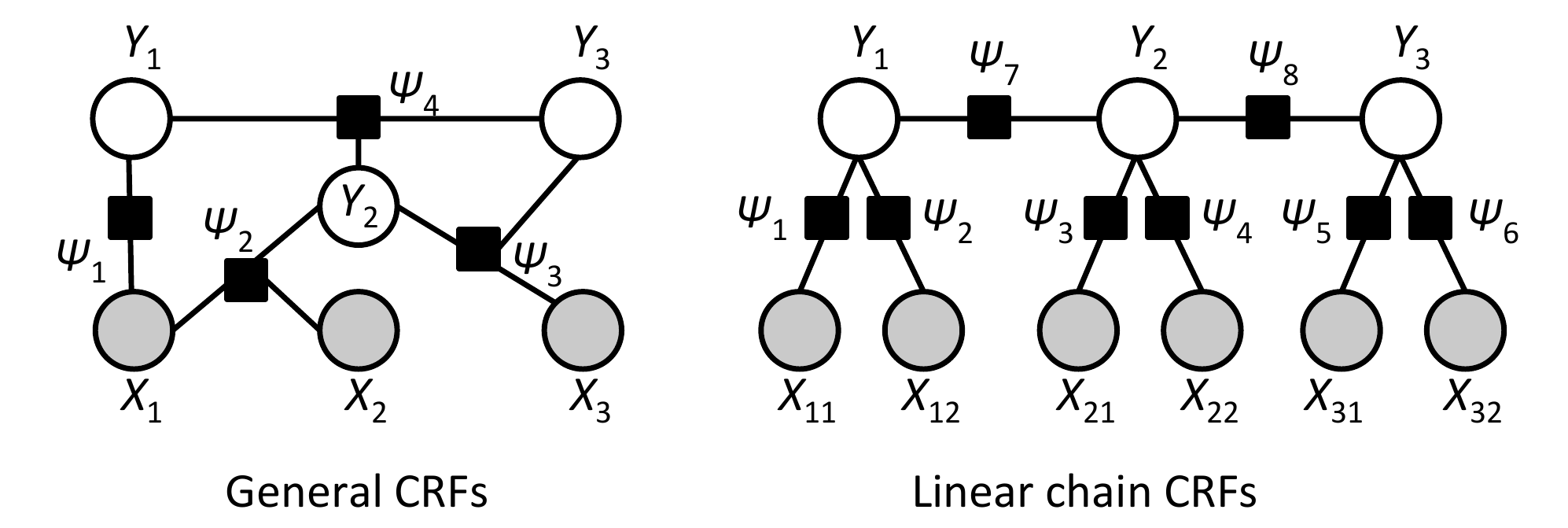}
\centering
\caption{Diagram of general CRFs and linear chain CRFs.}  
\label{fig:crf_diagram}
\end{figure}

Each factor $\Psi_a$ is a function of $\textbf{y}_a$ and $\textbf{x}_a$, which are subsets of the unobserved and observed variables respectively (i.e., $\textbf{y}_a \subseteq Y$ and $\textbf{x}_a \subseteq X$). The value of $\Psi_a$ is a non-negative scalar, which can be interpreted as a measure of how compatible this subset of assignment $\textbf{y}_a$ to the unobserved variables is with its dependent observations $\textbf{x}_a$. An example of a general CRF and its corresponding factor graph is shown on the left panel of Fig.~\ref{fig:crf_diagram}. In this graph, $\Psi_1$ depends on $X_1$ and $Y_1$, and $\Psi_2$ depends on $Y_2$, $X_1$, and $X_2$ for instance. Since there is no constraint to the underlying factor graph of CRF, we can see it is flexible in expressing various structures among data. 

With equation~\ref{eq:2_1}, inferring labels (i.e., unobserved variables) from observations can be modeled with a maximization problem: finding the label assignment \textbf{y} which maximizes the conditional probability $p(\textbf{y}|\textbf{x})$ given the observations \textbf{x}. However, solving this maximization problem in general CRFs is intractable \cite{109}. There are two usually adopted approaches to obtaining feasible solution. First, if we limit the underlying factor graphs of CRFs to several special cases, e.g., a chain or a tree, the exact inference can be solved in polynomial time. On the other hand, several algorithms can be used to obtain approximate inferences, e.g., Markov chain Monte Carlo sampler \cite{110}, and loopy belief propagation \cite{111}. Since a linear sequence is the most common and dominant structure for arranging topics in course materials, we choose the linear chain CRF in this thesis and design our algorithm based on it. Another benefit of a linear chain architecture is that it reduces the model complexity and mitigates overfitting. This is crucial, especially when annotated training data is hard to obtain, such as in our problems. 

In the right panel of Fig.~\ref{fig:crf_diagram}, we show an example of linear chain CRFs. Similar to general CRFs, $X$ and $Y$ also represent the observed and unobserved variables respectively, except that $Y_t$ are structured to form a chain. This chain structure adds a constraint to the probabilistic dependence expressed by the model that an unobserved variable $Y_t$ can directly depend on only the single previous unobserved variable $Y_{t-1}$ and several observations $\textbf{x}_t=\{X_{ts}\}_{s=1}^{N(Y_t)}$. Here $N(Y_t)$ denotes the number of observed variables depending on $Y_t$. 

With the linear chain structure, the conditional probability $p(\textbf{y}|\textbf{x})$ can be rewritten as following equation
\begin{equation} \label{eq:2_3}
p(\textbf{y}|\textbf{x})=\frac{1}{\textrm{Z}(\textbf{x})}\prod_{t=1}^{\mathrm{T}}\Psi_t(Y_t,Y_{t-1},\textbf{x})
\end{equation}
by replacing {$\Psi_a$}, the set of factors in G, with {$\Psi_t$}. Each factor $\Psi_t$ is a function of $Y_t$, $Y_{t-1}$ and $\textbf{x}$, and these factors represent the linear-chain factor graph. In real application $\Psi_t$ is usually set as the following form $\Psi_t(Y_t,Y_{t-1},\textbf{x})=\exp\{\sum_{k=1}^{\mathrm{K}}\theta_k f_k (Y_t,Y_{t-1},\textbf{x})\}$, and Equation~\ref{eq:2_3} is rewritten as 
\begin{equation} \label{eq:2_4}
p(\textbf{y}|\textbf{x})=\frac{1}{\textrm{Z}(\textbf{x})}\prod_{t=1}^{\mathrm{T}}\exp\{\sum_{k=1}^{\mathrm{K}}\theta_k f_k (Y_t,Y_{t-1},\textbf{x})\}.
\end{equation}
Here $f_k (Y_t,Y_{t-1},\textbf{x})$ is a feature function that researchers need to design based on domain knowledge, and $\theta = \{\theta_k\}_{k=1}^{\mathrm{K}}$ is the parameter set that has to be learned from training data. 
This chain structure is called the Markov property, which assumes the modeled stochastic process is memoryless, i.e., the prediction to the current unobserved variable depends only on the prediction to the previous one, and no other earlier prediction. Another popular model that assumes this property to hold is the hidden Markov models (HMM), and in fact this linear chain CRF can be interpreted as a generalized HMM, where the factor function $\Psi$ does not need to have a probabilistic interpretation as HMM does. With the memoryless property, the inference problem of linear chain CRF (as well as HMM) can be solved efficiently with a dynamic-programming algorithm \cite{109}.

In addition to the inference problem, another issue of applying CRFs to real tasks is parameter estimation, or training. The maximum likelihood criterion is typically used for estimating parameters: given the fully labeled training data $C=\{\textbf{x}^{(i)},\textbf{y}^{(i)}\}_{i=1}^\mathrm{N}$, where $(\textbf{x}^{(i)}, \textbf{y}^{(i)})$ is the $i$-th sample in the data, $\textbf{x}^{(i)}=(\textbf{x}_t^{(i)})_{t=1}^\mathrm{T}$ is a sequence of observations, and $\textbf{y}^{(i)}=(Y_t^{(i)})_{t=1}^\mathrm{T}$ is a sequence of labels corresponding to $x^{(i)}$, we estimate the model parameter $\theta$ with the maximum likelihood estimator $\hat{\theta}=\mathrm{arg⁡max}_{\theta}⁡ l(\theta)$. $l(\theta)$ is the objective function and equals $\sum_{i=1}^{\mathrm{N}} \mathrm{log}⁡p(\textbf{y}^{(i)}|\textbf{x}^{(i)};\theta)$ with $p(\textbf{y}^{(i)}|\textbf{x}^{(i)};\theta)$ as defined in Equation~\ref{eq:2_4}. However, in general the estimator does not have an analytic form. Therefore, a gradient ascent approach is adopted to obtain an approximate solution to this optimization problem (other approaches also exist but gradient ascent is most commonly used in practice). The algorithm for gradient ascent can be summarized as follows:

\begin{algorithm}
\caption{Gradient ascent algorithm}\label{GradAsc}
\begin{algorithmic}[1]
\State Randomly initialize the parameter set $\theta$
\State \textbf{repeat}
\State \hskip\algorithmicindent Compute the gradient of the objective function, $\nabla l(\theta)$
\State \begin{varwidth}[t]{\linewidth}
      \hskip\algorithmicindent Update the parameter set $\theta$ according to pre-defined learning rate $\rho$\par
        \hskip\algorithmicindent $\theta$ $\coloneqq$ $\theta$ + $\rho \nabla l(\theta)$ \par
        \hskip\algorithmicindent
      \end{varwidth}
\State \textbf{until} convergence criterion is achieved.
\end{algorithmic}
\end{algorithm}

This algorithm updates the estimated parameters along the direction where the objective function is increased most at each step. When the convergence criterion (e.g., the difference of estimation in two consecutive iterations is less than the pre-defined threshold) is achieved, the estimation is the trained model parameters. There are many variations of this algorithm, such as Newton's method, BFGS, and conjugate gradient \cite{109}. These variations attempt to improve convergence speed with different techniques but share the same compute-gradient-and-update idea.

These training and inference techniques are widely used in various applications of linear chain CRF. In Chapter 4, we will discuss how to apply the general model in our problem of linking discovery.

\subsection{Word embedding}
In order to apply statistical models to natural language content, we have to represent content in a form that the model accepts, i.e., numeric vectors. Here we give an introduction to the vector representations employed in this thesis.

The first representation is unigram embedding, or Bag of Word (BoW) embedding. In this simple embedding, a document is represented as a vector [N($w_1$), N($w_2$), ..., N($w_{|V|}$)], where $w_i$ is the $i$-th word in vocabulary V, and N($w_i$) is the score of $w_i$ in this document. The score can be the number of occurrences, the word frequency, or term frequency-inverse document frequency (TF-IDF) \cite{113}. The upside of unigram embedding is that this method is intuitive and easy to train. However, since each word is represented as an atomic unit in the vector and different words are encoded independently, the long-range lexical dependency, such as the context of a word, is missing in this representation.

To make the shallow and local representation embed lexical dependency in a longer range, we can adopt an $n$-gram model, which is an extension of unigram embedding, and each element in the vector represents a combination of $n$ words instead of a single word. Nonetheless, this model provides only limited added value. An $n$-gram model greatly increases the dimension of vector representation, since it exhaustively enumerates all possible combinations of $n$ words. Due to the curse of dimensionality, in practice we can only use a small $n$ in order not to overfit, especially when the size of training data is limited. Thus, the range of dependency this method can encode is still restricted.

We turn to word2vec embedding for our second representation with longer lexical dependency \cite{118}. Word2vec is a two-layer neural network. Its input is a text corpus and its output is the vector representation for each word in the corpus. As compared to the $n$-gram and unigram method, word2vec is a continuous language model, which means that each word is represented as a continuous vector. The upside of this representation is its capability of encoding semantic and syntactic dependencies among words. In the $n$-gram or unigram method, each word or combination of words is represented with an independent element in the vector, and the relations among words cannot be encoded efficiently; in word2vec, the neural network model is designed to discover and represent semantic and syntactic dependencies from patterns from words' context. For instance, based on the word2vec model trained on millions of Wikipedia pages, $v_{\mathrm{King}}-v_{\mathrm{Man}}+v_{\mathrm{Woman}} \approx v_{\mathrm{Queen}}$, and $v_{\mathrm{Apples}}-v_{\mathrm{Apple}}+v_{\mathrm{Car}} \approx v_{\mathrm{Cars}}$, where $v_i$ denotes the word2vec representation of word $i$. With word2vec embedding, the document representation can be simply obtained by averaging word vectors over the entire document. 

\begin{figure}[t]
\includegraphics[width=12cm]{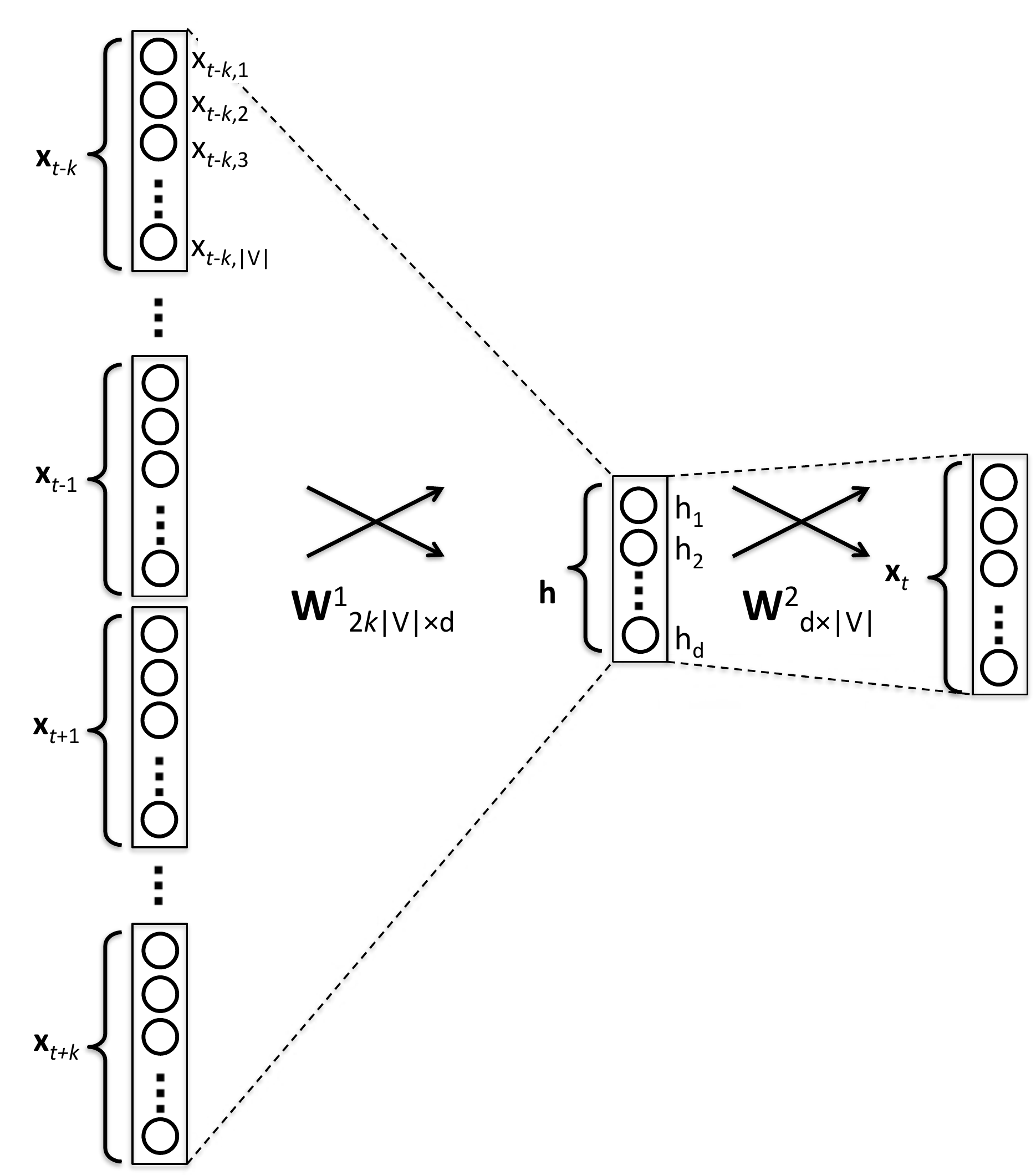}
\centering
\caption{Architecture of feedforward neural network employed to obtain word2vec embedding.}  
\label{fig:word2vec_diagram}
\end{figure}

The word2vec embedding is trained using a feedforward neural network model with architecture shown in Fig.~\ref{fig:word2vec_diagram} \cite{119}. In the figure, $\textbf{x}_t$ is an one-hot vector with its $i$-th element equal to $\delta(w_t=v_i)$; $\textbf{W}^1$ and $\textbf{W}^2$ are matrices of weights to be learned from a corpus; \textbf{h} is a vector of hidden layer projection obtained by transforming the hidden layer input with the sigmoid function $\sigma$; $k$ is the hyper-parameter deciding the size of context for this model to learn from. Here $\delta(∙)$ is an indicator function with $w_t$ as the $t$-th word in corpus and $v_i$ as the $i$-th word in vocabulary. These vectors and parameters are related to each other based on the following equation:
\begin{equation} \label{eq:2_5}
\textbf{x}_t=\textbf{W}^2\textbf{h}=\textbf{W}^2(\sigma(\textbf{W}^1[x_{t-k}^\mathrm{T},x_{t-k+1}^\mathrm{T}, ..., x_{t+k}^\mathrm{T}]^\mathrm{T})).
\end{equation}

This model can be interpreted as a classifier trained to predict a word based on its neighbors, typically using the Stochastic Gradient Descent (SGD) training algorithm. The algorithm is very similar to the one introduced in Section 2.3.2, except that, in each step, we update the estimated parameters $\textbf{W}^1$ and $\textbf{W}^2$ along the direction where the objective function is decreased most, and we use cross entropy as the objective function. After the training, the vector representation of $w_t$ is $\textbf{W}^1\textbf{x}_t$. In this way, the neural network can encode the long-range semantic and syntactic dependencies in vectors by discovering patterns from the context of words.

In fact, there are other common approaches to represent text documents as vectors that encode high-level lexical and semantic information: doc2vec \cite{114} and topic modeling \cite{115, 116, 117}. Doc2vec is a very similar algorithm to word2vec except that it learns representation for larger blocks of text directly, such as paragraphs or sentences. Topic modeling refers to a family of methods for discovering latent semantic structure and identifying the subsets of words co-occurring more frequently in documents of various "topics". In this thesis we choose not to use these representations. Doc2vec requires too much in-domain data for training. The learned "topics" in topic modeling are too broad for our problem. For instance, with topic modeling we can easily identify \textit{python}, \textit{complexity}, or \textit{object oriented programming} corresponding to topic \textit{computer science}, and \textit{standard deviation}, or \textit{hypothesis testing} belonging to \textit{statistics}. However, when it comes to distinguishing \textit{complexity} from \textit{programming}, since the two concepts are in different lectures, topic modeling usually introduces a lot of noise. Hence, we surmise that doc2vec and topic modeling are not suitable for our problem.

\section{Corpora}
Before starting to implement our proposed framework and investigating the effects on learning, we have to first decide which materials and MOOC subjects our system should be built upon. Today there are over four thousands MOOCs on the Web covering subjects from science to humanities. Types of materials and pedagogies adopted in these MOOCs are diverse. It is impractical to expect an exhaustive exploration of every condition. Thus, in this thesis we make a tradeoff between feasibility of experiments and generalizability of results. In the following we discuss the decisions we make and the rationale.

\subsection{Course subjects}
Experiments in this thesis use two MOOCs: Introduction to Statistics: Descriptive Statistics (Stat2.1x), and Introduction to Computer Science and Programming (6.00x). Stat2.1x was offered by University of California, Berkeley, from February to March in 2013 on edX \cite{24}, and 6.00x was offered by Massachusetts Institute of Technology (MIT), from February to June in 2013 on edX \cite{25}. Stat2.1x is an introduction to fundamental concepts and methods of statistics, which require basic high-school level Mathematics. 6.00x is aimed at undergraduate students with little programming experience, and discusses how to solve real problems with computational approaches and computer programming. Both MOOCs were very successful. Stat2.1x has over 47,000 registrants, and 6.00x has over 72,000 registrants. Due to the popularity of these two MOOCs and the growing interest in STEM education these years, we choose to focus investigation in this thesis on Stat2.1x and 6.00x. The popularity of underlying MOOC subjects makes findings in experiments more influential, representative and likely to be applied to different conditions. Furthermore, our familiarity with the topics is another plus.

In our following investigation, these two MOOCs serve different purposes. We use Stat2.1x for developing minimum viable product \cite{26} and 6.00x for the final evaluation. In system development, minimum viable product is an intermediate stage where a product with a minimum amount of features is built to gather information and user feedback about the product. In this stage, the goal is to validate product ideas from interaction with real users with minimum cost. This provides insights for further system development and greatly reduces risk as compared to implementing all features in the product at once. We believe Stat2.1x can serve this purpose well for two reasons. First, this MOOC is shorter (less than two months) than most of the others, but still contains necessary components and course materials. Therefore we can implement our framework on a complete MOOC more readily, e.g., labeling linking on fewer materials. Second, statistics is familiar and interesting to many, thus making it easier to recruit experimental subjects in our study. For these reasons, we select Stat2.1x for an intermediate validation of the benefit and scalability of \textit{educational content linking}. The role of Stat2.1x can be interpreted as a development set in a machine learning system.

With the validation and feedback, we improve our implementation on 6.00x and evaluate the resulting system in depth. In addition to aiming at answering fundamental questions such as whether linking is beneficial or scalable, we also explore advanced features, such as reproducibility, generalizability, and portability of the framework. We can interpret 6.00x as a test set in machine learning system. Using one MOOC for development and a different MOOC for testing makes the evaluation more credible and less subject to the criticism that we overfit our implementation to a particular MOOC.

There are other benefits of using 6.00x. Since this MOOC was offered by MIT, there are many more resources available to us. We can easily reach course staff and MIT students who have taken the corresponding course on campus for insightful understanding. Moreover, this MOOC and its corresponding residential course have been offered many times on edX and at MIT. These multiple offerings leave room for expanding our survey along various dimensions in the future.

\subsection{Course materials}
Within a MOOC, a wide range of course materials are available to learners, such as lecture videos, lecture slides, labs, textbook, discussion forum, course Wiki, quizzes and exams. Considering the development cost, again we choose a subset from these materials for experiments in this thesis: lecture videos, slides, and textbook for Stat2.1x, and the previous three materials together with discussion forum for 6.00x. There are several reasons for us to make this choice. First, these types of materials are common to many MOOCs nowadays and contain a large portion of learning content. Second, these materials have similar form over different course subjects. This fact makes the experiment easier to reproduce from MOOC to MOOC. In contrast, for example, quizzes and exams have diverse styles, from multiple-choice questions to computational problems to essay writing, and each course subject emphasizes on various styles. Thus it might be challenging to generalize experimental results to a variety of MOOCs. Third, these materials allow us to investigate various types of linking, from linking two types of materials composed by the same instructor that can be aligned in order properly, to linking two materials with very different creators and organization. One example of the first type is linking between lecture videos and slides, and examples for the second one are linking lecture videos to textbook or to discussion forum. We will explain these two types in detail in the next section. For these reasons, we believe our choice of materials can help us obtain generalizable and reproducible experimental results with reasonable cost.

Note that discussion forum is only chosen in the evaluation MOOC (i.e., 6.00x). This is because accessing data with personally identifiable information, such as forum posts, requires lengthy paperwork. This work should not be a part of development of minimum viable product.

In Table \ref{table:corpus_size} we summarize the quantity of these materials. The first column lists the number of video hours, slide pages, textbook sections, and discussion threads. Here we measure the size of the textbook by number of sections rather than pages, since the textbook used in Stat2.1x is a Web-based electronic book, and pages in this book are not properly defined. Furthermore, considering the cost of data annotation, we only used the threads posted under the lecture videos in our experiment; these 1,239 threads are about one tenth of the total posts in 6.00x. The second and third columns show the number of words in the available material and the count of unique words, respectively. Here video transcription is used for computing the number of words.

From this table we observe that the amount of content in 6.00x is much greater than the amount in Stat2.1x. This is another reason why we chose Stat2.1x to develop the minimum viable product. The much smaller corpus means a faster process of establishing linking among the course materials.

\begin{table}[]
\centering
\caption{Summarization of sizes of course materials used in this thesis.}
\label{table:corpus_size}
\begin{tabular}{|l|l|l|l|}
\hline
                  & Sizes        & Words & Vocabulary \\ \hline
\multicolumn{4}{|l|}{Stat2.1x}                        \\ \hline
Lecture video     & 7 hours      &  62k  &  1,743     \\
Lecture slides    & 157 pages    &  11k  &   785     \\
Textbook \cite{27} & 77  sections &  45k  &  1,825      \\ \hline
\multicolumn{4}{|l|}{6.00x}                           \\ \hline
Lecture video     & 21 hours     & 174k  &  3,086     \\
Lecture slides    & 498 pages    &  32k  &  1,952     \\
Textbook \cite{28} & 144 sections &  119k &  4,594     \\
Discussion forum  & 1,239 threads & 236k  & 6,772      \\ \hline
\end{tabular}
\end{table}

\chapter{Would linking help learning?}

This chapter investigates the first research question: if we are able to link course materials with human assistance, would it help learners? In MOOCs most instruction and knowledge acquisition happen with learning content delivery. Thus, in the previous chapter, we surmise that making materials more accessible by linking them together can enhance learning experience and outcomes. For example, when learners are confused at a specific point of the lecture, more accessible materials allow them to find useful content for resolving their confusion more easily. In this chapter, we explore the research question for supporting our theory with empirical evidence. To approach this question, we will provide an end-to-end study investigating the following issues

\begin{itemize}
  \item How to link course content with human assistance?
  \item How to present linking along with content to learners?
  \item How to measure the effect of linking on learning?
  \item Is linking helpful?
\end{itemize}

The study is conducted on two MOOCs: Stat2.1x and 6.00x, which are described in detail in Section 2.4. In the first MOOC we evaluate the idea with minimum input, and in the second MOOC we measure system performance in realistic conditions. In our experiment, we discover that, using human annotation, we can build an interface presenting course material with meaningful interconnection. Our interface is shown to be beneficial in supporting learners to complete their tasks: it allows learners to search for information more efficiently, retain more concepts using the same amount of time, and focus on informative learning content. Moreover, we contribute to the research community by providing a user study pipeline that can be conducted at scale and in a cost-effective way.

\section{Linking materials}

Linking is an abstract and general idea; however to implement a real system based on the idea, a concrete definition is required. Linking refers to the relations among objects and can typically be visualized as a graph diagram, with vertices representing the objects and edges for the linking. However, ordinary people are usually not comfortable interacting with a general graph diagram \cite{120}, since too many possible paths in the graph is confusing and overloads the human cognitive system \cite{59}; in order not to distract learners, most learning content, e.g., lectures, or textbook sections, is aligned in sequence. Therefore, we also limit our linking to a specific trunk-and-leaves architecture. In this section, we will discuss how to link course content with human assistance under this architecture.

\subsection{The Linking tree}

\begin{figure}[t]
\includegraphics[width=4cm]{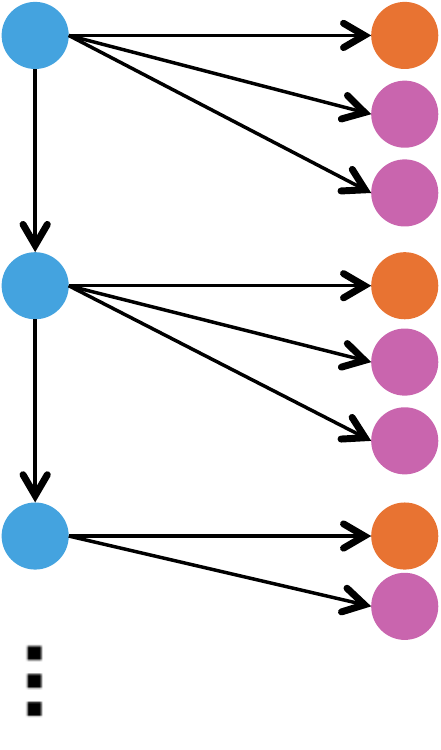}
\centering
\caption{Diagram of the trunk-and-leaves architecture (i.e., the linking tree).}  
\label{fig:trunk_diagram}
\end{figure}

Fig.~\ref{fig:trunk_diagram} illustrates the trunk-and-leaves architecture we limit the linking to, with blue nodes representing the trunk and the rest for the leaves. A node in the diagram represents a learning object, which will be defined in Section 3.1.2. The trunk visualizes the main flow of the courses and shows students a clear learning path they can follow. Each leaf node attaches to one object on the trunk, and represents a supplementary learning object for the corresponding trunk node.

In this thesis, we select lecture video sequences as the trunk of the tree, since most online or residential classes are centered around lecture or lecture videos. In the following, we will discuss how to obtain supplementary objects, i.e., the leaves, for each trunk node with human annotation.

\subsection{Homologous and heterologous linking}
We identify supplementary objects for each node on the trunk by discovering the relation between three pairs of course materials: lecture videos and slides, videos and textbook, as well as videos and discussions. In this thesis, instead of treating the entire video as an atomic element, we discover the relation on the level of the video segment. We surmise that the finer granularity is helpful in visualizing the in-video structure, such as subgoals, subtopics, or meaningful conceptual pieces; the structure improves learning and navigation by summarizing and abstracting low-level details as well as reducing learners' cognitive load \cite{22}. In order to achieve this level of granularity, we define a learning object as a segment of lecture video, a page in lecture slides, a textbook section, or a discussion thread.

\begin{figure}[t]
\includegraphics[width=11cm]{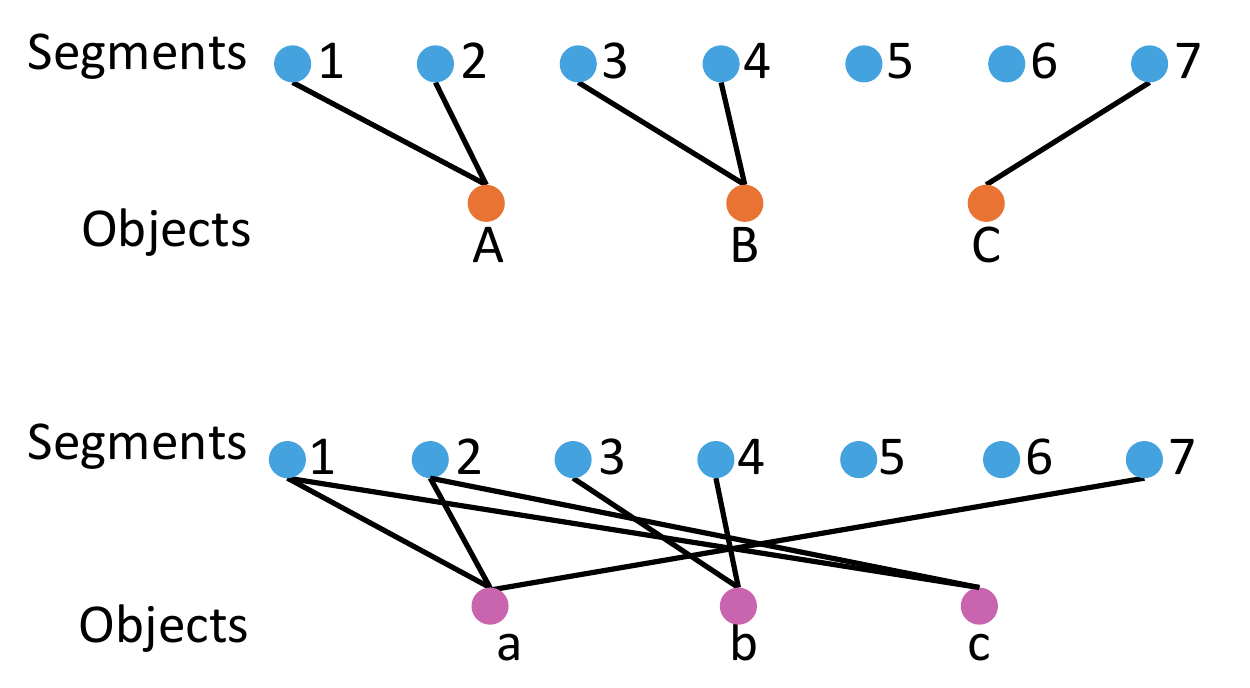}
\centering
\caption{Examples of homologous (upper panel) and heterologous linking (lower panel).}  
\label{fig:linking_type_exp}
\end{figure}

Before describing how to discover relations between materials, we first discuss two types of relations: the homologous and heterologous linking. The reason why we discuss these two types first is because their linking patterns are distinct, and the difference can greatly affect how to discover relations. In Fig.~\ref{fig:linking_type_exp}, we show examples of these two types of linking, with homologous in the upper panel and heterologous in the lower. In the figure two sequences of learning objects from two types of materials along with the relation between each object pair is illustrated. Here objects in the trunk (i.e., segments of lecture videos) are represented as blue nodes, and objects from another material (e.g., pages of slides, textbook sections or discussion threads) are in orange or pink. The indices of objects are also labeled (1 to 7, A to C, and a to c).

As shown in the figure, homologous linking is a many-to-one and monotonic (or order preserving) mapping between two sequences of learning objects. A monotonic mapping satisfies the following attribute:
\begin{equation} \label{eq:1}
\textrm{x}\leq_\alpha\textrm{y}~\textrm{implies}~f(\textrm{x})\leq_\beta f(\textrm{y})~\textrm{if}~f(\textrm{x})\textrm{,}~f(\textrm{y})~\textrm{is not}~\emptyset
\end{equation}
where x and y are learning objects in material $\alpha$, \textit{f}(x) and \textit{f}(y) are the objects in material $\beta$ and linked to x and y respectively, and $\emptyset$ is the empty set. x $\leq_\alpha$ y refers to the case that object x comes before object y in the material sequence $\alpha$; the precedence can be defined by the chapter/section/lecture indices or the thread posted time. Homologous linking mostly exists between two materials authored by the same person, e.g., between video segments and slides, since in this case topic arrangement in different materials usually follows the same ontology.

In contrast, heterologous linking refers to the case when mappings between two object sequences are many-to-many or not order preserving, as the example shown in the lower panel of Fig.~\ref{fig:linking_type_exp}. In this example, the precedence of objects in one sequence is not preserved after mapping these objects to objects in another sequence, and there are many crisscrosses when visualizing the mapping. Heterologous linking usually exists when the underlying two materials come from various authors. This is because the cognitive system in which humans interpret and store knowledge varies from person to person. It is very likely that various authors arrange topics and content in different ways.

The linking between lecture videos and forum discussions, or lecture videos and textbook can usually be classified as a heterologous relationship. For a textbook, its arrangement of chapters and sections can be totally different from the arrangement of lectures in a course. As for posts in a forum, if we sort them by created time, they can also be in a distinct order from the lectures. This is because every learner has various learning progress and pace; hence even at the same point in time, different learners may start discussions about distinct topics.

In fact, instead of a dichotomy, it is more precise to interpret the monotonic property as a spectrum, where the proportion of mapping that violates Equation \ref{eq:1} changes gradually from zero to one. For example, although both video-to-discussion and video-to-textbook linking are not order-preserved, there are usually more crisscrosses in the former. The reason why we choose to simplify the spectrum to two conditions, the homologous and heterologous linking, is because it is not practical to investigate every point on the spectrum. Since the proportion of order-preserved mapping between materials is highly correlated to the complexity of identifying supplementary objects, we design two methods to discover the relations for the two types respectively.

\subsection{Linking representation}
\begin{figure}[t]
\includegraphics[width=15cm]{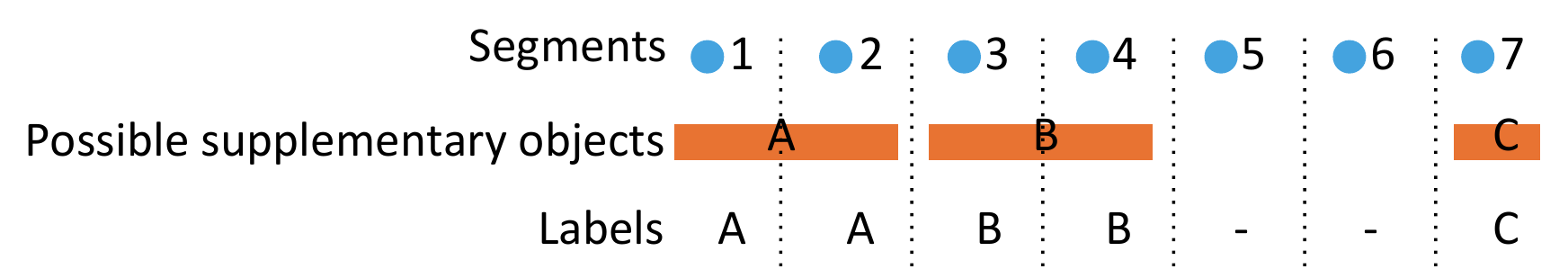}
\centering
\caption{Annotating homologous linking as an alignment problem.}  
\label{fig:homo_anno}
\end{figure}

Since homologous linking is a many-to-one and monotonic mapping, we formulate the relation discovery as an alignment problem. In Fig.~\ref{fig:homo_anno} we show how to annotate homologous linking based on this formulation to represent the relation configuration between materials illustrated in the left panel of Fig.~\ref{fig:linking_type_exp}. Given two sequences of materials, one is the trunk and the other is a set of candidates of the leaves, we identify the non-overlapping, sequential chunk of trunk nodes corresponding to each leaf in order, and label these nodes with the index of the leaf. Here, we define a chunk of trunk nodes as corresponding to a leaf when the former and the latter contains identical discussion to a concept. In addition, since the video transcription sentence is the only unit that can be obtained easily and is at a finer granularity than the entire video, we choose one sentence as a video segment (i.e., a node on the trunk).

\begin{figure}[t]
\begin{adjustwidth}{-3cm}{-3cm}
\begin{center}
\includegraphics[width=17cm]{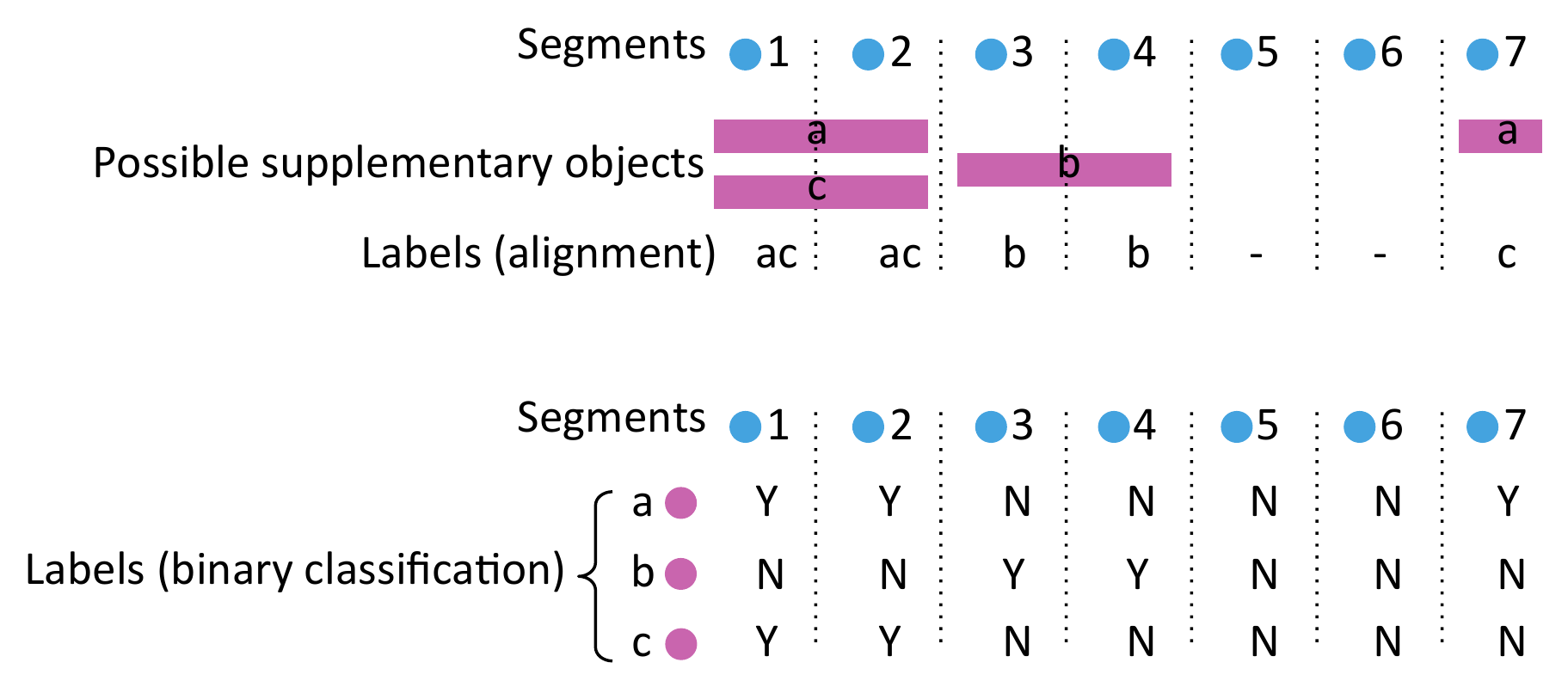}
\end{center}
\end{adjustwidth}
\caption{Annotating heterologous linking as an alignment problem (upper panel) or a binary classification problem (lower panel).}  
\label{fig:hetero_anno}
\end{figure}

We can adopt the same formulation for heterologous linking, as shown in the upper panel of Fig.~\ref{fig:hetero_anno}. However, in this case, since the aligned chunk of trunk nodes does not have to be sequential, and the chunks for different leaves could overlap, identifying these chunks is much more complicated than in the homologous case. Furthermore, the possibility of one video segment aligned to multiple leaves makes the problem become a multi-label classification one, which increases the complexity of designing an automated method to infer the relation. Consequently, we propose another formulation for heterologous linking, as shown in the lower panel of Fig.~\ref{fig:hetero_anno}.

In this alternative formulation, we divide the entire relation identification problem into several sub-problems by considering each leaf independently. That is, in each sub-problem, our goal is to discover the relation between the sequences of trunk nodes and a separate leaf. Each sub-problem can be interpreted as a binary classification task, where every trunk node is classified as related (denoted as "Y" in the figure) or non-related (denoted as "N") to the leaf. In this way, we can conquer the entire problem by solving many much simpler sub-problems.

We can also adopt a transcription sentence as a node on the trunk. However, the workload of relation discovery can be too heavy for the human because we have many sub-problems to solve. Since in the two MOOCs investigated in this thesis we have both homologous and heterologous linking in the corpus, in implementation we first annotate the former linking, and merge the sequential chunk of sentences that are aligned to the same leaf as a new video segment (for clarity, in the following we refer to this video unit used in heterologous linking as "video vignette", and use "video segment" for a general purpose, e.g., sentence in homologous linking or vignette in heterologous). These vignettes inferred from the alignment are used as trunk nodes in the following heterologous linking to reduce the workload of annotators. Besides, we define a trunk node as related to the leaf if the concept contained in the trunk node is equivalent to, an instance of, or a part of the leaf. Here, we choose a more lenient definition as compared to the "correspondence" defined in homologous linking, because in the heterologous case, content is usually organized in various manners and it is less likely to find identical mapping in the two underlying materials.

To sum up, in this thesis we link MOOC materials with the following steps:
\begin{enumerate}
  \item Do homologous linking of video transcription sentences (trunk nodes) against lecture slides (leaves). The homologous linking is formulated as an alignment problem.
  \item Group transcription sentences linked to the same slide together, and define each group as a "video vignette".
  \item Do heterologous linking of video vignettes against textbook sections and discussion forum posts. The heterologous linking is formulated as a binary classification problem. 
\end{enumerate}

\subsection{Annotation tasks}

\begin{figure}[t]
\begin{adjustwidth}{-3cm}{-3cm}
\begin{center}
\includegraphics[width=17cm]{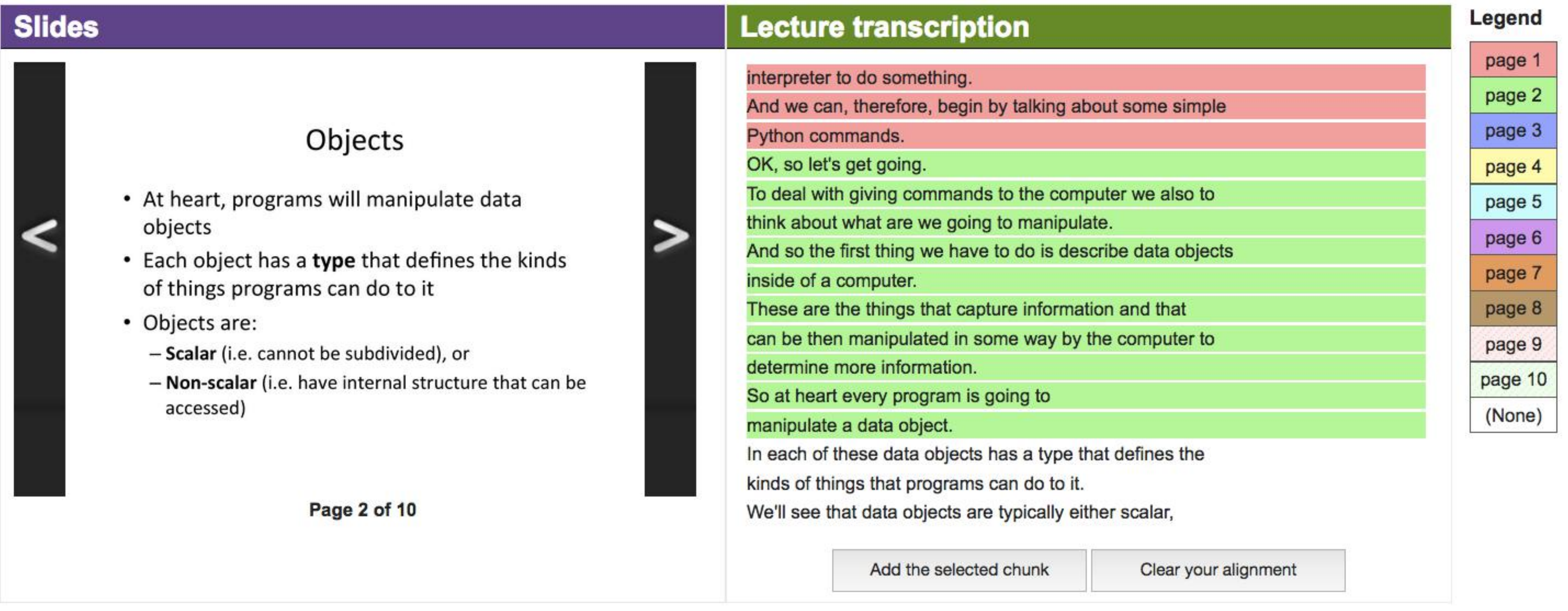}
\end{center}
\end{adjustwidth}
\caption{The website we build to collect homologous linking (i.e., the alignment between lecture video transcription and slides).}
\label{fig:homo_web}
\end{figure}

For these two types of linking, we design two websites to collect human annotation. In Fig.~\ref{fig:homo_web}, a screenshot of the website used for homologous linking is shown. Since, in this thesis, the only homologous linking investigated is the alignment between lecture video transcription and slides, we thus design the interface to present each time a transcription of a lecture video and a deck of slides from the same lecture in parallel. In the website, a human annotator first selects a slide page by clicking "<" (previous page), or ">" (next page). Then the annotator clicks and drags on the sentences he/she intends to align to the selected slide, and clicks on the "Add the selected chunk" button to confirm the alignment. After the confirmation, sentences aligned to different pages of slides are highlighted with different background colors, which are listed on the rightmost side of the screen. For instance, in this figure the first three sentences are aligned to the first slide, and the following 10 sentences are aligned to the second. The interface also provides a "Clear your alignment" button for annotators to clear confirmed alignment. Note that in this interface we do not show the lecture video, because we intend to simplify the workflow of this annotation task, and make our annotators focus on the transcription sentences.

\begin{figure}[t]
\includegraphics[width=13cm]{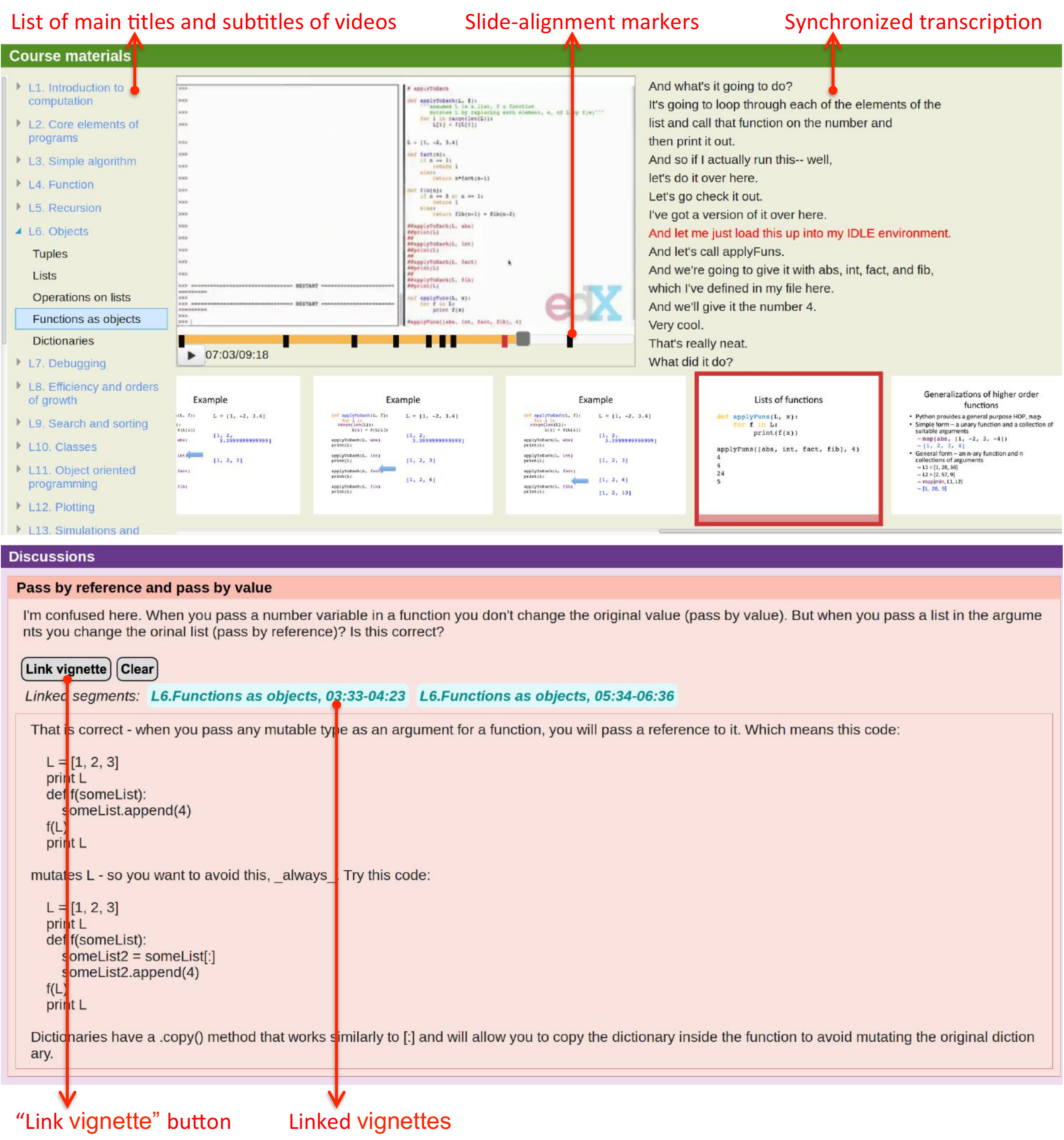}
\centering
\caption{The website we built to collect heterologous linking (i.e., the binary classification task of deciding whether a pair of video vignette and discussion thread or a pair of video vignette and textbook section is relevant).}  
\label{fig:hetero_web}
\end{figure}

We also design another website for annotating heterologous linking, and as an example we show its screenshot in Fig.~\ref{fig:hetero_web}. In implementation we investigate two mappings, i.e., lecture videos to discussion forum and videos to textbook, for the heterologous condition; therefore, we also design the interface to present content of videos and discussions (or textbook) side by side. As shown in the figure, in the upper half of this website lecture videos from the entire course are presented. Annotators can access these videos by clicking on the main title (which is a title shared among several lecture videos, and presented in blue text here) and subtitle (which is a title specific to each video, and presented in black text) of each video listed on the left hand side of the screen. In addition to a lecture video, here we also provide the aligned video transcription on the right and the thumbnails of aligned slides below the video. The transcription is synchronized with the video based on the time code extracted from the video subtitle file. The alignment between a video and its slides is inferred from human annotation in the homologous task as well as the time code of transcription sentences; we show this alignment information by rendering black markers on the video scrubber, and each marker represents the beginning of a video vignette that is aligned to one slide. As compared to the website for homologous linking, here we present video along with many relevant materials (e.g., the aligned slides and transcription) simultaneously, such that annotators can have a comprehensive understanding of each video vignette, which is the unit we work on in the heterologous task.

In the lower half of this website, a discussion thread (or a textbook section) is shown to annotators. An annotator has to select the relevant video vignettes from the entire course, and link these vignettes to the thread by clicking the "Link vignette" button. Linked vignettes are also shown on the screen with text in cyan. Since a video vignette $v_i$ can be defined by two markers (i.e., the marker representing the beginning of a video vignette that is aligned to slide \textit{i} and the next marker for slide \textit{i}+1), the annotator can simply select $v_i$ by dragging the video scrubber to any place between the two markers.

We then recruit human subjects to annotate linking in our corpus with the two websites. In Stat2.1x, the linking between lecture videos and slides (homologous) and the linking between videos and textbook (heterologous) are annotated; as for 6.00x, in addition to the two pairs of materials above, the linking between videos and discussions is also labeled. For the statistics course, in order to expedite the development process, we employ three annotators (including the author) who are graduate students or postdoctoral researchers with expertise in statistics for the labeling. Each annotator spent 5 hours in labeling the homologous task and 8 hours in the heterologous one. A majority voting is applied on the labeling results of the three annotators to obtain the final linking annotation used in this thesis.

For understanding the consistency among annotators, we compute Cohen's kappa score \cite{127} to measure the inter-annotator agreement. The kappa score can be written as the following equation:
\begin{equation} \label{eq:2}
\kappa = \frac{p_a - p_c}{1 - p_c}.
\end{equation}
In this equation, $p_a$ is probability of agreement among annotators observed in samples (i.e., in the corpus), and $p_c$ is the theoretical probability of chance agreement. In Stat2.1x, the kappa scores are 0.867 and 0.599 for the homologous and heterologous task respectively. According to several arbitrary guidelines, these scores show almost perfect and moderate agreement among annotators in the two tasks respectively \cite{121, 122}. Besides, the lower score in the heterologous linking also reflects that the underlying task is more complicated than the homologous one.

As for 6.00x, one of our goals is to establish a more realistic pipeline with its materials. Thus, instead of researchers, we choose to recruit online workers from AMT for the homologous linking, and teaching assistants in both the edX and MIT offering of 6.00x for the heterologous tasks. Online workers are employed here for homologous linking since they are an economic choice for data annotation with satisfactory quality \cite{83}, especially when the underlying task is simple. In contrast, we choose to recruit teaching assistants for the heterologous task, because, in the designed annotation workflow, annotators have to be familiar with the entire course before they can select relevant videos for each discussion (or textbook section) efficiently. This task requires annotators to spend a much longer period of time to ramp up. However, since on the crowdsourcing task-matching platform workers usually have thousands of tasks to choose from at the same time, the returning worker rate is much lower as compared to recruiting teaching assistants for annotation. The low returning rate means we have to spend a large portion of time in training new workers to be familiar with the content, and with a lesser portion of experienced workers yielding quality output.

We created a total of 945 HITs (i.e., Human Intelligence Tasks)\footnote{On AMT, each HIT is a self-contained task a worker can perform and receive a reward after completing it}. on AMT for aligning 105 video-slide pairs, with nine workers on each pair and a reward of \$0.25 for each HIT. 100 workers participate in the annotation; the mean and standard deviation of total time spent for each worker are 35.5 minutes and 63.4 minutes respectively. Majority voting over the nine labeling results in each pair of video and slides is also taken to obtain the alignment used in this thesis.

For the labeling of 144 textbook sections to lecture videos, we recruited four teaching assistants from the 6.00x course offered at MIT; the labeling of 1,239 discussion threads was done by two teaching assistants from the MIT 6.00 course and five teaching assistants from the edX version. Each section or discussion thread is labeled by three different annotators for the majority consensus process. These annotators spent 7.5, 7.5, 5, and 3 hours in the textbook task, and 16, 14, 10, 10, 6, 6, and 2 hours for labeling the forum. We pay these annotators at the rate of \$45 per hour.

We also computed kappa scores for these annotations. For the video-to-slide, video-to-textbook, and video-to-discussion linking, the scores are 0.810, 0.761, and 0.434 respectively. We are satisfied with these results because all of the scores also show almost perfect or moderate agreement among annotators \cite{121, 122}. Comparing these numbers to the scores obtained in Stat2.1x, we find the online workers can also yield consistent annotation as researchers do in homologous linking (cf. 0.810 \textit{vs}. 0.867); annotators are more consistent in linking textbook to video sequence (cf. 0.761 \textit{vs}. 0.599), presumably because the author of the textbook used in 6.00x is also one of the lecture instructors. Therefore, it is easier to identify the related learning objects from two material sequences. Besides, linking discussions is undoubtedly the most complicated task since the learner-generated content is noisier and less organized than educator-generated ones. This fact reflects on its lowest inter-annotator consistency. Even so, the 0.434 kappa score still shows fair agreement among teaching assistants.

\section{Presenting linking to learners}
With the linking among course materials annotated, we then design an interface to visualize the annotated relation while learners access the course content. Our ultimate objective is providing learners guidance to make learning content more accessible and to help them find supportive materials more efficiently when they are in need, such as confused. After surveying relevant literature \cite{22, 53, 56, 59, 64, 104, 105, 123}, we identify three high-level goals that inform our design. 

\textbf{Supporting relational navigation among materials.} Many observations suggest that, in the current MOOC platforms, it is difficult for learners to identify related materials \cite{53, 56, 64, 104, 105}. Thus many interactions cannot be achieved, such as skipping redundant forum posts, or navigating from a specific point of lecture to further discussions in forum and detailed explanation in textbook. To support these needs of navigation, we leverage the annotated linking among learning materials. We design our interface to visually illustrate the material relation. The visualization guides learners and allows them to jump back and forth among relevant content.

\textbf{Providing easy access to different conceptual pieces within a lecture video.} Previous research has shown that presenting videos along with sub-goals helps people learn better, since the sub-goals can abstract away low-level details and reduce the cognitive load of learners \cite{22, 59, 123}. Since the lecture slides are usually the skeleton of a lecture, and each slide can be interpreted as a conceptual piece or a sub-goal of this lecture, we design the interface to visualize the alignment between slides and videos. In this way, videos aligned to different conceptual pieces can be accessed efficiently.

\textbf{Minimizing distraction while providing guidance.} The additional navigation and guidance introduce new elements to the interface. Thus learners have to learn about how to manipulate the new interaction, which is a distraction for learners and can overload their cognitive system. Since the distraction has a negative effect on learning \cite{59}, we also design our interface to minimize the disturbance. Specifically, we borrow many design decisions from mainstream MOOC platforms to make interacting with our interface intuitive. In the following, we introduce our interface and how we design it to achieve the three goals in detail.

\begin{figure}[t]
\includegraphics[width=14cm]{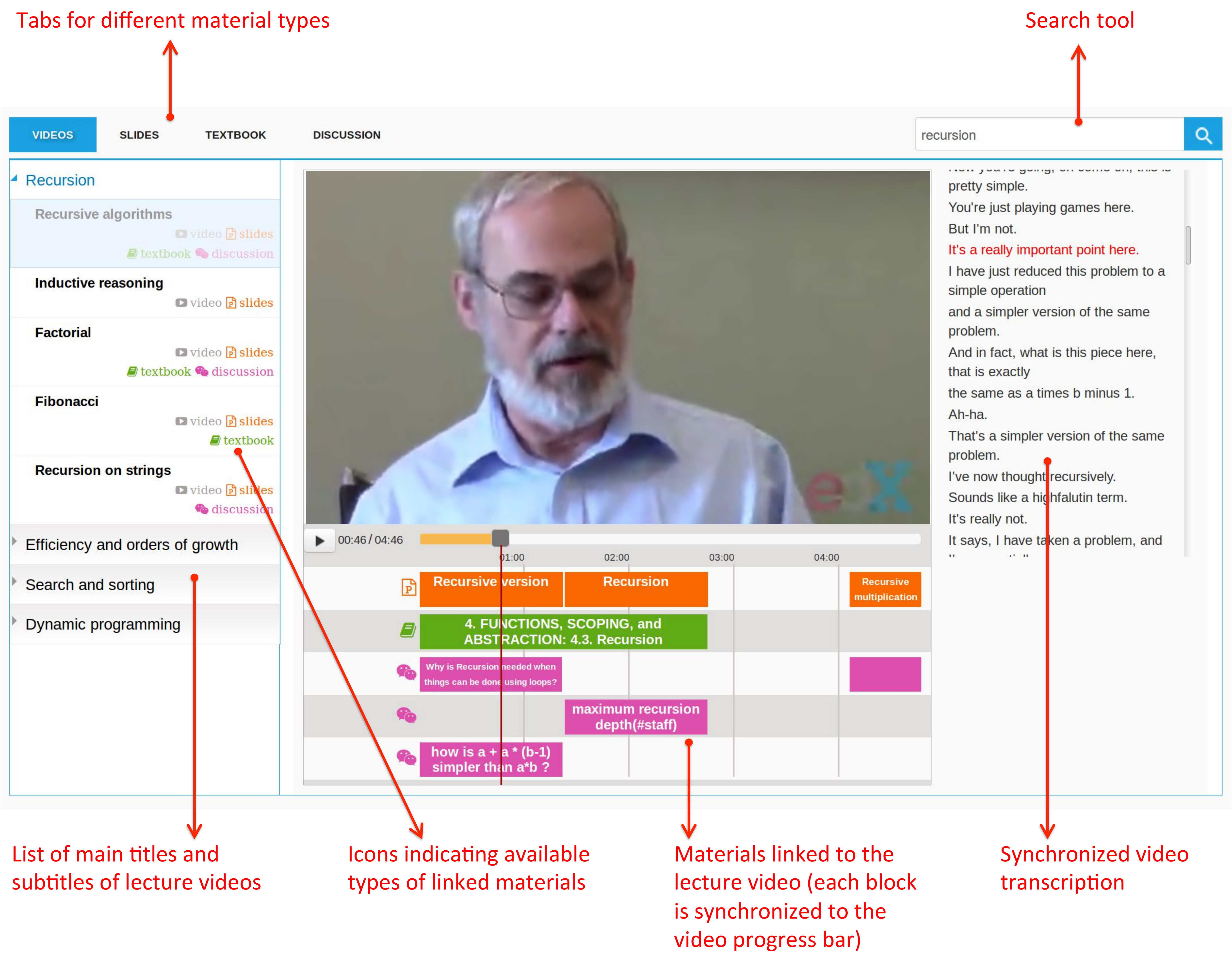}
\centering
\caption{The implemented interface to present learning content and linking information simultaneously. In this interface, course materials are retrieved by entering queries in the provided search tool. The retrieved materials are listed according to their types (in the top left corner of this screenshot) and their original positions in each material sequence (e.g., lecture or chapter indices). Titles of listed materials are shown on the left hand side, and content selected from the list is in the middle. If the selected content is a video (i.e., the trunk), linked supplementary objects (i.e., the leaves) are also displayed as orange, green, or pink blocks under the video scrubber.}  
\label{fig:linking_interface}
\end{figure}

In Fig.~\ref{fig:linking_interface}, a screenshot of the interface presenting content and linking simultaneously is shown. In the interface, there are four main components: key-term search, material list, content presentation and linking visualization. To start interacting with this interface, the user has to enter the topic he or she intends to learn in the search field. Our server retrieves learning materials relevant to the entered topic by 1) stemming the search query for query expansion, 2) enumerating \textit{n}-grams (\textit{n} equals one to five) in the expanded query, 3) scoring each lecture video, slides, textbook section, and discussion thread with the number of matched \textit{n}-grams, and 4) returning the materials with \textit{N} (we set \textit{N} to 60 in the following experiment) highest scores. We provide a search tool in this interface, instead of simply presenting content of the entire course, because search is a common and mature technique that helps users narrow down candidate documents and obtain desired information.

The returned materials are listed based on their types (i.e., video, slides, textbook, and discussion) and their original position in each material sequence (e.g., index of chapter). The material sequence accessed by selecting the "video" tab (i.e., the list of videos' main titles and subtitles on the left hand side of the screen) is the main flow, or the trunk, of returned content. In addition to videos, returned slides, textbook sections and discussion threads (i.e., the leaves) are also attached to relevant videos according to the linking annotation; orange, green, and pink icons for these supplementary contents are appended to the titles of corresponding videos to illustrate the available types of materials. The returned slides, textbook sections and discussion threads that are not related to any videos are listed under the corresponding tabs next to the "video" tab. 

Considering the learnability, we design the layout of the interface to resemble the arrangement in most prevalent MOOC platforms such as edX. By borrowing the design decisions made by the professional user experience teams in these platforms, we are able to make our interface intuitive. The intuitiveness allows learners to concentrate on learning the content, instead of learning how to use the website. This fact can not only enhance learning experience, but also reduce noise when we measure how linking affects learning in a user study. Besides, we choose to preserve the original flow of materials when presenting returned content, instead of listing materials by their relevance scores. Preserving the original flow allows us to visualize the context and prerequisite dependency among returned materials, which is crucial for achieving meaningful learning \cite{29}.

\begin{figure}[t]
\begin{adjustwidth}{-3cm}{-3cm}
\begin{center}
\includegraphics[width=17cm]{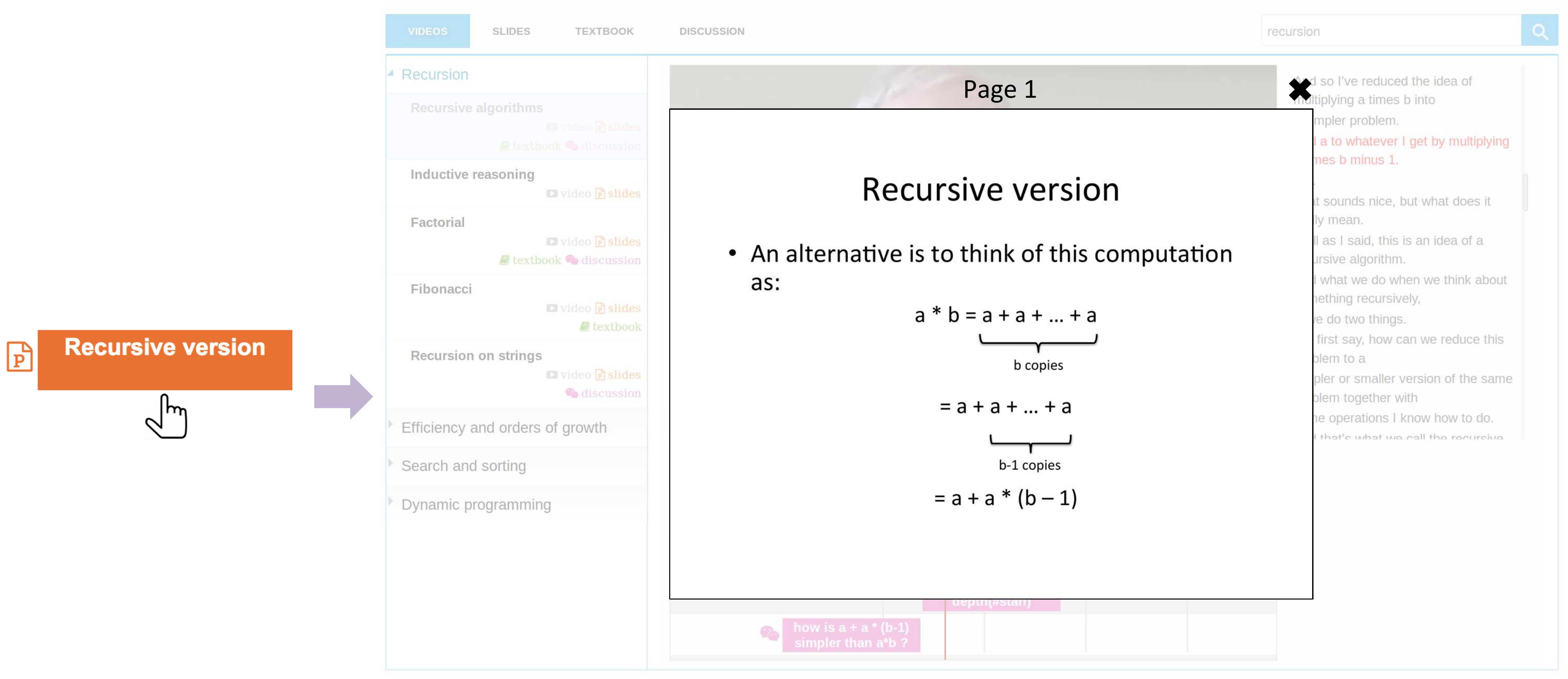}
\end{center}
\end{adjustwidth}
\caption{By clicking on any of the colored blocks under video scrubber, content of the linked supplementary object represented by this clicked block will be rendered in a lightbox. In this figure, we provide an example illustrating how a linked slide is displayed in our interface after its corresponding block was clicked.}  
\label{fig:sup_content_acc}
\end{figure}

After selecting a video from the list, the learning content along with linking information is shown in the middle. As illustrated in the figure, a lecture video and the synchronized transcription are presented. Under the video scrubber, several orange, green, and pink blocks are rendered. These colored blocks are synchronized with the video progress bar. Each colored block corresponds to one linked supplementary object (i.e., one slide, one textbook section, or one discussion thread), and the span of the block represents the video vignette that is linked to the underlying object. As shown in Fig.~\ref{fig:sup_content_acc}, by clicking each colored block the corresponding object will be displayed in a lightbox. As for the returned slides, sections, and threads that are not linked to any video, learners can also access the content by selecting material title from the list under the corresponding tab; resulting content will also be displayed in the middle of the website. 

We surmise that this interface can enhance learning experience by helping learners access relevant information and identify underlying sub-topics of videos. As compared to a conventional video player where only the video scrubber is provided, the synchronized object identifiers serve as recommendation that might be useful for learners at different points in their learning path. For instance, if a learner is watching a lecture video and he/she is confused at a specific point in the video, with our interface this learner can access easily the detailed explanation in the textbook from linking attached to this video vignette; if the learner wants to learn more about a concept mentioned at some point in the video, the linked forum threads might provide further discussions. Furthermore, since relevant materials are linked and placed together under the video scrubber, it is easier to identify redundant learning content, such as duplicated questions in the forum; therefore this interface can make navigation much more efficient.

Additionally, the lecture slides are typically the skeleton of a lecture; each slide can be interpreted as a sub-topic or a sub-goal in the lecture. Thus, by aligning slides to a lecture video, we divide the video into several conceptual pieces, where each piece corresponds to one sub-goal or sub-topic. In the designed interface, the alignment is visualized. Hence learners can visually identify the structure of the lecture, and navigate to different sub-topics easily. 

\begin{figure}[t]
\includegraphics[width=14cm]{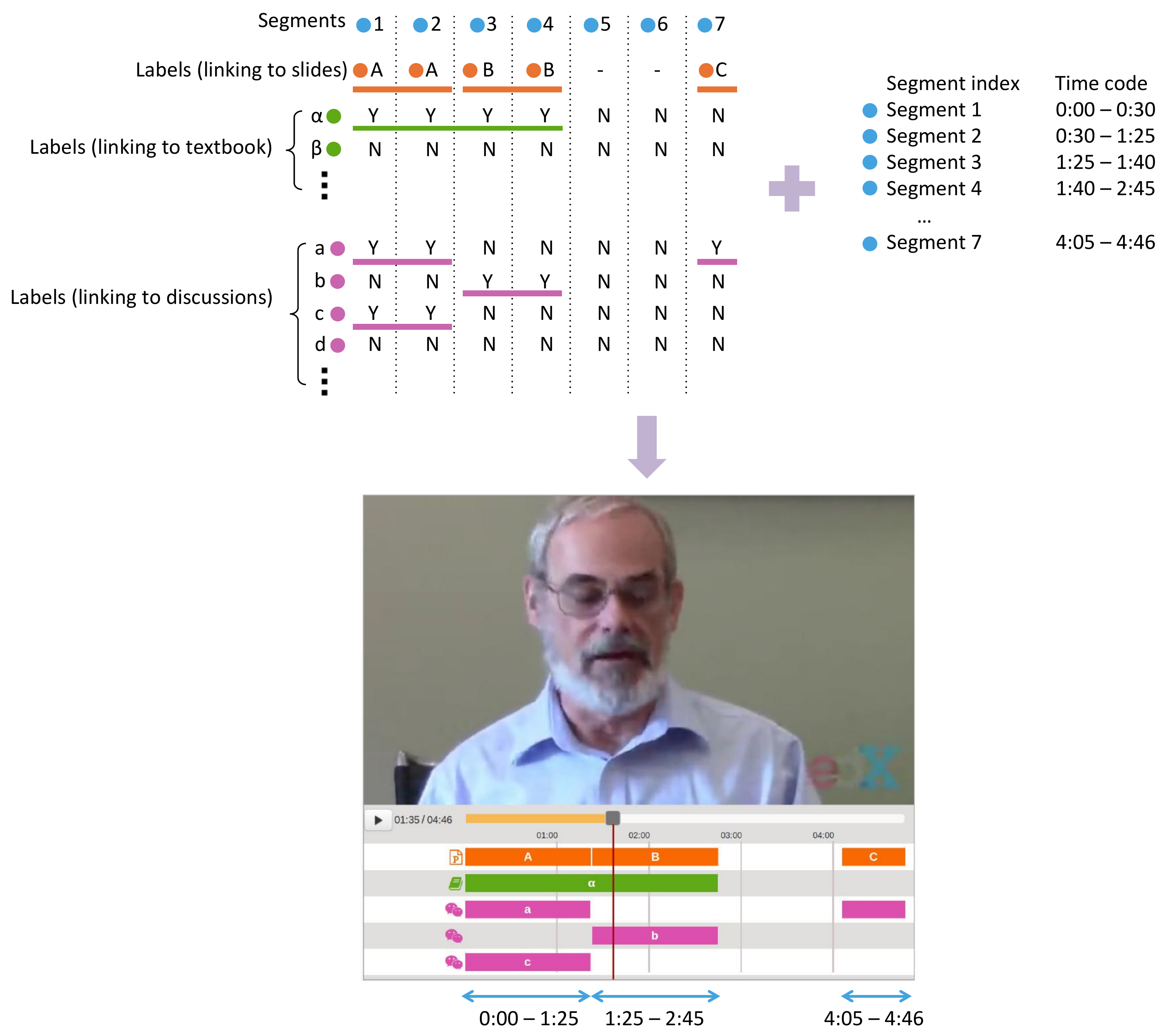}
\centering
\caption{With the annotated linking from video segments (i.e., sentences or vignettes) to slides, textbook, and discussions, as well as the time code of each segment, the synchronized linked objects under the video scrubber can be rendered. In this example, each page of slides is indexed with A, B, and C; each textbook section is indexed with $\alpha$, $\beta$, and so on; each discussion thread is indexed with a, b, c, and so on.}  
\label{fig:linking_to_interface}
\end{figure}

The remaining part that should be discussed is how to obtain the synchronized object identifiers below the video scrubber. As shown in Fig.~\ref{fig:linking_to_interface}, this can be done easily with the linking annotation described in Section 3.1. With the time code extracted from the video subtitle file, for each supplementary object we can obtain the beginning and ending time code for the segment of transcription sentences that is linked to this object. The video player with the synchronized linked objects can then be rendered with the time information.

\section{Comparative study}
To answer the research question "if we are able to link course materials with human assistance, would it help learners?", we assessed the learning effect of presenting linking to learners. Specifically, we conducted a comparative study, in which we presented experimental subjects with interfaces with or without linking, and measured their performance on designed learning tasks. We focused our study on three aspects:

\begin{itemize}
  \item How do learners search for desired learning content when linking is presented?
  \item How does linking affect learners in integrating and memorizing information within a fixed period of time?
  \item How does linking affect different cohorts of learners?
\end{itemize}

\subsection{Study design}
We adopted a between-subjects design for our study, where each learner was randomly assigned to either the \textit{linking} interface (i.e., the interface described in Section 3.2) or a baseline interface that stripped off all inter-material relations from the \textit{linking} interface. We will introduce the baseline interface in detail in Section 3.3.2.

We designed two learning task scenarios, "information search" and "concept retention," for learners to perform with their assigned interface. Learners' performance in these tasks was analyzed to investigate the learning effect of linking around the three aspects described above.

\begin{itemize}
  \item \textbf{Information search} tasks involved finding corresponding learning content to a given problem. In each of these tasks, a learner is randomly assigned to a problem sampled from the courses' quizzes. This learner then has to use the assigned interface to find a piece of learning content that explains how to solve the problem. A learning content piece can be a specific moment in a lecture video, a page of slides, a textbook section, or a discussion thread (only in 6.00x). This scenario emulates a situation where a learner attempts to find informative content while he/she faces a problem. 
  \item \textbf{Concept retention} tasks require learners to remember, understand, and integrate concepts relevant to a given topic. In each task, we randomly gave a learner a topic sampled from the courses, and gave 10 minutes for this learner to learn about the topic with the assigned interface. After the learning stage we asked this learner to write a short essay that includes as many concepts as he/she can remember as possible. In the writing stage, this learner is not allowed to access the interface and learning content. We set the time limit in order to evaluate how efficiently learners can browse through the materials, as well as capture and remember the high-level information. This scenario emulates the condition where learners attempt to have an integral and high-level understanding of a topic with a limited amount of time. 
\end{itemize}

Typically, researchers may want to apply intervention straight in a course, so that they can measure directly the effect of intervention. In contrast, in this thesis we choose to focus our investigation around the two designed learning scenarios, and explore learners' navigation behavior. We made this decision because learning involves complicated mental processes from motivation and memorization to understanding and problem solving. It may be too elusive to ascertain all of these processes in one set of experiments. Exploring the effect of linking in a course may introduce too many variables and noises, and obfuscate the advantage brought by the intervention. Therefore, as suggested in previous study \cite{125}, we concentrate our investigation on how linking can help material navigation. If we are able to show a positive effect of linking in this subset of learning processes, with the abundant literature discussing the correlation between navigation and learning \cite{22, 56, 59, 64, 123, 124}, the benefit of linking in learning is self-evident. 

\begin{figure}[t]
\begin{adjustwidth}{-3.5cm}{-3.5cm}
\begin{center}
\includegraphics[width=18.5cm]{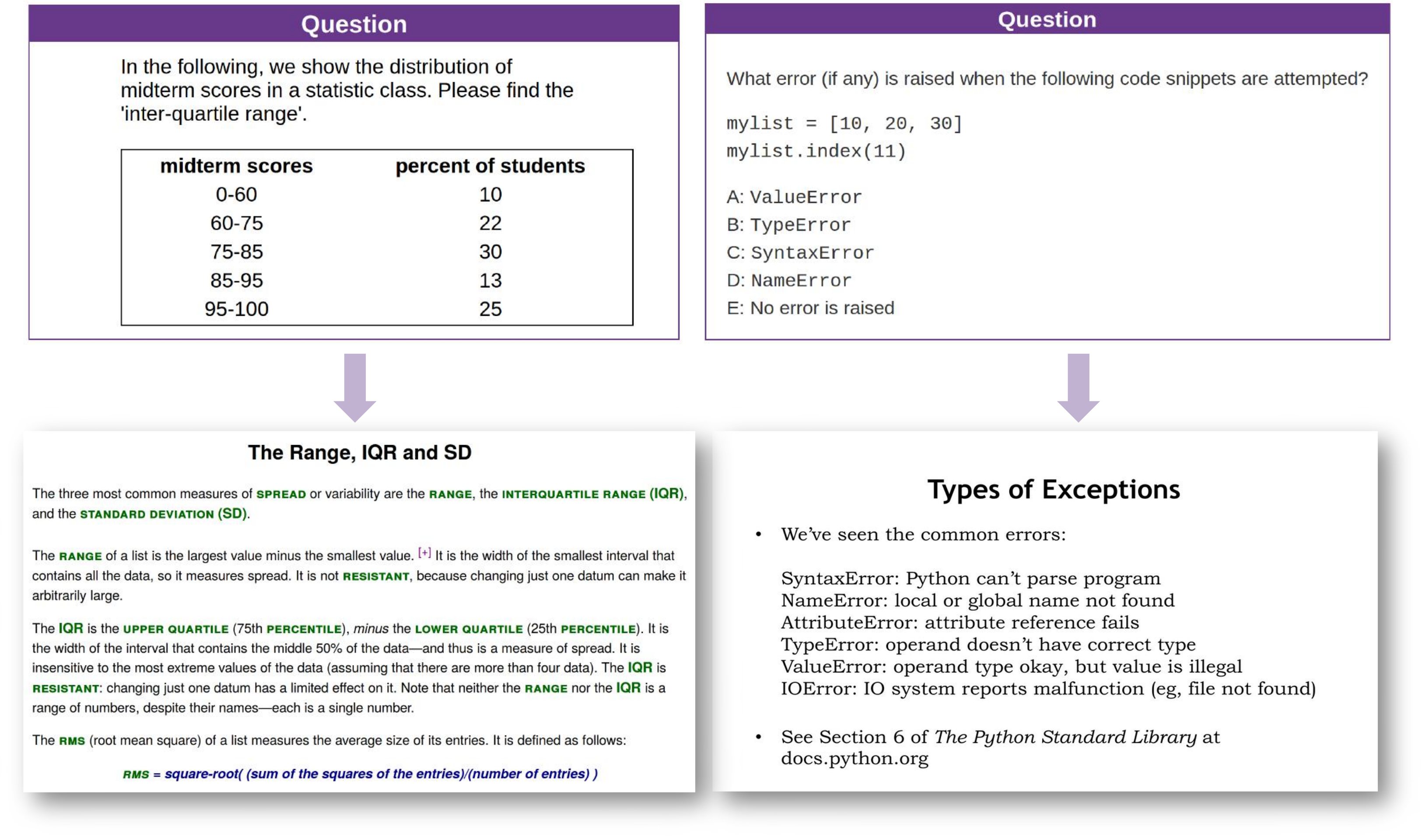}
\end{center}
\end{adjustwidth}
\caption{The first row shows two sampled problems used in the information search learning scenario. For each of the problems, a piece of learning content that is accepted as the answer is shown in the second row; a textbook section (the content with title "\textit{The Range, IQR and SD}")) and a page of slides (title: "\textit{Types of Exceptions}") is displayed respectively. In this figure, the left hand side is a problem-answer pair for Stat2.1x and the right is from 6.00x.}  
\label{fig:prob_exp}
\end{figure}

\begin{figure}[t]
\begin{adjustwidth}{-3cm}{-3cm}
\begin{center}
\includegraphics[width=17cm]{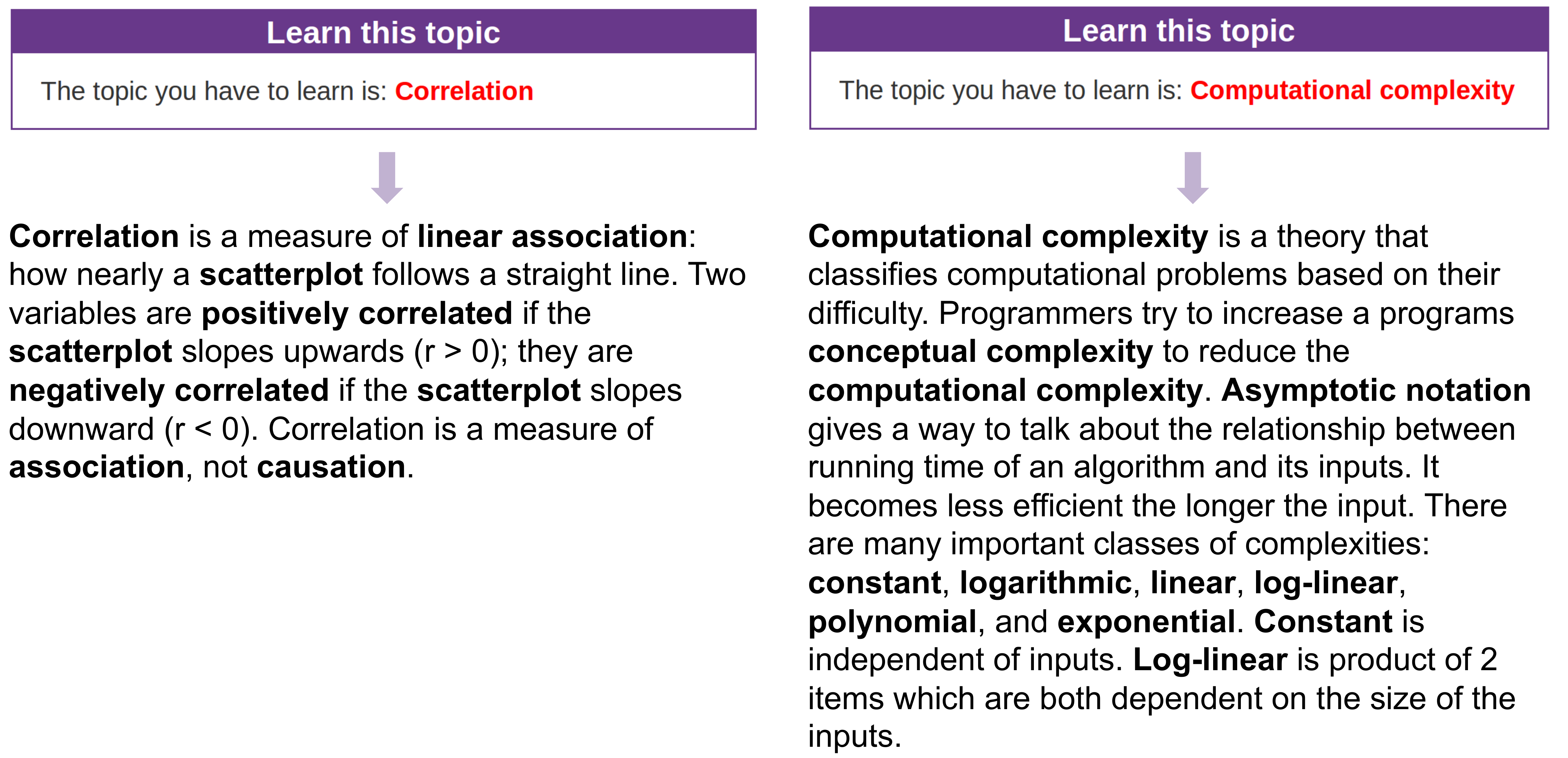}
\end{center}
\end{adjustwidth}
\caption{The first row shows two sampled topics used in the concept retention scenario. For each topic, an essay submitted by a learner in our user study is also shown as an example. We also highlight concepts in essays in bold font. In this figure, the left hand side is a topic-essay pair for Stat2.1x and the right is for 6.00x.}  
\label{fig:topic_exp}
\end{figure}

For these two scenarios, we sampled 10 problems and topics respectively in each of the two MOOCs (i.e., Stat2.1x and 6.00x) we investigated. In the sampling we emphasize the first half of each of these two courses, because lectures from the latter are usually more advanced, complicated and required prerequisite knowledge learnt earlier in the course. With the emphasis on foundational lectures, we attempt to reduce noise introduced by diverse prior knowledge learners may have. In Fig.~\ref{fig:prob_exp}, we show two sampled problems from each of the MOOCs along with examples of learning content pieces we accept as answers. In Fig.~\ref{fig:topic_exp}, two examples of sampled topics along with one learner's submission respectively are given. Concepts in these submitted essays are highlighted in bold font. We list the entire sets of sampled problems and topics for each MOOC in Appendix \textcolor{red}{?}. 

\subsection{Baseline}
In the comparative study, we have to implement a baseline interface, and investigate whether assigning learners with either baseline or \textit{linking} interface affects their performance in accomplishing tasks. Thus, we design the \textit{null} interface.

\begin{figure}[t]
\begin{adjustwidth}{-3cm}{-3cm}
\begin{center}
\includegraphics[width=17cm]{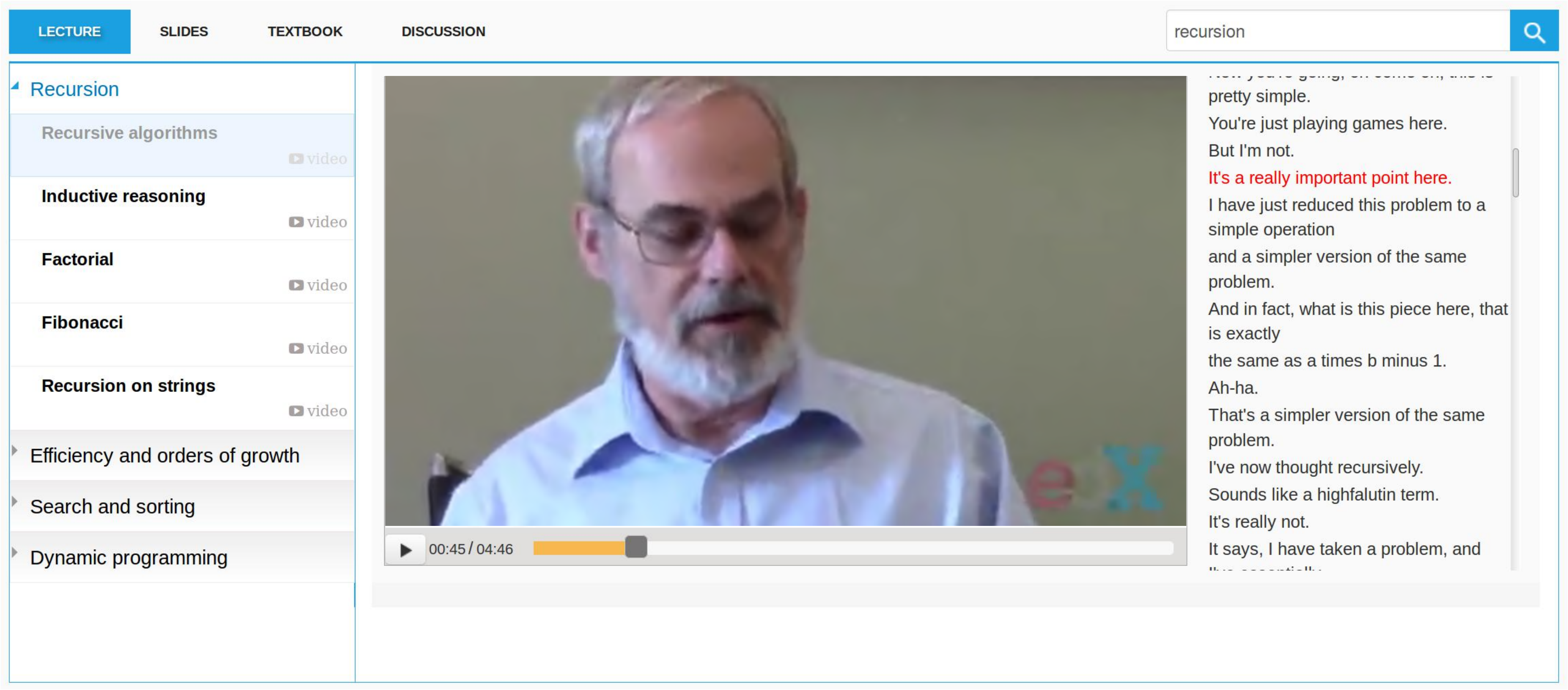}
\end{center}
\end{adjustwidth}
\caption{The implemented \textit{null} interface that serves as one of our baselines. This interface still includes the same components for key-term search, material list, and content presentation as the \textit{linking} interface does. The layout and visual design are also kept identical. The only difference is that we strip away the linking visualization, and there is no synchronized supplementary learning object under each lecture video (i.e., the trunk).}  
\label{fig:null_inter}
\end{figure}

In Fig.~\ref{fig:null_inter}, a screenshot of the \textit{null} interface is shown. In this interface the only difference is that the visualization is not presented. The visual layouts as well as components for key-term search, material list, and content presentation that are designed in the \textit{linking} interface are retained. As illustrated in this figure, in the \textit{null} interface users also start by entering search queries; then the retrieved materials are also listed according to their types (i.e., the panels of material types listed in the top left corner) and their original positions in each material sequence (e.g., lecture or chapter indices, as shown on the left hand side of the figure); the learning content selected from the sequence is rendered in the middle. However, every material type is presented independently and no relational information is provided, e.g., the linked supplementary objects (or the leaves) are no longer rendered under lecture videos (i.e., the trunk). By comparing \textit{linking} to this \textit{null} interface, we can investigate how offering learners the information of relation among learning content affects their behavior.

\subsection{Experiment subjects}
As for experimental subjects, we chose to recruit online workers from AMT \cite{23}. Generally, we may want the subject pool to be learners who are actually taking the course. However, in our case, online workers are a good approximation to enrolled learners, because we measure their microscopic behavior in accomplishing specific learning tasks where clear instruction is provided. In these tasks, online workers have very similar goals as learners, such as finding desired information as quickly as possible or learning more high-level concepts within a limited amount of time. Therefore, they interact with our interface much like enrolled learners do. Besides, although these workers are monetarily driven, as shown in the following quote from a worker's feedback.

\begin{quotation}
\textit{"I really like this HIT. I hope I am doing them well for you as intended. I want to thank you as well, because I'm actually learning quite a bit about computer programming and I really like the lectures and how they are organized, every time the 10 minutes are up, I'm kind of disappointed because I feel like I was just getting started learning about a subject I'm interested in."}
\end{quotation}

\noindent
Our HITs also motivate workers intellectually, and attract many who want to learn about the two courses. Moreover, performing a live experiment in an actual MOOC is expensive and time consuming. In contrast, the abundant online labor pool and the diverse demographics of workers ensure that we can access users of various backgrounds at large scale, and simultaneously with reasonable cost in time and money. This fact allows us to investigate the third focused aspect, "how linking affects different cohorts of learners," to understand the usefulness of the proposed framework in a heterogeneous learner body. With all the upsides, we thus choose these micropayment online workers as our experimental subjects.

\subsection{Experiment scale}
In Table~\ref{table:study_size}, the scale of our experiment is summarized. In Stat2.1x, for each scenario we deployed 2,000 HITs on AMT, i.e., two interfaces (\textit{null} and \textit{linking}), 10 problems or topics, and 100 HITs accomplished by 100 unique online workers for each pair of problems/topics and interface. The reward of each HIT is \$0.35 and \$1.00 for the information search and concept retention scenario respectively. As listed in the table, a total of 497 and 751 unique AMT workers participated in each of the two scenarios. These numbers are different from 2,000 because we allowed each worker to solve more than one problem or topic (in contrast each subject can only work with one assigned interface under the between-subject design). The experiment took four months to complete.

The scale of experiment in 6.00x is also shown in the table. Here 2,000 HITs were also deployed for each scenario. In the two scenarios, 393 and 631 workers participated respectively. Observing the slow complete rate of the experiment in Stat2.1x, we increased the reward of each HIT to \$1.50 and \$2.00 for the two scenarios. Since the involved topics are more complicated in 6.00x, the potential and qualified workers for our tasks are fewer. This is another reason why we decided to provide larger monetary incentive. This experiment took two and a half months to complete.

\begin{table}[]
\centering
\caption{Summarization of sizes of comparative study in Stat2.1x and 6.00x.}
\label{table:study_size}
\begin{tabular}{|l|c|c|c|c|}
\hline
                   & \multicolumn{2}{c|}{Number of tasks} & \multicolumn{2}{c|}{Number of unique workers} \\ \hline
                   & Stat2.1x           & 6.00x           & Stat2.1x                & 6.00x               \\ \hline
Information search & 2,000              & 2,000           & 497                     &     393                \\ \hline
Concept retention  & 2,000              & 2,000           & 751                     &     631                \\ \hline
\end{tabular}
\end{table}

\section{Results}
With the experimental setup and deployment described above, we then measure subjects' performance in accomplishing learning tasks, to explore the effect of linking on learning from three aspects (i.e., how does linking affect search, affect integrating or memorizing information, and affect different cohorts). For studying how linking affects various cohorts of learners, in our tasks we also require subjects to fill in a background survey. Based on information the subjects provided in the survey, we tease out three demographic factors that may influence their performance - their highest accomplished degree, their previous experience in online courses, and previous exposure to relevant topics/courses. 

To study the effect of these factors, we divided subjects with three different criteria (i.e., whether or not they have had exposure to statistics or Python programming language, have taken MOOCs previously, and have at least a bachelor's degree). In Table~\ref{table:n_tasks_stat}, we list the numbers of completed tasks classified by each criterion for each learning scenario (i.e., information search and concept retention) and interface (i.e., \textit{null} and \textit{linking}) in the study of Stat2.1x. As we can see here, about seven out of ten and six of ten of the participants in the two scenarios reported that they have some prior knowledge in statistics (the second and third row of the table); only about a quarter of subjects in these scenarios have attended some MOOCs before (the fourth and fifth row of the table); slightly more than half of the participants have a bachelor's or higher degree (the sixth and seventh row of the table). We also broke down the completed tasks in 6.00x and summarized the result in Table~\ref{table:n_tasks_600}. The data shows that we have slightly more tasks contributed by subjects with previous exposure to MOOCs and with at least a bachelor's degree, but less tasks completed by subjects with experience in the course topics (i.e., the Python programming language). On top of these divisions we measure learning performance among each cohort to investigate the effect of linking on search behavior as well as integrating and memorizing information.

\begin{table}[]
\centering
\caption{
Number of tasks completed by each cohort for each learning scenario (i.e., information search and concept retention) and interface (i.e., \textit{null} and \textit{linking}) in the study of Stat2.1x.}
\label{table:n_tasks_stat}
\begin{tabular}{|ll|cc|cc|}
\hline
                                                  &     & \multicolumn{2}{c|}{Information search} & \multicolumn{2}{c|}{Concept retention} \\ \hline
                                                  &     & \textit{null}     & \textit{linking}    & \textit{null}    & \textit{linking}    \\ \hline
\multicolumn{2}{|l|}{Overall}                           & 1,000             & 1,000               & 1,000            & 1,000               \\ \hline
\multicolumn{1}{|l|}{\multirow{2}{*}{Statistics}} & Yes & 714               & 704                 & 594              & 597                 \\ \cline{2-2}
\multicolumn{1}{|l|}{}                            & No  & 286               & 296                 & 406              & 403                 \\ \hline
\multicolumn{1}{|l|}{\multirow{2}{*}{MOOCs}}      & Yes & 295               & 249                 & 205              & 287                 \\ \cline{2-2}
\multicolumn{1}{|l|}{}                            & No  & 705               & 751                 & 795              & 713                 \\ \hline
\multicolumn{1}{|l|}{\multirow{2}{*}{$\geq$Bachelor}}   & Yes & 573               & 522                 & 549              & 519                 \\ \cline{2-2}
\multicolumn{1}{|l|}{}                            & No  & 427               & 478                 & 451              & 481                 \\ \hline
\end{tabular}
\end{table}

\begin{table}[]
\centering
\caption{Number of tasks completed by each cohort for each learning scenario (i.e., information search and concept retention) and interface (i.e., \textit{null} and \textit{linking}) in the study of 6.00x.}
\label{table:n_tasks_600}
\begin{tabular}{|l|l|cc|cc|}
\hline
\multicolumn{2}{|l|}{\multirow{2}{*}{}} & \multicolumn{2}{c|}{Information search} & \multicolumn{2}{c|}{Concept retention} \\ \cline{3-6} 
\multicolumn{2}{|l|}{} & null & linking & null & linking \\ \hline
\multicolumn{2}{|l|}{Overall} & 1,000 & 1,000 & 1,000 & 1,000 \\ \hline
\multirow{2}{*}{Python} & Yes & 455 & 536 & 443 & 409 \\ \cline{2-2}
 & No & 545 & 464 & 557 & 591 \\ \hline
\multirow{2}{*}{MOOCs} & Yes & 384 & 397 & 319 & 315 \\ \cline{2-2}
 & No & 616 & 603 & 681 & 685 \\ \hline
\multirow{2}{*}{$\geq$Bachelor} & Yes & 607 & 623 & 617 & 540 \\ \cline{2-2}
 & No & 393 & 377 & 383 & 460 \\ \hline
\end{tabular}
\end{table}

\subsection{How linking affects search}
In this section, we investigate learners' performance in the information search scenario. Here, we computed two metrics: average searching time and average accuracy. The first metric evaluates how fast each subject identified a piece of learning content (i.e., a specific moment in a lecture video, a page of slides, a textbook section, or a discussion thread) as answer to the assigned problem and submitted HIT; the second metric measures whether the identified content can indeed answer the problem. To measure the accuracy, for each problem the learning content pieces that are valid answers are labeled. In Stat2.1x, three annotators who are graduate students or postdoctoral researchers with expertise in statistics did the labeling; as for 6.00x, we recruited three teaching assistants from the same class offered at MIT to obtain the annotation. One thing to be noted is that, when a worker identified a specific moment in a lecture video as the answer, we accept this submission as correct only if it deviates from any of our labeled answers by less than one minute. With these metrics, we attempt to understand how linking affects learners' behavior when they are trying to find learning content.

\begin{table}[]
\centering
\caption{Learner performance in the information search scenario in the study of Stat2.1x. Performance is evaluated by the average searching time and average accuracy metrics, and measured within various cohorts using different interfaces.}
\label{table:search_res_stat}
\begin{tabular}{|l|l|cc|l|cc|l|}
\hline
\multicolumn{2}{|l|}{}            & \multicolumn{3}{c|}{Average searching time (seconds)}             & \multicolumn{3}{c|}{Average accuracy (\%)}                    \\ \hline
\multicolumn{2}{|l|}{}            & \textit{null}             & \multicolumn{2}{c|}{\textit{linking}} & \textit{null}         & \multicolumn{2}{c|}{\textit{linking}} \\ \hline
\multicolumn{2}{|l|}{Overall}     & 206                       & \multicolumn{2}{c|}{152}              & 69.2                  & \multicolumn{2}{c|}{69.5}             \\ \hline
\multirow{2}{*}{Statistics} & Yes & 166                       & \multicolumn{2}{c|}{147}              & 71.1                  & \multicolumn{2}{c|}{70.5}             \\ \cline{2-2}
                            & No  & 295                       & \multicolumn{2}{c|}{160}              & 64.9                  & \multicolumn{2}{c|}{67.1}             \\ \hline
\multirow{2}{*}{MOOCs}      & Yes & 166                       & \multicolumn{2}{c|}{139}              & 72.0                  & \multicolumn{2}{c|}{70.6}             \\ \cline{2-2}
                            & No  & 225                       & \multicolumn{2}{c|}{154}              & 68.2                  & \multicolumn{2}{c|}{68.9}             \\ \hline
\multirow{2}{*}{$\geq$Bachelor}   & Yes & 198                       & \multicolumn{2}{c|}{163}              & 70.7                  & \multicolumn{2}{c|}{70.6}             \\ \cline{2-2}
                            & No  & 208                       & \multicolumn{2}{c|}{136}              & 67.5                  & \multicolumn{2}{c|}{68.5}             \\ \hline
\end{tabular}
\end{table}

Table~\ref{table:search_res_stat} summarizes learner performances in the information search scenario in the study of Stat2.1x. Performance is evaluated by the average searching time (columns 1 and 2) and average accuracy (columns 3 and 4) metrics, and measured within cohorts having various backgrounds (row 1 for overall subjects; rows 2 and 3 for whether subjects have prior knowledge in statistics; rows 4 and 5 for whether subjects have attended MOOCs before; rows 6 and 7 for whether they have at least a bachelor's degree) and using different interfaces (columns 1 and 3:  \textit{null}; columns 2 and 4: \textit{linking}). As mentioned above, experiments based on online workers suffer from spammers. To control the quality of workers' submissions, in each learner cohort we discard submissions with top and bottom 5\% search time. This mechanism can filter cases such as workers trying to cheat on the system by randomly selecting a piece of learning content, or workers leaving their computers during tasks.

\begin{figure}[t]
\includegraphics[width=9cm]{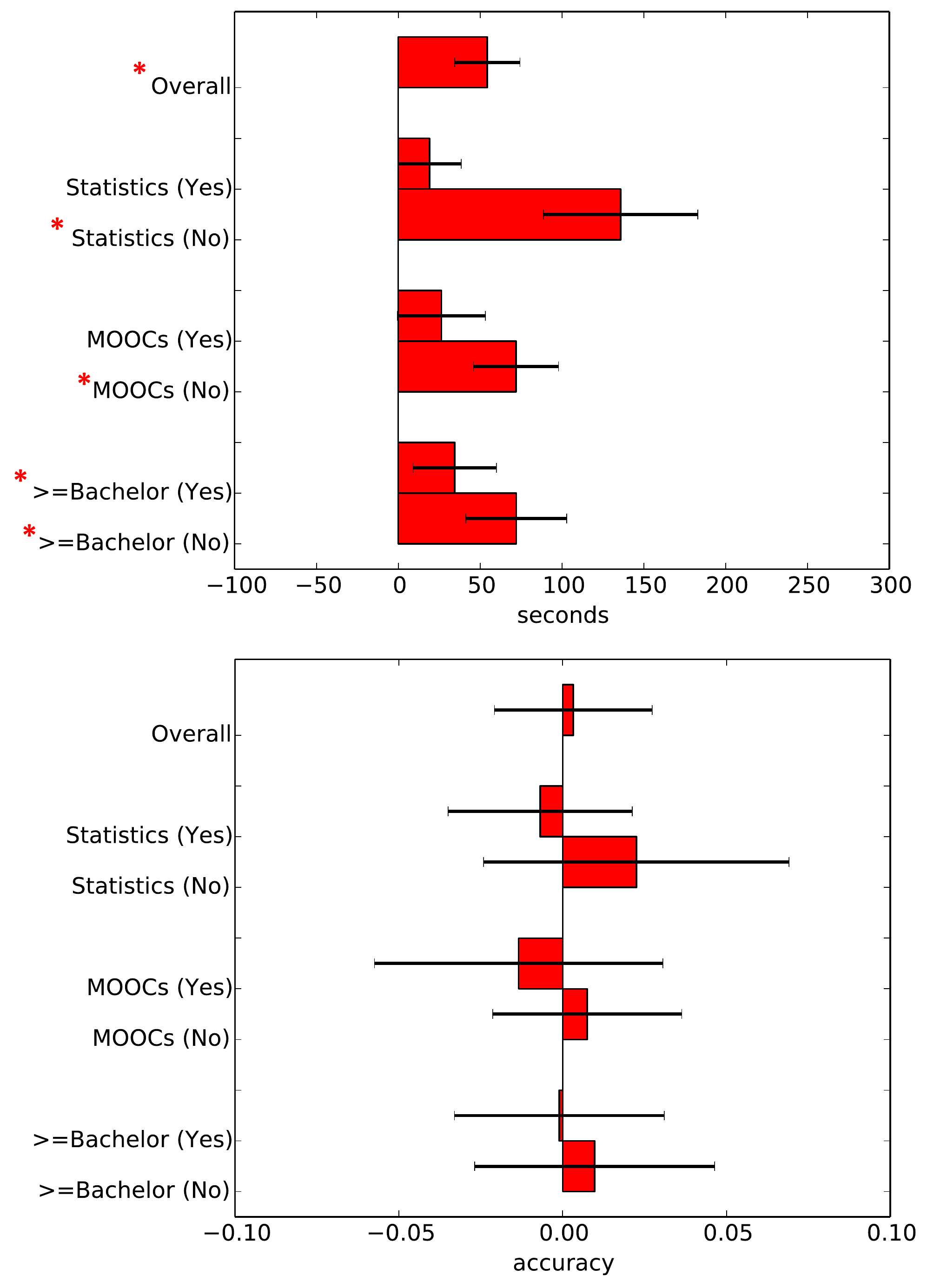}
\centering
\caption{The improvement in search time (upper panel) and accuracy (lower panel) when \textit{linking} interface is used. Learning performance improvement is measured in the study of Stat2.1x. The 95\% confidence intervals (shown as error bars) and significance test results (marked with red asterisk if the difference is statistically significant) are also provided.}
\label{fig:diff_search_stat}
\end{figure}

To examine how providing linking information affects learners in search, we focus on the performance difference between subjects using each of the interfaces. These differences are plotted in Fig.~\ref{fig:diff_search_stat}. For consistency, the length of each bar represents the improvement of a given metric when deploying the \textit{linking} interface as compared to deploying the \textit{null}. Thus, the upper panel corresponds to the average time using \textit{null} interface subtracted by the time using \textit{linking}. In contrast, the lower panel is computed by subtracting the accuracy when using \textit{null} from the accuracy when using \textit{linking}. In the figure, learner cohorts are aligned in the same order as in the table. In addition to the values of differences, the 95\% confidence intervals are also presented. Furthermore, the differences that are statistically significant (we adopt a one tailed, two-sample t-test for the average search time and a one-tailed, binomial proportion test for the average accuracy; significance level is set to 0.05) are marked with red asterisk.

Focusing first on row 1 of Table~\ref{table:search_res_stat}, as well as the first bar in the upper and lower panel of Fig.~\ref{fig:diff_search_stat}, we see that the overall search time is reduced by 36\% (or 54 seconds) when using the \textit{linking} interface (cf. 206 \textit{vs.} 152), and this reduction is statistically significant. In contrast, there is no significant difference in task accomplishing accuracy for using the two interfaces. The result shows that subjects can search for desired information much faster without sacrificing search accuracy, and it provides evidence to our surmise that the linking framework benefits educational content navigation. 

Table~\ref{table:search_res_stat} and Fig.~\ref{fig:diff_search_stat} also include individual results for the three demographic groups. In all of the six cases the \textit{linking} interface yields less search time (with a reduction from 19 seconds to 135 seconds), and the time reduction is statistically significant in four out of the six cases, i.e., subjects without prior knowledge in statistics, without prior exposure to MOOCs, and with/without a bachelor's degree or higher. To interpret these results, we can classify subjects who are less familiar with the course materials, less experienced with MOOC, and less educated as na\"{\i}ve learners. This is because subjects with less familiarity have to take more time and effort to fill the holes of prerequisite knowledge before they can understand a new topic. MOOC learners tend to be self-learners and desire to constantly enrich themselves with learning by utilizing any available resource. In contrast, subjects with no experience in MOOCs are more likely to be passive learners or less comfortable with learning from online materials. As for educating, its purpose is not only to teach students specific knowledge, but also to teach how to learn. Thus, there is a higher chance that less educated subjects have less experience in learning.

In the results we can observe that the \textit{linking} interface yields greater time reduction for novice subjects. This is perhaps not surprising. As pointed out by Kirschner et al. \cite{59}, due to the lack of learning experience and comprehensive understanding of the underlying course topics, typically novice learners cannot properly explore learning content on their own. Without providing guidance to these learners, their cognitive system can be overloaded by new topics they have to learn and prerequisite knowledge holes they have to fill; thus, they are more likely to feel frustration and struggle. As compared to the baseline, our \textit{linking} interface visualizes the linking among pieces of learning content, supports relational navigation among materials, and provides easy access to each sub-goal or sub-concept within a lecture video. These features serve as various guiding functions and help learners navigate through learning content. Therefore, learners can find information more efficiently, and greater improvement is observed in novices since they are those learners who are more likely to struggle, need more support, and can benefit more from the guidance.

As for the search accuracy, although the performance difference between the two interfaces varies from -1.4\% to 2.2\% in various cohorts, none of these discrepancies is statistically significant. Our results indicate that in this experiment linking has little impact on task accuracy. This could be due to the fact that the difference between the two interfaces is about whether the inter-material relation is visualized or not, and the two interfaces are built upon the same set of learning content as well as identical search mechanism. Since Stat2.1x is a rather small MOOC that contains only 7-hour lectures spanning 5 weeks, and we only use a limited set of materials (i.e., videos, slides, and textbook) to build this minimum viable product, it is achievable for learners to find correct pieces of learning content with reasonable time and patience.

From these results in the study of Stat2.1x we conclude that, by having the educational content linked and visualizing the inter-material relation with our \textit{linking} interface, learners can find desired information more efficiently without sacrificing the search correctness. Moreover, among the studied cohorts of learners, novices can benefit more from the provided guidance. These observations show one of the possible ways that linking can help learning.

With the encouraging result, we further expand the horizon of study by exploring another MOOC, 6.00x. Since learning is dependent on the course subject, with the study in 6.00x we attempt to investigate whether the benefit of linking shown above is topic-dependent, or the improvement could be a more general fact. If we can provide evidence showing that linking can yield similar improvement in some learning factors with a different course, it is stronger proof for the benefit of linking to be generalizable to various topics. Additionally, in the study of 6.00x we also attempt to evaluate our linking framework in a more realistic condition. Thus, here materials from online forums are also provided in the interfaces; data annotation (the linking among materials and the correct pieces of learning content for each problem in the user study task) is done by teaching assistants instead of researchers themselves.

Table~\ref{table:search_res_600} summarizes learner performances in the information search scenario in the study of 6.00x. Similar to the study of Stat2.1x, performance is evaluated by the average searching time (columns 1 and 2) and average accuracy (columns 3 and 4) metrics. These metrics are measured within cohorts having various backgrounds (rows 1 to 7) and using different interfaces (columns 1 and 3: \textit{null}; columns 2 and 4: \textit{linking}). We employ similar dividing criteria for the background (i.e., prior knowledge, experience in MOOCs, and highest degree). Besides, the same quality control mechanism (i.e., discarding submissions with top and bottom 5\% search time) is utilized to minimize the noise from spammers. 

\begin{table}[]
\centering
\caption{Learner performance in the information search scenario in the study of 6.00x. Similar to the study above, performance is evaluated by the average searching time and average accuracy metrics, and measured in various cohorts using different interfaces.}
\label{table:search_res_600}
\begin{tabular}{|l|l|cc|cc|}
\hline
\multicolumn{2}{|l|}{\multirow{2}{*}{}} & \multicolumn{2}{c|}{Average search time (seconds)} & \multicolumn{2}{c|}{Average accuracy} \\ \cline{3-6} 
\multicolumn{2}{|l|}{} & \textit{null} & \textit{linking} & \textit{null} & \textit{linking} \\ \hline
\multicolumn{2}{|l|}{Overall} & 443 & 349 & 87.7 & 89.5 \\ \hline
\multirow{2}{*}{Python} & Yes & 419 & 323 & 90.3 & 90.3 \\ \cline{2-2}
 & No & 463 & 378 & 85.6 & 88.6 \\ \hline
\multirow{2}{*}{MOOCs} & Yes & 427 & 336 & 88.0 & 89.4 \\ \cline{2-2}
 & No & 454 & 357 & 87.6 & 89.5 \\ \hline
\multirow{2}{*}{$\geq$Bachelor} & Yes & 472 & 359 & 89.5 & 91.5 \\ \cline{2-2}
 & No & 399 & 331 & 85.1 & 86.2 \\ \hline
\end{tabular}
\end{table}

\begin{figure}[t]
\includegraphics[width=9cm]{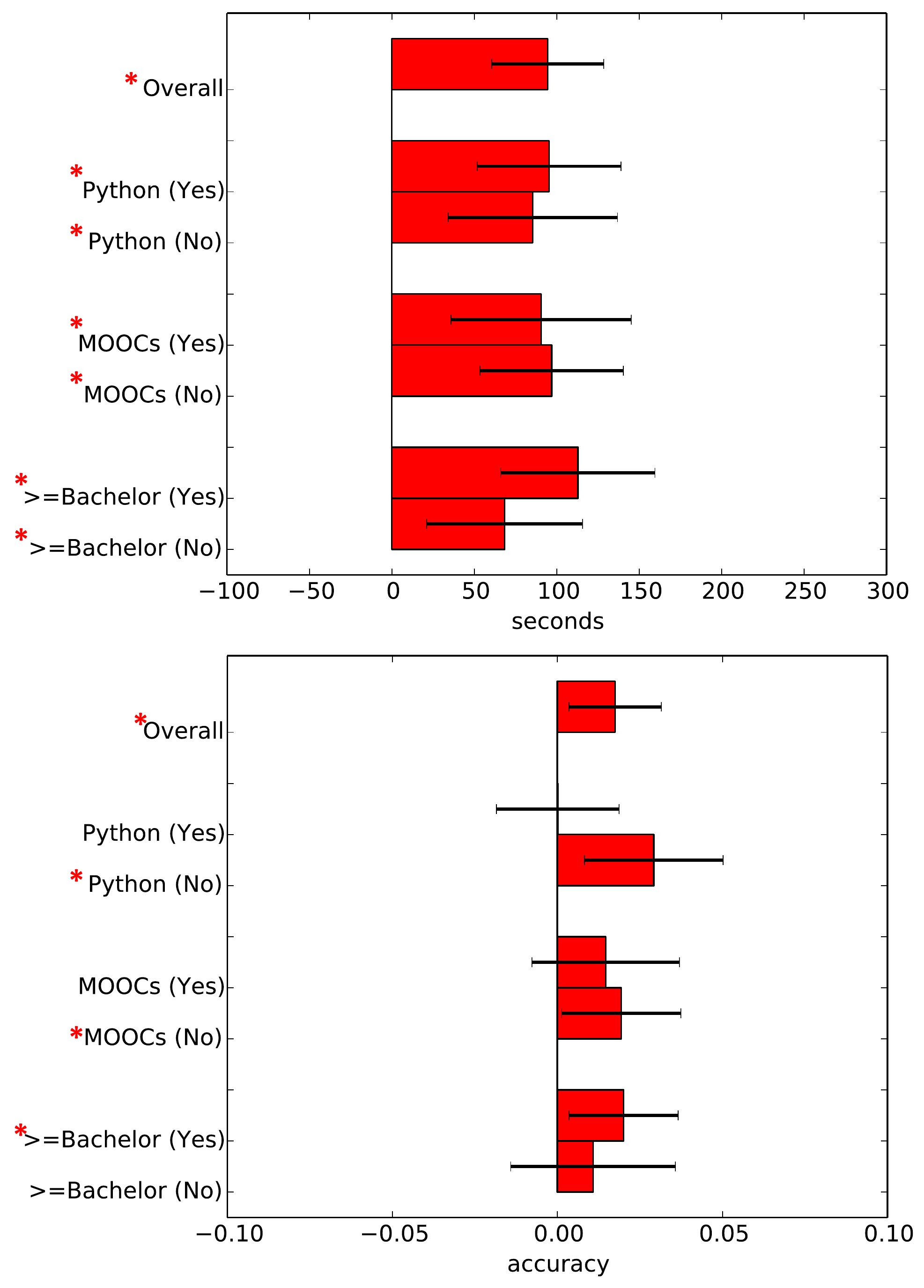}
\centering
\caption{The improvement in search time and accuracy when the \textit{linking} interface is deployed. Learning performance is measured in the study of 6.00x. The 95\% confidence intervals and significance test results are also provided.}
\label{fig:diff_search_600}
\end{figure}

To examine the benefit brought by linking, here we also focus on the performance difference when various interfaces are deployed, and plot the differences in Fig.~\ref{fig:diff_search_600}. Similar to the study in Stat2.1x, values of the bars in this figure represent the time reduction (upper panel) and accuracy increase (lower panel) achieved by deploying the \textit{linking} interface. Improvement in different cohorts is also displayed in the same order as in the table. Besides, the 95\% confidence intervals (the error bars) as well as whether the differences are statistically significant (marked with red asterisk if significant) are also indicated.

From Table~\ref{table:search_res_600} and Fig.~\ref{fig:diff_search_600}, the first thing we can observe is that when the \textit{linking} interface was deployed, each cohort of experimental subjects took significantly less time in accomplishing the tasks. As for the accuracy of completed tasks, a statistically significant improvement from the \textit{linking} interface is found in the entire group of subjects, subjects without prior experience in Python language, subjects without prior exposure to MOOCs, and subjects with at least a bachelor's degree. No significant difference is measured in the other cohorts.

The observations show that our previous conclusion, that, when linking is presented, learners can find desired information more efficiently without sacrificing search correctness, can be reached in the study of a different course. This result strengthens our claim that the idea of linking benefits learning. In addition, we actually see improvement in search accuracy here in several cohorts. This presumably results from an increased amount of learning materials available. In 6.00x the length of the course is three times greater than the length of Stat2.1x, and the discussions are also available for learners. We hypothesize that when learners have more material to navigate through, visualizing the relations between materials will have a larger effect on their ability to find information.

To validate this conjecture, we conducted a regression analysis between search time used in tasks and the accuracy improvement from linking. It would support our hypothesis if we can find evidence showing that when learners spent longer on their tasks, the \textit{linking} interface yielded larger accuracy improvements (note that in the study here subjects spent twice as much time as the time they used in the study of Stat2.1x). We design the regression analysis by first sorting each of the 1,000 tasks in the \textit{null} group and in the \textit{linking} group separately according to the search time. Each set of tasks is then divided into 10 equal sized batches from tasks using the least amount to the most amount of time. We average the search time over the $i^{th}$ batch of the two sets (i.e., tasks using null or linking) as the value of the independent variable of sample $i$ in the regression; we subtract the accuracy of the $i^{th}$ batch of tasks using the $null$ interface from the accuracy of the $i^{th}$ batch using $linking$ as the value of the dependent variable.

\begin{figure}[t]
\includegraphics[width=9cm]{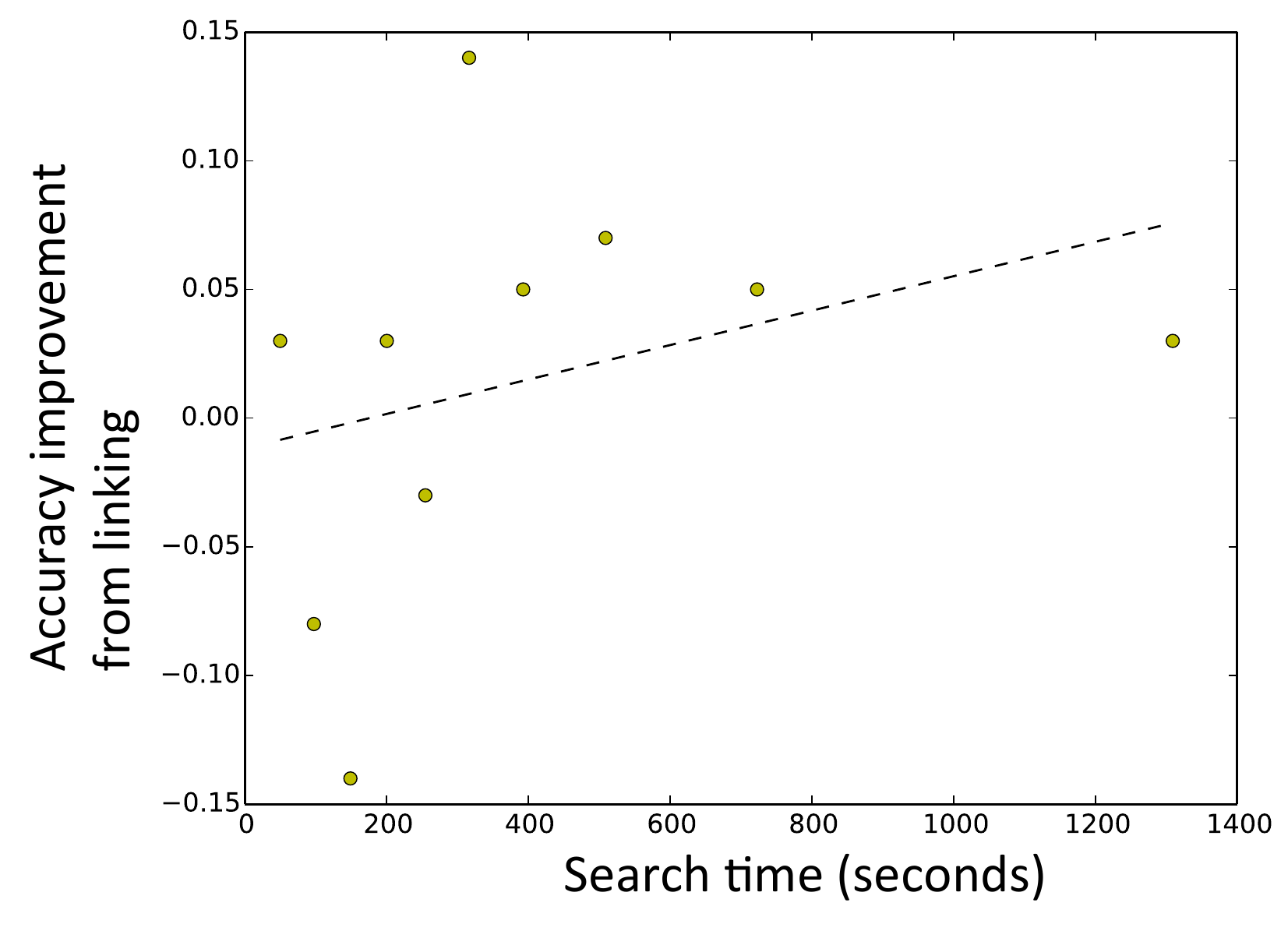}
\centering
\caption{The first order regression model (dashed line) relates the average search time (in seconds) of task batches (horizontal axis) to the accuracy improvement yielded by deploying the linking interface (vertical axis).}
\label{fig:regression}
\end{figure}

The result of our regression analysis is plotted in Fig.~\ref{fig:regression}. Clearly our regression model has a positive slope, which shows that search time and improved accuracy are positively correlated. This observation supports our previous conjecture. However, we must note that the $p$-value of a hypothesis test that the slope is positive is 0.17, which is higher than the usually used significance level of 0.05.

The second observation that can be made from Table~\ref{table:search_res_600} and Fig.~\ref{fig:diff_search_600} is that both na\"{\i}ve and advanced learners benefit from the $linking$ interface in terms of reduced task completion time. We believe this finding is related to the subject matter in the user study. 6.00x covers a wide range of advanced topics such as algorithms, complexity, computational problem solving, and Python language programming. As compared to Stat2.1x, the statistic course contains material which is typically taught systematically in a high school class, most people are usually familiar with only part of the topics in 6.00x. For instance, a computer scientist might know algorithms and theory of computation but might not use Python; a data analyst might be familiar with using Python to analyze data, but might not be an expert in algorithms. Thus, some subjects classified as advanced learners based on our categorization could be beginners in the topics used in some of the learning task\footnote{This can be illustrated with some feedback we received in this user study. One said, \textit{"I was already familiar with most of the concepts except for dynamic programming and program complexity. I was a computer science major 30 years ago. Back then they were teaching IBM 360/370 assembly language and FORTRAN 77. I code now in C, Python, PHP, SQL, shell scripts and elisp for my own little projects now and again, although those are too small to warrant much attention for dynamic or complexity considerations."} Another participant mentioned, \textit{"I already know much about Python, but I find out new things doing these! I don't think I ever really understood the use of recursion until I completed this task."}}. Hence we believe this may be why a more uniform improvement over cohorts is observed.

In conclusion, the results in the study of 6.00x provide more evidence supporting the benefit of linking in learning. With the $linking$ interface, learners can also find desired information faster. The improvement in learning performance is observed in both search time and accuracy, and in more cohorts of subjects. These results also imply the potential of applying linking to various course subjects for better learning. In section 3.5, we will analyze learners' click-through patterns in order to provide more understanding and examples about why linking yields better performance, but in the next section, we will first attempt to analyze the benefit of linking from another measure: information memorization.

\subsection{How linking affects information memorization}
As compared to "finding the desired information", concept retention is a more complicated scenario involving searching, integrating, and memorizing knowledge. In this section we investigate whether linking can enhance learners' performance in this complex condition. To evaluate the performance we compute one metric: number of unique key-terms. This metric measures the information richness in the paragraphs submitted by subjects, and thus it reflects how many concepts relevant to the assigned topic learners can retain after a fixed-length learning stage. We adopt a rather simple metric (as compared to other metrics that evaluate and grade essays \cite{130}) and inform subjects how their submissions will be evaluated, in order to give learners a concrete goal and simplify the learning tasks. We attempt to tease out from the tasks the factors that are not closely related to material navigation (e.g., the fluency or wording in the essay). Furthermore, other complicated metrics are usually subjective and hard to be generalized to various domains (e.g., requiring many manually graded essays for an automated grading algorithm to learn from), and thus they do not align with our purpose. 

For computing our key-term metric, we only need to label a set of relevant terms for each topic used in the concept retention scenario. For the labeling, we first designate the glossary in each textbook used in Stat2.1x and 6.00x as the set of candidate terms. The same annotators recruited in Section 3.4.1 (i.e., the information search scenario) were also asked to label the relevant topics for each term in the candidate set. With the annotation, we can compute the metric simply by counting how many terms were covered in the essay (since multiple words can sometimes refer to the same word stem, e.g., cats, catty, and cat, in practice we first conduct word stemming \cite{131} to reduce the derived or inflected words in the essay to their word stem before counting the key-terms). This metric allows us to understand how linking affects learners when they are trying to acquire and remember high-level information about a topic.

\begin{table}[]
\centering
\caption{Learner performance in the concept retention scenario in the study of Stat2.1x. Performance is evaluated by the number of unique key-terms in submitted essays and measured within various cohorts using different interfaces.}
\label{table:retention_res_stat}
\begin{tabular}{|l|l|cc|l|l|l|l|}
\hline
\multicolumn{2}{|l|}{}            & \multicolumn{6}{c|}{Number of unique key-terms}           \\ \hline
\multicolumn{2}{|l|}{}            & \textit{null}     & \multicolumn{5}{c|}{\textit{linking}} \\ \hline
\multicolumn{2}{|l|}{Overall}     & 4.39              & \multicolumn{5}{c|}{4.91}             \\ \hline
\multirow{2}{*}{Statistics} & Yes & 4.71              & \multicolumn{5}{c|}{5.11}             \\ \cline{2-2}
                            & No  & 3.98              & \multicolumn{5}{c|}{4.60}             \\ \hline
\multirow{2}{*}{MOOCs}      & Yes & 4.83              & \multicolumn{5}{c|}{5.14}             \\ \cline{2-2}
                            & No  & 4.27              & \multicolumn{5}{c|}{4.77}             \\ \hline
\multirow{2}{*}{$\geq$Bachelor}   & Yes & 4.73              & \multicolumn{5}{c|}{5.23}             \\ \cline{2-2}
                            & No  & 3.98              & \multicolumn{5}{c|}{4.60}             \\ \hline
\end{tabular}
\end{table}

Table~\ref{table:retention_res_stat} summarizes learner performances in the concept retention scenario in the study of Stat2.1x. Performance is evaluated by the number of unique key-terms contained in submitted essays. As in the information search scenario, the evaluation is also computed within cohorts having various backgrounds (rows 1 to 7) and using different interfaces (column 1: $null$; column 2: $linking$). We found that there are workers trying to cheat on the tasks by copying and pasting paragraphs they found online (e.g., Wikipedia) in their essays. Therefore we utilized an open online plagiarism checker \cite{126}. This checker segments paragraphs to be checked into sentences, searches these sentences on Google, and reports plagiarism if some highly similar documents are found on the Web. With this checker, we identify the spammers, reject their results, and control the quality of the experiment.

\begin{figure}[t]
\includegraphics[width=9cm]{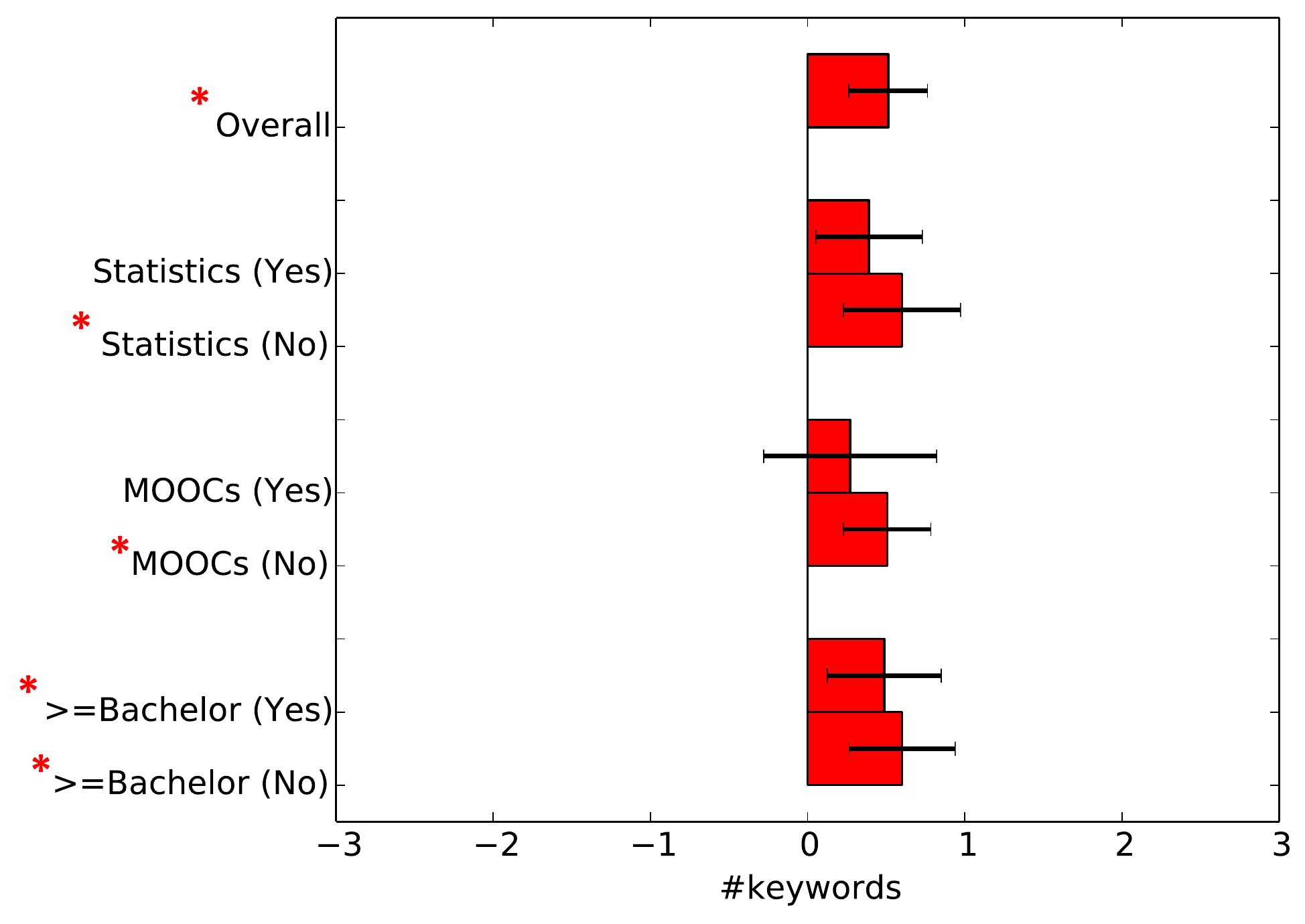}
\centering
\caption{The improvement in the number of unique key-terms contained in submitted essays when $linking$ interface is used. Learning performance is measured in the study of Stat2.1x. The 95\% confidence intervals and significance test results are also provided.}
\label{fig:diff_retention_stat}
\end{figure}

To focus on the performance difference when various interfaces were deployed, we also visualize the improvement from the $linking$ interface measured in each subject cohort in Fig.~\ref{fig:diff_retention_stat}. That is, the length of each bar represents the average number of unique key-terms when $linking$ is deployed subtracted by the number when $null$ is used. Additionally, the 95\% confidence intervals (the error bars) as well as whether the differences are statistically significant (marked with red asterisk if significant) are also indicated. Here, we adopted a one tailed, two-sample t-test for significance test and set the significance level to 0.05.

The first row of Table~\ref{table:retention_res_stat} and the first bar in Fig.~\ref{fig:diff_retention_stat} show that, overall, subjects are able to mention a greater number (12\%) of key-terms when using the $linking$ interface (cf. 4.39 $vs$. 4.91), and the difference is statistically significant. Looking over the rest of Table~\ref{table:retention_res_stat} and Fig.~\ref{fig:diff_retention_stat}, we observe that there is a similar trend to that in the information search scenario, where the $linking$ interface yields improvement over each cohort of subjects, and in four out of the six cases (i.e., subjects with no prior knowledge in statistics, without prior exposure to MOOCs, and with/without a bachelor's degree or higher) the differences pass the significance test. Furthermore, it seems that the novices also benefit more from linking than advanced learners (e.g., all the three na\"{\i}ve cohorts show statistically significant improvement).

These results reveal another aspect of benefit linking may provide. In the course materials, there are usually many learning pieces relevant to a topic; some of them are complementary and some might be redundant. With the visualized inter-material relation, complementary content can be identified easily and thus be better utilized to reinforce learning. For example, while watching the lecture video, subjects can refer to the aligned slides to understand the lecture at the concept level, as well as to the linked textbook sections or posts for detailed discussions. The identical learning content can also be skipped easily. Furthermore, the visualization helps learners better plan their learning path within the limited-length learning session, and avoid exploring irrelevant or secondary content to the assigned topics. These features made possible by visualizing linking can also be interpreted as the guidance which leads learners navigating through learning content when accomplishing assigned tasks. Therefore, subjects, especially novices, can access knowledge more efficiently in the learning session, and retain more key-terms when they write down what they can remember. In section 3.5 we will provide more evidence to our claim here.

We also investigated whether this aspect of benefit can be generalized to various course subjects in a more realistic condition. Hence, similar to the information search scenario, we further studied a different MOOC, 6.00x, as well as explored in an expanded material set (i.e., forum discussions were additionally used) and data annotation pipeline (i.e., teaching assistants were recruited as annotators).

\begin{figure}[t]
\includegraphics[width=9cm]{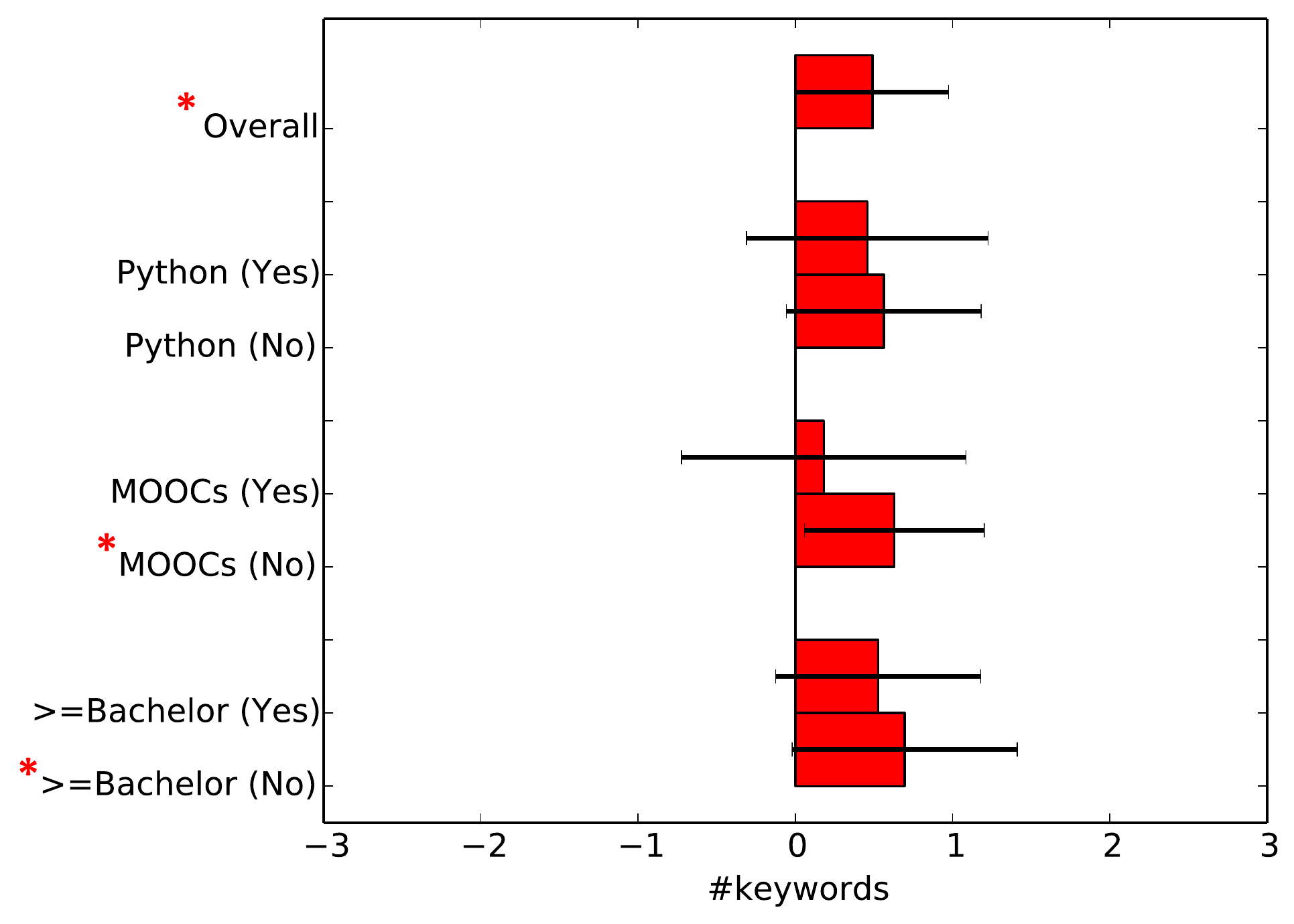}
\centering
\caption{The improvement in the number of unique key-terms when the $linking$ interface is deployed. Learning performance is measured in the study of 6.00x. The 95\% confidence intervals and significance test results are also provided.}
\label{fig:diff_retention_600}
\end{figure}

Table~\ref{table:retention_res_600} summarizes learner performances in the concept retention scenario in the study of 6.00x. Similarly, performance is evaluated by the average number of unique key-terms in the submitted essays, and measured within various cohorts (rows 1 to 7) using different interfaces (column 1: $null$; column 2: $linking$). In this study, the same quality control mechanism using plagiarism checking is employed to filter out the noise from spammers. Furthermore, to focus on the benefit brought by linking, in Fig.~\ref{fig:diff_retention_600} we plot the performance difference when various interfaces were deployed (i.e., number of key-terms when $linking$ is assigned subtracted by the number when the $null$ interface is used). In addition to the differences, the 95\% confidence intervals (the error bars) as well as the significant test result (marked with red asterisk if significant) are also indicated in the figure.

\begin{table}[]
\centering
\caption{Learner performance in the concept retention scenario in the study of 6.00x. Similar to the study of Stat2.1x, performance is evaluated by the number of unique key-terms in submitted essays and measured within various cohorts using different interfaces.}
\label{table:retention_res_600}
\begin{tabular}{|l|l|cc|}
\hline
\multicolumn{2}{|l|}{\multirow{2}{*}{}} & \multicolumn{2}{c|}{Number of unique key-terms} \\ \cline{3-4} 
\multicolumn{2}{|l|}{} & \textit{null} & \textit{linking} \\ \hline
\multicolumn{2}{|l|}{Overall} & 8.07 & 8.56 \\ \hline
\multirow{2}{*}{Python} & Yes & 8.64 & 9.09 \\ \cline{2-2}
 & No & 7.64 & 8.20 \\ \hline
\multirow{2}{*}{MOOCs} & Yes & 8.37 & 8.55 \\ \cline{2-2}
 & No & 7.93 & 8.56 \\ \hline
\multirow{2}{*}{$\geq$Bachelor} & Yes & 8.60 & 9.13 \\ \cline{2-2}
 & No & 7.21 & 7.91 \\ \hline
\end{tabular}
\end{table}

From Table~\ref{table:retention_res_600} and Fig.~\ref{fig:diff_retention_600}, first we can observe that when the $linking$ interface was deployed, subjects in each cohort mention a greater number of key-terms. This observation is identical to the results in Stat2.1x. However, we can find that the improvement is statistically significant in fewer cohorts (i.e., in the entire group of subjects, subjects without prior exposure to MOOCs, and subjects without a bachelor's degree). We might be observing fewer cohorts showing significant improvement because of the advanced topics used in this user study. As discussed above, our stratification of learners, which is inherited from the study of Stat2.1x, might not be able to tell novices from advanced learners. Therefore learners in each cohort are too diverse to perform consistently. This claim is supported by the much larger standard deviations (i.e., longer error bars) observed in Fig.~\ref{fig:diff_retention_600} than the ones in Fig.~\ref{fig:diff_retention_stat}.

These results also strengthen our previous claim that, when linking is shown in the interface, learners can access information more efficiently and retain more key-terms in their summary of assigned topics. The improvement not only suggests another benefit of linking in learning, but also implies the possibility of applying our framework in various course subjects. However, these observations only support our hypothesis that linking is helpful in learning, but cannot explain why. Without knowing the reason, we cannot utilize this linking pedagogy in suitable conditions. Thus, in the next section, we analyze the learners' click log in order to provide an explanation.

\section{Click log analysis}
From the user study results discussed above, we can summarize that, by presenting the linking among learning materials, learners can access course materials more efficiently and perform better in learning tasks. However, we are also curious to discover how the linking changes learners' navigation behavior and why the change yields improvement in accomplishing our learning tasks. Therefore, we examine the click log\footnote{In our study of 6.00x, in addition to the submitted answers (i.e., the selected learning object or the essay summarizing the assigned topic), we also record how learners interact with our interfaces. Whenever a learner initiates an event to search a query or click on any learning object, the event along with the triggered time is stored in our server.} generated when learners attempted the tasks. We utilize the log to extract the following three metrics:

\begin{itemize}
  \item \textbf{Number of search queries} used in each learning task.
  \item \textbf{Number of learning objects}\footnote{Note that because of the way we recorded the event, here the definition of a learning object is slightly different from the rest of this thesis. A learning object in this section refers to a lecture video, a page of lecture slides, a textbook section, or a discussion thread.} surveyed in each task.
  \item \textbf{Time spent} (measured in seconds) in each surveyed learning object.
\end{itemize}
In each learning scenario (i.e., search and retention), we compute the three metrics averaging over the tasks using each of the two interfaces (i.e., $null$ and $linking$). The results are summarized in Table~\ref{table:click_log_analysis}.

\begin{table}[]
\centering
\caption{In information search and concept retention scenario of the study of 6.00x, the three metrics (number of search queries used to accomplish a task, number of learning objects surveyed in each task, and the spent time in each learning object) when various interfaces were deployed are computed. The averages ($\mu$) and standard deviations ($\sigma$) of the three metrics are listed here.}
\label{table:click_log_analysis}
\begin{tabular}{|l|l|cc|cc|}
\hline
\multicolumn{2}{|l|}{\multirow{2}{*}{}} & \multicolumn{2}{c|}{Information search} & \multicolumn{2}{c|}{Concept retention} \\ \cline{3-6} 
\multicolumn{2}{|l|}{} & \textit{null} & \textit{linking} & \textit{null} & \textit{linking} \\ \hline
\multicolumn{2}{|l|}{\#Search queries ($\mu$, $\sigma$)} & (2.9, 2.9) & (2.7, 2.8) & (1.6, 1.2) & (1.4, 0.9) \\ \hline
\multicolumn{2}{|l|}{\#Learning objects ($\mu$, $\sigma$)} & (10.9, 9.7) & (7.7, 7.5) & (11.8, 8.7) & (7.8, 6.6) \\ \hline
\multicolumn{2}{|l|}{Spent time per object ($\mu$, $\sigma$)} & (32.3, 73.4) & (35.1, 64.9) & (46.0, 88.4) & (70.6, 115.8) \\ \hline
\end{tabular}
\end{table}

Comparing first the numbers in the two scenarios, we observe that in the information search scenario, learners tend to use more search queries, survey more learning objects, and spend less time on each object. Considering the average time learners have to spend in interacting with the interfaces in the two scenarios (6.6 minutes in information search in average and 10 minutes in concept retention), the difference in numbers of queries and learning objects between the two scenarios is even larger. We surmise that the discrepancy results from the nature of the two scenarios. In the search scenario, learners have to only identify the objects which contain information for answering the assigned problems; however, they need to decide which search queries to use. As for the retention scenario, learners have to digest and remember information in the content, but it is obvious that they should use the assigned topics or relevant terms as the queries. Thus, in the search scenario, learners are inclined to survey more queries and learning objects, but spend less time on each of the queries or objects.

Then we juxtapose the metrics of each interface within the two scenarios. We find that, as compared to using the $null$ interface, when $linking$ is deployed, experimental subjects tend to use fewer search queries (information search: 2.7 $vs$. 2.9, concept retention: 1.4 $vs$. 1.6), survey fewer objects (information search: 7.7 $vs$. 10.9, concept retention: 7.8 $vs$. 11.8), but spend more time on each object (information search: 35.1 $vs$. 32.3, concept retention: 70.6 $vs$. 46.0). We believe this observation explains our user study results. The observation suggests that, when linking is visualized, learners are able to identify those learning contents which are more informative for the assigned topics or problems; in contrast, when $null$ is deployed, learners have to try more queries and learning objects. Thus, with our $linking$ interface, learners can spend more time in understanding relevant information for the search or retention tasks, and achieve better performance. The observation here and the user study results can also be related by reduced cognitive load, which has a positive effect on learning \cite{59}. When linking is presented, it is easier for learners to filter out less useful learning objects, which alleviates learners' cognitive load in understanding the materials.

To support our conjecture, we further investigate how relevant the learning objects learners surveyed are to their assigned tasks. In order to measure the relevance, we utilize the labeled valid learning pieces which were used in evaluating whether the selected learning content is correct in the information search scenario. With the click log recorded in this search scenario, we measure the percentage of surveyed learning objects which contain at least one valid learning piece for the assigned problem.

The mean and standard deviation of percentage is 0.33 and 0.26 for tasks using the $null$ interface, and 0.50 and 0.33 for tasks using the $linking$ one. We can observe that when $linking$ is deployed, learners tend to survey more objects containing valid learning pieces. This fact supports our previous claim that with the relation among learning materials presented, learners are able to filter out learning objects which are less likely to contain useful content, and focus on informative objects. Thus, better learning outcomes are achieved.

To further illustrate how linking can help learners identify useful information, we visualize two sampled search paths recorded from two subjects when they were completing assigned tasks using different interfaces. We sampled the two paths based on the following criteria. First, in the paths, subjects surveyed the same number of learning objects as the interface average (i.e., 10 objects for the path recorded in the task using the $null$ interface and 7 objects for the $linking$ one). Second, learners surveyed the same number of objects that contain at least one valid learning piece as the interface average (i.e., 3 informative objects for both interfaces). Third, the two paths were recorded from tasks assigned with the same question.

\begin{figure}[t]
\includegraphics[width=15cm]{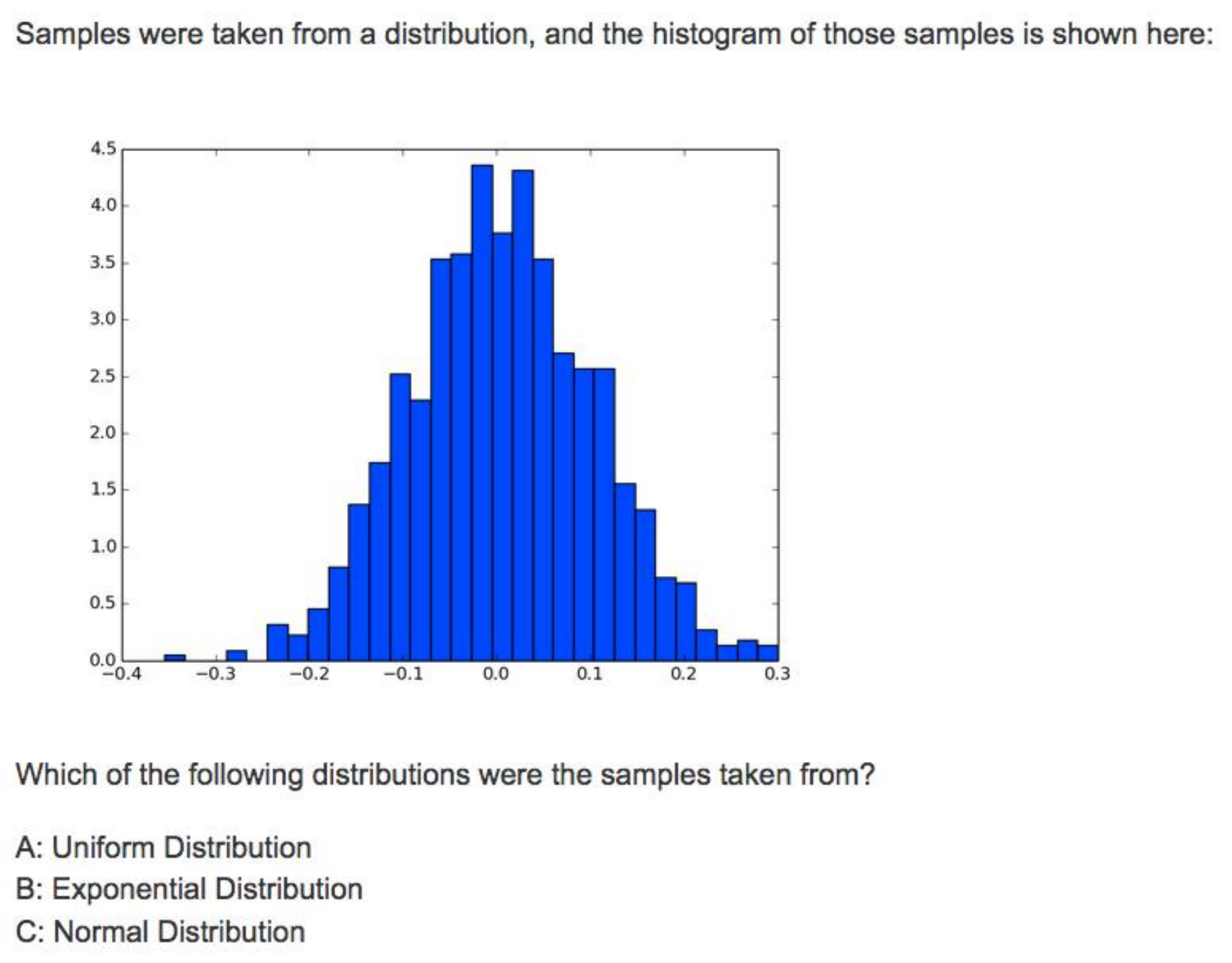}
\centering
\caption{The question asked in the tasks where we recorded the two sampled search paths.}
\label{fig:task_q}
\end{figure}

\begin{figure}[t]
\includegraphics[width=14cm]{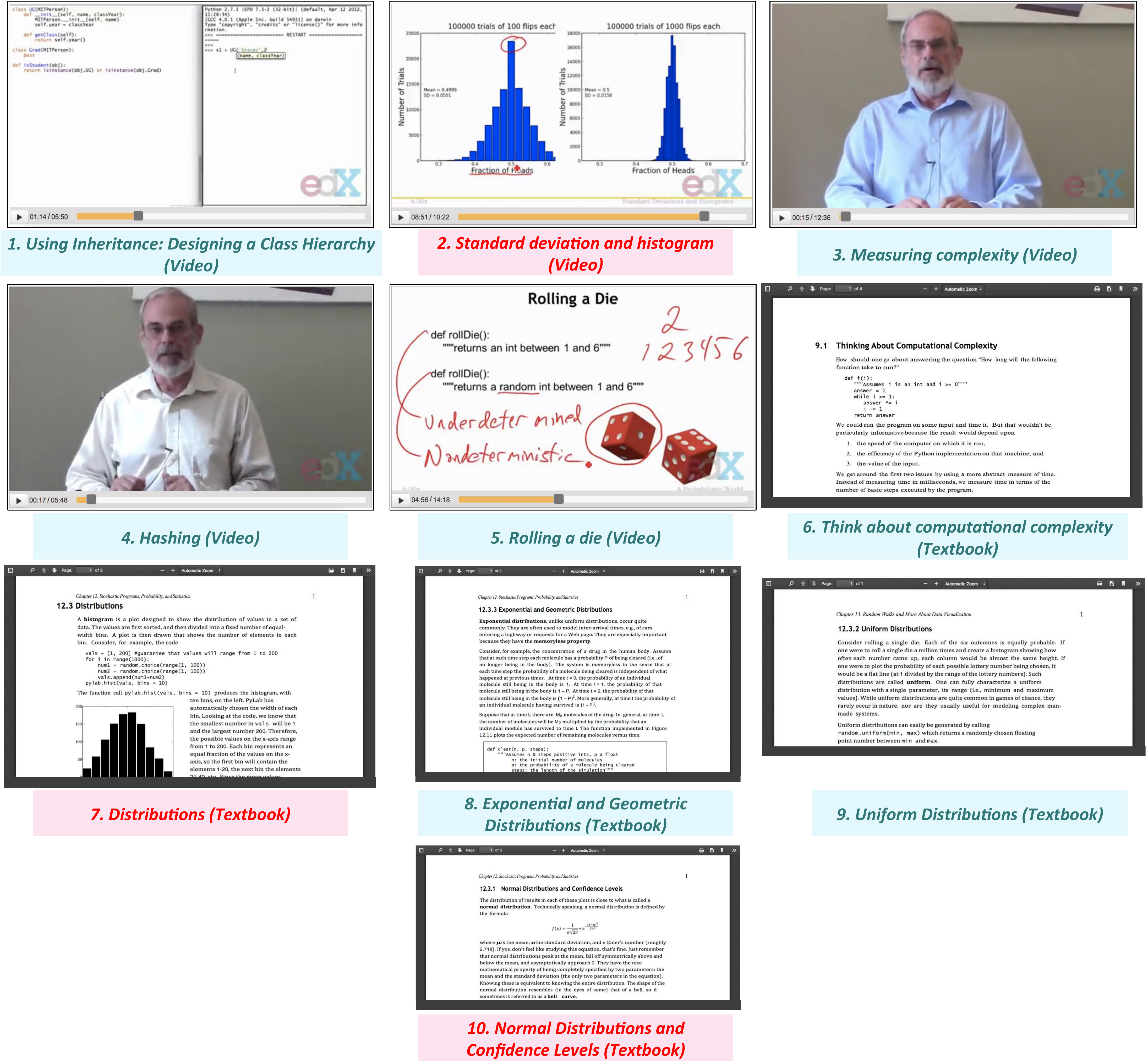}
\centering
\caption{The sampled search path recorded when a subject used the $null$ interface to complete the assigned task. In this path this subject surveyed 10 objects and 3 of them contain valid learning pieces. The 3 objects are indicated with their titles in red; the titles of the rest objects are put in cyan. These numbers equal the average of the $null$ interface.}
\label{fig:null_path}
\end{figure}

\begin{figure}[t]
\includegraphics[width=14cm]{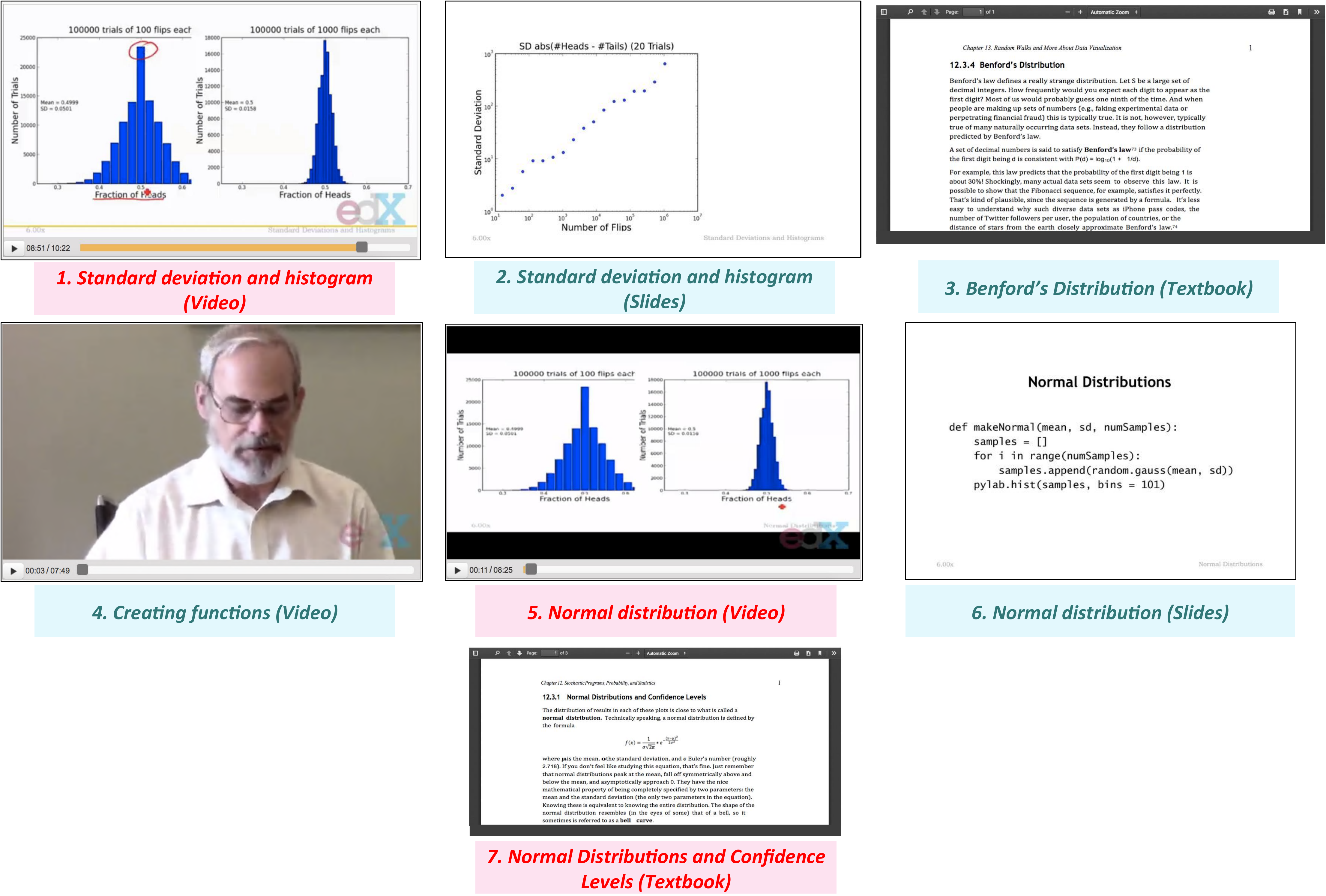}
\centering
\caption{The sampled search path recorded when a subject used the $linking$ interface to complete the same task as in Fig.~\ref{fig:null_path}. In this path this learner surveyed 7 objects and 3 of them contain valid learning pieces. These numbers equal the average of the $linking$ interface.}
\label{fig:linking_path}
\end{figure}

In Fig.~\ref{fig:task_q}, we show the question the two sampled search paths corresponded to. This question is about the normal distribution. In Fig.~\ref{fig:null_path} and \ref{fig:linking_path}, two search paths observed in tasks using the $null$ and $linking$ interface respectively are presented. In these paths, screenshots and titles of surveyed learning objects are listed according to the visited order; the material type of each object is also indicated. Furthermore, the titles of objects containing valid learning pieces are put in red; the titles of other objects are in cyan.

From these two paths, we first notice that the learner who was assigned with the $linking$ interface has significantly better survey quality. Most objects this learner selected are relevant to the topic of the normal distribution. In contrast, the search path of the learner using the $null$ interface seems unplanned. This observation agrees with our previous claim that learners can benefit from linking since they are able to more easily filter out less informative materials and focus on the useful ones. Furthermore, we note that the learner using the $linking$ interface switched between different types of materials more frequently. This fact also strengthens our assertion that when linking is presented, learners are capable of utilizing complementary information from various types of materials to reinforce their learning.

In this section, we analyzed the click log recorded in the user study of 6.00x, in order to explain why linking yields better learning performance in our experiment. We found that, when the linking among materials is presented to learners, they survey fewer queries and objects as well as spending more time in each object. Moreover, the surveyed object is more informative and therefore learners can perform better in tasks. However, we should note that the differences in metrics computed here are not statistically significant. Hence, further stratification of samples is required to obtain stronger evidence.

\section{Conclusions}
This chapter explores our first research question: can manually generated linking help learning? We start by defining two types of linking: homologous and heterologous linking. We formulate the annotation of these two linking types as alignment and binary classification problems respectively, and demonstrate how the annotation can be done by researchers, online workers, and course staff. Then, we implement a $linking$ interface that can present learning materials and the annotated linking among them simultaneously. After that, we conduct a large-scale user research study with the two selected STEM courses, statistics and programming language, to investigate the question.

Our user research shows that, this $linking$ interface enables learners to search for desired learning content more efficiently and retain more concepts more readily. By analyzing the click log recorded when learners using various interfaces in the user research, we observe that, presenting both content and linking at the same time helps learners focus on informative learning materials, and thus potentially reduces learners' cognitive load. These results support the notion that manual linking can indeed improve learning outcomes. 

\chapter{Can we link automatically?}
In this chapter, we investigate methods to link courseware automatically. We showed in the previous chapter that linking can assist learners navigating through course materials, and help learners find supportive learning content when they are in need, e.g., confused. The visualization of relation among content also provides guidance to mitigate learners' cognitive load. Hence, learning experience and outcome can be enhanced.

However, from our experience in developing a linking system, annotating relation information requires deep and comprehensive understanding of the course subject, and the labeling itself is time-consuming. Moreover, even though the linking is generated for the current class, obtaining the relation annotation for future offerings requires a lot of redundant work - since in a new class offering, more forum posts come in (but many of them are duplicated), and several lectures may be re-organized, we have to maintain and repeat labeling on the updated materials. Thus, implementing the educational content linking framework manually is cost-prohibitive, not efficient, and not scalable. In order to allow the proposed framework to be more easily deployed and in a more general condition, especially when the size of learning materials is enormous, we investigate whether linking of learning content can be generated by machines.

In the following investigation of automated linking methods, we focus on three issues: 
\begin{itemize}
  \item How to design an automated linking method?
  \item How closely can the automatically generated linking approach to human annotation?
  \item Can automated linking still benefit learners?
\end{itemize}
For the first issue, we will discuss how to formulate linking generation as a prediction/classification problem, as well as how to solve the problem with machine learning and human language technologies. For the second issue, we will compare machine generated linking with ground truth established in advance by a human (i.e., the linking we collected in the previous chapter). The comparison can be made easily and shed light on the capability of implemented automated methods. However, such evaluation is not perfectly precise, since it measures the similarity between two generated linkings, instead of our ultimate goal: learning outcome improvement. Furthermore, typically there are many configurations of linking which might benefit learners. Computing the similarity to a gold standard is a biased metric which ignores all the other beneficial possibilities. Therefore, we also explore the third issue by conducting a user study on the automated linking, with a pipeline similar to that designed in the previous chapter. Although conducting a user study is more costly and time consuming, it evaluates directly the benefits of machine generated linking on learners.

We explore the three issues on the same two MOOCs: Stat2.1x and 6.00x. In our experiment, we find that, although there are some differences between the machine and human generated linking, they only result in slight negative effects on learning. Moreover, the interface driven by automatic linking (denoted as the auto linking interface in the following) still allows learners to achieve better performance in their tasks. Furthermore, we analyze the difference pattern, and conclude that anecdotally most disagreements between human annotators and our contributed linking algorithm do not make much difference to learners. In the rest of this chapter, we describe our exploration in detail.

The study discussed above was conducted by comparing an interface presenting linking information to a baseline that implements the conventional strategy for delivering learning materials online (i.e., the null interface). Since MOOC is an emerging research field in education, researchers and educators have constantly been integrating innovative pedagogies into the design of MOOC. We are curious how effective our proposed linking framework is as compared to these state-of-the-art techniques. Therefore, in this chapter we conduct another user study to compare our linking/auto linking interface to the interface currently deployed on edX (denoted as the edx interface). Our results show that learners still achieve better performance in the explored learning tasks with our manually and automatically generated linking.

\section{Problem formulation}
To design our automated algorithm, we choose to focus on the natural language content in course materials and formulate the linking generation task as a sequential tagging problem. Natural language is an integral part of education for knowledge transferal; thus we believe an algorithm based on the understanding of natural language in learning content can be more generalizable to different courses. The sequential tagging formulation is as follows: for nodes on the trunk of a linking tree, we view them as a sequence of documents, and represent them as a sequence of input feature vectors $\textbf{x}=(\textbf{x}_1,\textbf{x}_2,...,\textbf{x}_\mathrm{T})$. Determining which supplementary objects should be linked to these nodes can be interpreted as predicting a sequence of labels $\textbf{y}=(Y_1,Y_2,...,Y_\mathrm{T})$ given $\textbf{x}$. Here $Y_i$ represents the linking configuration of trunk node $i$, e.g., the index of an aligned slide or whether a given discussion thread is linked to node $i$ or not.

We adopt the sequential tagging formulation, because the ordering and context is informative when modeling learning materials. Many theories in cognitive science of learning suggest that, to achieve meaningful learning\footnote{The state where the newly acquired knowledge is fully understood and ready for future used in different circumstances.}, humans have to match the information they learn to how their minds are structured, and integrate the new information with their prior knowledge and existing cognitive system \cite{18, 29}. Guided by these theories, learning content is typically structured in a sequentially dependent manner to help students acquire knowledge. Thus, we surmise that the contextual dependency can be helpful in linking prediction, e.g., it could be more probable that neighboring video segments are linked to similar or identical sets of supplementary learning objects. 

There are many other applications of this sequential tagging formulation in the natural language understanding domain, including part-of-speech (POS) tagging \cite{136}, semantic tagging \cite{137}, and machine translation \cite{134}. These applications work on the granularity of words (i.e., each token is a word) and attempt to interpret the meaning of each token. Since natural language is usually interpreted in sequence by a human, modeling the context is also beneficial in understanding the syntax and semantics. In our formulation, a learning object, which can be a video segment, a slide, a textbook section, or a post thread, is adopted as a token, and we model the contextual and lexical dependency upon this larger unit with similar formulation.

Due to the abundance of applications, many machine learning models were proposed to solve the problems, including linear-chain CRF, Hidden Markov Models (HMM), general graphical models, and long short-term memory (LSTM) recurrent neural networks \cite{135}. In this thesis, we also adopt linear-chain CRF for our linking problem; HMM can be interpreted as a linear-chain CRF with generative modeling\footnote{Linear-chain CRF is a discriminative model.}; as compared to linear-chain CRF or HMM. A general graphical model removes the limitation of the Markov property and thus has higher complexity\footnote{Higher complexity means that the underlying model has more free variables and more ability to represent dependency in data.}; LSTM is a model with even higher complexity and non-linearity, and it is widely used to express complicated data dependencies.

We choose linear-chain CRF for several reasons. First, the size of our corpus is relatively small as compared to many other natural language applications. This is due to the difficulty of data annotation: labeling the relations among learning objects requires deeper understanding of the content as compared to annotating the POS tags or semantics of each word. Our corpus size cannot afford the training of complicated models such as LSTM or general graphical models because of the hazard of overfitting. Second, as compared to non-linear models such as LSTM, a linear model makes it easier to understand the results. For instance, each weight in the model can be directly interpreted as the importance of a corresponding feature. The interpretation can facilitate subsequent system development. Third, a discriminative model is preferred in our tasks, since it allows us to incorporate new features into the model without making unnecessary assumptions about the underlying probabilistic distribution\footnote{A generative model such as HMM is based on a full probabilistic model. It has to model the probabilistic distribution of all variables, including the observed (i.e., the features or observations) and unobserved (i.e., the labels) ones. The distribution can be learned from a corpus from scratch but the learning usually requires too many data samples. We can also make assumptions to initialize or limit the distributions to a specific form for the learning to be feasible (especially when we do not have enough data). However, making assumptions is error-prone. In contrast, a discriminative model only models the dependency between observations and the unobserved variables that should be inferred. In such a model we do not have to learn the entire distribution or make unnecessary assumptions. Therefore, it is much easier to augment a discriminative model with new features (i.e., introducing new variables and dependencies into the model).}. This property allows the automated algorithm to be extended easily to various features, which is favorable because there are usually many information modalities in course materials. In the following, we will discuss how to predict linking with CRF under this formulation. 

\section{Sequential tagging with CRF}
As we discussed in Section 3.1, we categorize the relations among learning materials into homologous and heterologous linking, and adopt different annotation methodologies (i.e., the alignment and classification). We follow the same categorization and design our automated linking algorithm for the two types of relation respectively.

In the homologous case, we apply the linear-chain CRF model to solve the alignment problem. First, we have two types of materials to be linked: one is the trunk and the other is a set of candidates of leaves. In this thesis, we designate the trunk as the lecture videos in the course; thus the candidates in this homologous case are the lecture slides corresponding to each video. For each lecture $i$, the video transcription sentences form the sequence of input feature vectors $\textbf{x}^{(i)}=(\textbf{x}_t^{(i)})_{t=1}^{\mathrm{T}}$ in the CRF (i.e., $\textbf{x}_t^{(i)}$) is the $t^{\mathrm{th}}$ transcription sentence). Then we apply CRF to predict a sequence of labels $\textbf{y}^{(i)}=(Y_t^{(i)})_{t=1}^{\mathrm{T}}$ given $\textbf{x}^{(i)}$. Here the value of unobserved variable $Y_t^{(i)}$ is the index of slide aligned to $\textbf{x}_t^{(i)}$; that is, $Y_t^{(i)} \in \{1,2,...,S^{(i)}\}$ and $S^{(i)}$ is the number of slides used in lecture $i$. In this way, we transform the alignment task into a problem of inferring the value (i.e., the label or the index of the aligned slide) of $Y_t$ from observation \textbf{x}, and solve this inference problem with CRF.

With the linear chain structure of CRF, the model not only learns the dependence between observation \textbf{x} (i.e., the content in each learning object) and alignment \textbf{y}, but also the contextual dependence (i.e., dependence between $Y_{t-1}$ and $Y_t$) over the sequence of objects. Hence, the pattern of order-preserved mapping can be learnt during the model training. The learnt patterns function as probabilistic rules affecting the prediction of alignment (e.g., if sentence $t-1$ is aligned to slide $s$, it is more likely that sentence $t$ is aligned to slide $s$ or $s+1$, but it is impossible for that sentence to be aligned to slide $s' \in \{1,2,...,s-1\}$). These rules model the order-preserving characteristic of homologous linking.

For heterologous linking, we also apply linear-chain CRF to the binary classification problem. Similar to the homologous case, we also have a sequence of learning objects from the trunk (i.e., the lecture videos in this thesis), and another learning object sequence as the candidates of leaves (i.e., textbook sections or forum posts in the following implementation). The video transcription is still the input. However, in contrast to using a sentence as a token, here we adopt a video vignette (i.e., a sequence of sentences aligned to the same page of slides in the previous alignment task) as an item\footnote{As explained above, we use larger units here to reduce the number of tokens, since each supplementary object is considered separately in this problem, which greatly increases the computation time of the model in training and inference. Furthermore, a video vignette is a more comparable unit to supplementary objects used here, which are either textbook sections or forum posts.}. Besides, instead of modeling a sequence of candidate objects at a time, every object is considered separately, and the CRF is employed to predict binary labels - whether the considered object is linked or not. That is, in this task the input of CRF is $\textbf{x}^{(ij)}=(\textbf{x}_t^{(ij)})_{t=1}^{\mathrm{T}}$ for lecture $i$ and supplementary object (e.g., textbook section or post) $j$; $\textbf{x}_t^{(ij)}$ is the collection of transcription sentences of lecture $i$ which are aligned to slide $t$ in that lecture. The CRF predicts a label sequence $\textbf{y}^{(ij)}=(Y_t^{(ij)})_{t=1}^{\mathrm{T}}$, where $Y_t^{(ij)} \in \{0, 1\}$ represent whether video vignette $t$ is linked to supplementary object $j$ or not.

With the linear chain architecture, in this binary classification task we still model the dependency of linking configurations among neighboring learning objects on the trunk, but with a loosened constraint. Our model here can still learn the linking pattern from a sequence of video vignettes. For instance, for neighboring vignettes, they are more likely to share the same relationship (i.e., linked or not linked) to a supplementary object. However, since each supplementary object is considered independently, no dependence across supplementary objects will be learnt. This modeling of data dependency agrees with our understanding of heterologous linking. As discussed above, heterologous linking is not order-preserved. Thus, certain learning objects arranged closely in one material may imply little about the arrangement of their linked objects in another material. For example, whether learning object $\mathrm{A}_i$ is linked to object $\mathrm{B}_j$ might have little to do with the event that $\mathrm{A}_{i-1}$ is linked to $\mathrm{B}_{j-1}$ due to the variation of object arrangement across materials\footnote{Here A and B represent two types of learning materials; $i$ and $j$ are the indices of learning objects in the two materials respectively.}. Since such dependency is of little information in predicting linking, modeling the dependency across supplementary and trunk objects simultaneously simply increases the risk of overfitting (due to the increased model complexity) and introduces noise.

In contrast, the linking configuration of a sequence of trunk learning objects to a supplementary one does correlate. This is because topic continuity in educational material is crucial for learners to better digest the content. It is unlikely that the lecturer switches topics abruptly or frequently. Thus, for instance, whether learning object $\mathrm{A}_i$ is linked to object $\mathrm{B}_j$ is dependent on the event that $\mathrm{A}_{i-1}$ is linked to $\mathrm{B}_j$. As compared to our model for homologous tasks, this model design is more reasonable for heterologous linking.

\section{Feature extraction}
Information in MOOC materials is multimodal. Text, vision, audio content, or even the click-log can be useful for linking. To represent the diverse data in a uniform way that can be learned by the CRF model, we have to design the feature function set in equation~\ref{2-eq:2_4}. These features should be informative and can help the inference of labels. In the following, we discuss the design of features for our linking task.

\textbf{Lexical similarity features.} Since natural language is the integral part of education and knowledge transferal, we design our first feature set, lexical similarity features, based on the text content of course materials. These features are designed based on the assumption that the similarity between two learning objects is correlated with whether the two objects are linked. In the alignment task, lexical features can be written as:
\begin{equation} \label{eq:4_1}
f_{yk}(Y_t,Y_{t-1},\textbf{x})=\mathrm{cos\_sim}(\Phi(\textbf{x}_{t+k}),\Phi(y))\textbf{1}_{\{Y_t=y\}},
k~\in~\{-K, -K+1, ..., K\}~and~y~\in~\{1,2,...,S\}
\end{equation}
Here \textbf{1} is an indicator function and K is a hyper-parameter deciding the length of context considered in the model. $\mathrm{Cos\_sim}(\textbf{x}_{t+k},y)$ is the cosine similarity between the vector representation (defined by $\Phi$) of video transcription sentence $t+k$ and the supplementary learning object $y$. In the alignment task, $y$ is the page index of lecture slides. Thus learning object $y$ is the $y^{\mathrm{th}}$ page in the slides. 

As for the binary classification task, the lexical features we extracted are  
\begin{equation} \label{eq:4_2}
f_{k}(Y_t,Y_{t-1},\textbf{x})=\mathrm{cos\_sim}(\Phi(\textbf{x}_{t+k}),\Phi(lo)),~k~\in~\{-K, -K+1, ..., K\}
\end{equation}
Since in this task each supplementary learning object is considered separately, the lexical features compute the cosine similarity between the vector representation of this supplementary object $lo$ and video vignette $t+k$ (i.e., sentences aligned to slide $t+k$ in the previous task).

To compute the cosine similarity, we have to define the vector representation, $\Phi$, of a document (i.e., a video sentence, a video vignette, or a supplementary learning object). The first adopted is a bag of words (BoW) representation. In this representation, we compute the TF-IDF score of each word in the document, and transform the document to a vector where each dimension corresponds to the score of a unique word. Second, we adopt the word2vec representation, which is introduced in detail in Section 2.3.3. Word2vec is a continuous language model trained with a neural network to compute the word probability based on the word's context in the corpus. After the model is trained, each word is represented as a vector in a continuous space by collecting the neural network weights corresponding to that word. With the word level embedding, each word in a document is transformed first to its word2vec representation, and the document vector is computed by averaging these word vectors. As opposed to the BoW model, the long-term semantic and syntactic regularities in language can be learned in the word2vec embedding. Thus we believe our linking algorithm can understand learning objects from different aspects with the two vector representations.

\textbf{Transition features.} As discussed above, we intend to learn the contextual dependence of linking configuration with CRF. Thus, we design the second feature set - transition features:
\begin{equation} \label{eq:4_3}
f_{yy'}(Y_t,Y_{t-1},\textbf{x})=\textbf{1}_{\{Y_t=y\}}\textbf{1}_{\{Y_{t-1}=y'\}},~
y,y'~\in~Y
\end{equation}
where $Y$ is the set of labels. The assumption behind these features is that the inference of a linked object for two consecutive video segments (i.e., sentences or vignettes) is dependent. These features are typically used in applications of CRF to encode temporal dependencies, and here they allow our CRF model to learn the temporal patterns of linking configuration.

\textbf{Visual features.} Lecture videos are usually the center of MOOCs. In addition to the human language content such as video transcription, the visual channel can also provide rich information to understand the materials and infer the linking. For example, studies show that scene changes in educational videos affect learners' watching behavior and usually coincide with structural breaks of videos \cite{128, 140, 141}. Hence, we design a set of visual features to extract this useful information: 
\begin{equation} \label{eq:4_4}
f_{y}(Y_t,Y_{t-1},\textbf{x})=\mathrm{frame\_distance}(t)\textbf{1}_{\{Y_t=y\}},~
y~\in~Y
\end{equation}
Here we define $\mathrm{frame\_distance}(t)$ as the Euclidean distance between video frames corresponding to the beginning and end of video segment $t$. The time code information of sentences is encoded in video subtitles, which are typically provided in MOOCs to enhance material accessibility. If there are no subtitles, we can still obtain the time code by aligning the audio signal with lecture transcription or perform automatic speech recognition.

Since video frames are represented by the color of each pixel, we also have to transform this information to vectors for computing the distance. We investigate two vector representations: HSV (hue, saturation, and value) histogram and horizontal projection. HSV histogram is widely used in tasks such as scene detection and represents a frame by its color distribution \cite{129}. HSV histogram of a video frame is obtained by first transforming the RGB color value of each pixel in the frame to the HSV space. Then the three coordinates (i.e., H, S, and V) of the HSV space are discretized into a number of bins. The number of pixels in each bin is counted to compute a three-dimensional histogram, and the histogram is flattened to a vector as the HSV representation of the frame. HSV histogram is popular since it models the mechanism of human color perception. Furthermore, as opposed to simply representing frames as vectors of color of each pixel, the histogram method is much more robust to noise.

However, the HSV representation fails to capture some distinct characteristics of educational videos. For instance, in MOOCs, it is usually the case that the entire video consists of shots of slides or scenes with very similar colors. The HSV descriptor cannot effectively distinguish among these slides or scenes. Thus, we implement the second descriptor, horizontal projection. To extract this descriptor, for a frame with $m$ by $n$ pixels in the HSV space, we first represent the frame with three $m$ by $n$ matrices corresponding to the three coordinates (i.e., H, S, and V). For each matrix we add up the intensity of each row to project the visual content along the horizontal direction and obtain an $m$ by 1 vector. The horizontal projection of a frame is the concatenation of vectors from the three matrices. Since many informative contents in educational videos are presented horizontally (e.g., bullets in slides), this tailored descriptor can describe video frames in a more pedagogically meaningful way.

One thing that should be noted is that in Equation \ref{eq:4_4}, these features only depend on the frame distance and label of segment $t$. This is problematic since the frame distance between the beginning and end of sentence $t$ has little to do with the label of the sentence, but is highly dependent on whether labels of sentence $t$ and $t-1$ are different. With the current label set, the dependence between label transition and frame distance cannot be represented by the features. 

\begin{figure}[t]
\includegraphics[width=11cm]{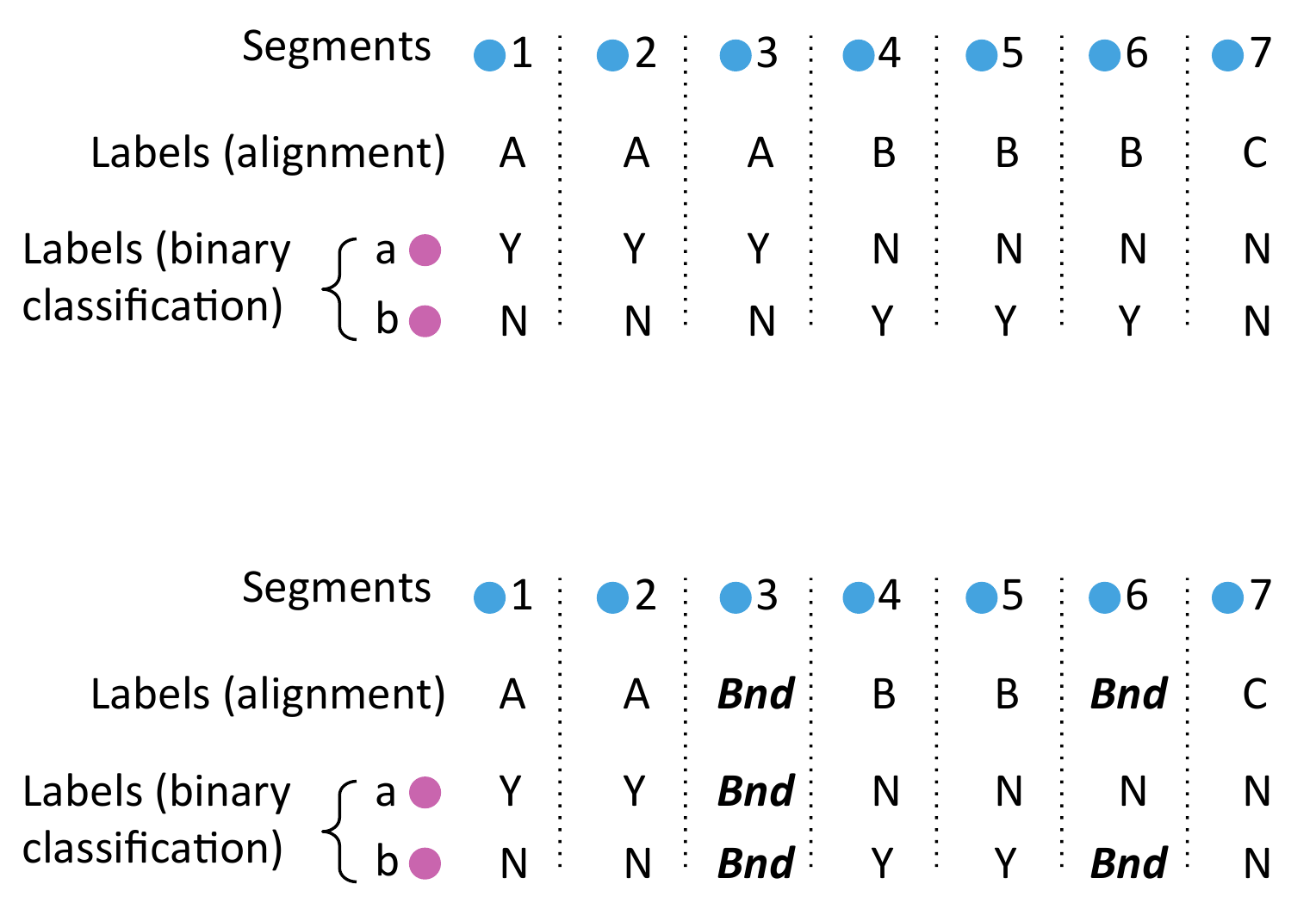}
\centering
\caption{Examples of representing the same linking configuration with two different label sets. In the upper panel, the original label set described in Section 4.2 is used. A, B, and C denote the aligned slide index; Y and N denote whether the considered object is linked. In the lower panel, the original label set along with a boundary label (denoted as \textit{'bnd'}) is used.}
\label{fig:boundary_label}
\end{figure}

Thus, we add another label, boundary, to the original label set. In Fig.~\ref{fig:boundary_label}, we illustrate how this additional label works by comparing the same linking configuration represented with two label sets. The upper panel of Fig.~\ref{fig:boundary_label} corresponds to the original label set described above, where the value of the label is the page index of the linked slide (alignment task) or whether the considered supplementary object is linked (binary classification task). In the lower panel, an additional boundary label (represented as \textit{'bnd'} in the figure) is also employed to denote the condition where the two consecutive segments are linked to various slides or have different relations to the considered object. With the additional boundary label, the dependence between label transition and frame distance can be encoded in the visual feature functions, and the extended label set has the same ability to express linking configuration as the original set. Thus, we adopt the extended label set for incorporating visual channel information to our automated linking algorithm. 

Another possible solution to encode the dependence between label transition and frame distance is using feature functions depending on both $Y_t$ and $Y_{t-1}$. However, this solution increases the number of features and makes the model more likely to overfit when a small training corpus is used. Thus we choose the extended label set as the solution.

In this thesis, we only use the frame distance features to encode visual information in course materials. There are other methods to understand the visual channel, e.g., optical character recognition (OCR), and semantic understanding of images \cite{77}. However, OCR cannot provide much additional information on top of other resources such as slides or video transcription. Current image semantic understanding can only extract shallow information such as \textit{there is a man writing on the black board} or \textit{the woman is talking}. Thus, although these methods are getting popular recently, we believe they are not suitable or mature enough for our tasks, and choose not to investigate them here.

\textbf{Meta-data features}. Some description of learning content is also informative in inferring linking. Thus we extract meta-data features to encode such information. In this thesis, two types of meta-data features are used: position and learner tagging. In the following we describe the two features respectively.  

The following is the equation of position features in the alignment task:
\begin{equation} \label{eq:4_5}
f_{y}(Y_t,Y_{t-1},\textbf{x})=\mathrm{exp}(|\frac{t}{\mathrm{T}}-\frac{y}{S}|),~y~\in~\{1,2,...,S\}
\end{equation}
In this feature set, the difference between the relative position of a video transcription sentence and a slide in their original sequence is encoded. The relative position is computed by dividing the index of the sentence or slide by the number of sentences or slides in the lecture. The motivation of this feature set is self-evident: if a sentence is mentioned at the beginning of the lecture, this sentence is more likely to be aligned to the first couple of slides.

We also extract similar features for the binary classification task:
\begin{equation} \label{eq:4_6}
f(Y_t,Y_{t-1},\textbf{x})=\mathrm{exp}(|\frac{t}{\mathrm{T}}-\frac{index(lo)}{\mathrm{N}}|)
\end{equation}
In this case $t$ is the index of video vignettes in the entire sequence of lecture videos, and T is the number of video vignettes in the entire course. \textit{index(lo)} is the index of the considered supplementary learning object, and N is the number of supplementary objects in the course (e.g., number of textbook sections or discussion threads). The textbook and forum is sorted and indexed based on the section number and thread creation time respectively.

Learner tagging is another set of meta-data features we utilized. In the MOOC platform where we collected learning content for experiments in this thesis, learners are allowed to post discussions under each lecture video \cite{5}. We record this tagging information and create a function $lo(t)$. This function maps the $t^{\mathrm{th}}$ video vignette to a set of discussion threads which were posted under the video this vignette belongs to. With this function, we extract our learner tagging features as follows:
\begin{equation} \label{eq:4_7}
f(Y_t,Y_{t-1},\textbf{x})=\textbf{1}_{\{lo \in lo(t)\}}
\end{equation}
where $lo$ represents the discussion thread we considered each time. Although learners sometimes post irrelevant discussions under the video, such as chatting with each other, we believe this feature set is helpful for our automated linking system by narrowing down the possibly linked vignettes to each thread. Note that the tagging information is only available in discussions with the current MOOC platform; thus the learner tagging features are only deployed in the task of linking videos and discussions.

As described above, we can observe that features we utilized here are diverse in their forms and value ranges. For all these feature functions, we only assume the existence of dependence between certain observations and labels; no assumption of distributions for these dependencies is required. Thus, with our CRF models it is easier to add new features and extend the automatic linking algorithm as compared to using a generative method such as HMM.

\section{Evaluation: similarity to human labeling}
We then evaluate our linking algorithm with materials in the two MOOCs introduced in Section 2.4: Stat2.1x and 6.00x. In these two MOOCs, we have lecture videos, slides, and textbook in Stat2.1x; as for 6.00x, the previous three types of materials along with forum posts are used. Thus, we investigate automated homologous linking between lecture videos and slides in the two MOOCs. For heterologous case, we study video-to-textbook-section linking in Stat2.1x, and video-to-textbook-section along with video-to-discussion-thread linking in 6.00x. Since in the heterologous case the video vignette is used as the linking unit, in the experiment we take a two-pass procedure: we first train a CRF to align video sentences to slides. Then we utilize the best alignment result in the development set to obtain the video vignette (i.e., assign the sentences aligned to each slide as one video vignette). With the video vignettes we then train CRFs to predict linking between vignettes and textbook sections/forum discussions.

As discussed at the beginning of this chapter, two issues are explored in the evaluation: 1) how closely can the automatically generated linking approach to human annotation? 2) Can automated linking benefit learners? In this section, the first issue is investigated. Here we compare automated linking results to humans' labeling collected and discussed in Section 3.1.4, and compute F1 scores to measure the similarity. As for the second issue, we will explore it in the next section.

In the evaluation, instead of simply splitting the corpus into training and testing set, we adopt a 5-fold cross validation technique. Specifically, we partition our materials into five equal sized batches. Every time, we choose one batch for testing. The remaining four batches are used for training and hyper-parameter selection. We iterate the training-testing procedure five times with different batches as test sets, and average the test set F1 scores to evaluate the model performance. We adopt a cross validation technique here, because we want to present the entire course with machine predicted linking. If we split the corpus into training and testing, the machines can only predict linking in a portion (i.e., the test set) of the courseware - it is meaningless to predict linking in the training set since the information of human annotation is already used for training the model.

We first investigate the performance of machine generated linking in Stat2.1x. In Table~\ref{table:stat21_res} we summarize the F1 scores of automated linking systems using various models and features in the homologous (i.e., linking between video sentences and slides) and heterologous (i.e., linking between video vignettes and textbook sections) tasks. To obtain comparable evaluation metrics, in both tasks F1 scores are computed at the sentence level. Therefore in the heterologous case we first map the vignette-level linking results to the sentence level before computing the F1 scores.

\begin{table}[]
\centering
\caption{The F1 scores (\%) of automated linking systems in Stat2.1x using various models (logistic regression and CRF) as well as lexical (BoW stands for bag of words and word2vec for the neural network word embedding) and visual (HSV stands for HSV histogram and HP for horizontal projection) features. Performance of both homologous (i.e., linking between video sentences and slides) and heterologous (i.e., linking between video vignettes and textbook sections) tasks is listed. In the table, the parentheses after word2vec denote that the HP visual features are only deployed in the homologous task.}
\label{table:stat21_res}
\begin{tabular}{|c|l|cc|}
\hline
\multicolumn{2}{|c|}{Linking systems (model, feature)} & Homologous & Heterologous \\ \hline
Logistic regression       & BoW                        & 74.9       & 29.6         \\ \hline
\multirow{3}{*}{CRF}      & BoW                        & 81.3       & 45.2         \\
                          & BoW + HSV                  & 84.9       & 44.6         \\
                          & BoW + HP                   & 85.8       & 44.8         \\ \hline
\multirow{2}{*}{CRF}      & word2vec (+ HP)            & 86.2       & 45.1         \\
                          & BoW + word2Vec (+ HP)      & 86.7       & 47.2         \\ \hline
\end{tabular}
\end{table}

In this table we study how visual and lexical features affect linking performance by using various vectorization techniques to encode learning objects for computing video frame distances and lexical similarities. Here, HSV histogram (denoted as HSV) and horizontal projection (denoted as HP) descriptors are investigated for frame representation; bag of words (denoted as BoW) and neural network embedding (denoted as word2vec) are studied for representing text content. Since text content is an integral part in education for knowledge transferring, and BoW is the most widely used text vectorization in natural language processing, in our study we start with linking systems using BoW for lexical features. The word2vec and visual features are gradually introduced to the systems for investigating their potential benefit.   

In addition to visual and lexical features, we also investigate the benefit of using transition features (i.e., modeling the contextual dependency) in predicting linking. The effect of transition features is studied by comparing the CRF algorithm with a baseline model, logistic regression. In logistic regression, the alignment and linking of each video segment is predicted separately and no contextual dependency is modeled. Thus by comparing logistic regression and CRF systems using identical visual and lexical feature sets, the potential benefit of transition features in linking performance can be studied. 

As for the meta-data features, we utilize them as default features in every linking system we report, since these features have been shown to be beneficial in general in a variety of applications \cite{132, 133}. In the experiment here with Stat2.1x, position features are the meta-data features we deploy in the linking system.

In this table first we can see that the logistic regression model (row 1) performs significantly worse than other systems. When an identical feature set (except transition features) is used, CRF outperforms logistic regression by 6.4\% in homologous linking and 15.6\% in the heterologous task (c.f., row 1 and 2). This observation shows the benefit of formulating linking as a sequential tagging problem: by treating the video segments as a data sequence and introducing transition features into the system, we can model the temporal patterns of linking configuration. As compared to simply modeling the content similarity in materials such as in logistic regression, the contextual information allows the model to better understand how the topics in courseware are organized and dependent on one another. Therefore, the transition features yield better linking performance in both tasks.

Then on top of lexical features computed from BoW embedding, we investigate how the additional visual features affect the performance of CRF linking systems (c.f., row 2 to 4). In homologous linking, both HP and HSV features yield improvement. In order to explain why visual features are helpful in this task, we analyze our course materials. We find that lecture videos are usually dominated by colloquial speech, and about only 23\% of sentences in the video transcription contain key terms (the terms in the textbook glossary) which are useful for identifying underlying concepts. This number shows that the lexical information is sparse. Besides, the lexical similarity features are sometimes noisy at the transition of two slides or topics. For instance, the lecturer might conclude the previous topic or connect two topics with a story. Thus, information from verbal content is often insufficient for inferring linking. In contrast, the forte of our visual features is encoding information pertinent to scene changes, which can provide complementary clues to the lexical features for aligning slides and videos. Thus, combining both features yields better performance. Furthermore, comparing row 3 and 4 we find that HP outperforms HSV in the linking performance. We believe that HP is more suitable for our tasks since it is tailored to educational videos where many contents are presented horizontally.

As for the heterologous task, we find both visual features yield no improvement. This is presumably because information encoded in these visual features is mostly about when the scenes are changed, and has little to do with inferring the linking between textbook sections and video transcription. Even if there is any useful information, it is very likely that the information has already been encoded in the alignment we used for obtaining video vignettes. Thus, providing visual features in this task may simply introduce irrelevant or redundant information to the model, and thus no improvement is observed.

We then study the benefit of word2vec embedding on top of our best systems up to now (i.e., row 4 in the homologous linking and row 2 in the heterologous task). In row 5 we replace the lexical similarity features computed from BoW document embedding with features from word2vec embedding. In the table, the parentheses after word2vec denote that the HP visual features are only deployed in the homologous task. As compared to the original best systems, a similar linking performance is achieved (c.f., 85.8\% to 86.2\% in the homologous linking and 45.2\% to 45.1\% in the heterologous task). In row 6, we integrate the lexical features computed from both BoW and word2vec embeddings, and further performance improvement is observed (c.f., 86.2\% to 86.7\% in homologous linking and 45.1\% to 47.2\% in heterologous task).

To explain the improvement, we analyze the two lexical feature sets. We found that values of similarity features computed from BoW embedding are sparse, i.e., most values are zero. This is presumably because each word in BoW representation is treated as an atomic unit; each word's corresponding element in the vector works independently from others. Thus, if two documents have no overlapping words, the cosine similarity between them is zero. Similarity features computed from this embedding encode information about how many overlapping words two learning objects have, which is highly correlated with whether the two objects are linked. However, sometimes different terms are chosen for expressing the same meaning in various objects or materials. The BoW embedding fails in dealing with such conditions and tends to give false negative, i.e., two objects are relevant but their similarity is zero or under-estimated.

On the other hand, we found that similarity values computed from word2vec embedding are much smoother. In the word2vec representation, words and documents are represented as vectors in a continuous space, where dimensions work jointly to encode different semantic or syntactic regularities. Semantically or syntactically related words could be distributed closely in the continuous space, and it is possible that the meaning of unseen words is reconstructed by their relevant terms. The breaches caused by mismatched wordings in two documents are alleviated. However, more false positives could be found, i.e., two objects are irrelevant but their similarity is high. Based on these findings, we believe BoW and word2vec embeddings provide complementary encodings of text content, and therefore integrating similarity features computed from the two yields better performance.

\begin{table}[]
\centering
\caption{The F1 scores (\%) of automated linking systems in 6.00x using various models (logistic regression and CRF) as well as lexical (BoW and word2vec) and visual (HP) features. Performance of both homologous (i.e., linking between video sentences and slides) and heterologous (i.e., linking between video vignettes and textbook sections/discussion threads) tasks is listed.}
\label{table:600_res}
\begin{tabular}{|c|l|c|cc|}
\hline
\multicolumn{2}{|c|}{\multirow{2}{*}{Linking systems (model, feature)}} & Homologous                                                 & \multicolumn{2}{c|}{Heterologous}                                                                                              \\ \cline{3-5} 
\multicolumn{2}{|c|}{}                                                  & \begin{tabular}[c]{@{}c@{}}Videos\\ to slides\end{tabular} & \begin{tabular}[c]{@{}c@{}}Videos to\\ textbook\end{tabular} & \begin{tabular}[c]{@{}c@{}}Videos to\\ discussions\end{tabular} \\ \hline
Logistic regression                & BoW                                & 63.3                                                       & 66.0                                                         & 31.3                                                            \\ \hline
\multirow{3}{*}{CRF}               & BoW                                & 71.7                                                       & 69.3                                                         & 32.1                                                            \\
                                   & BoW + HP                           & 73.7                                                       & -                                                            & -                                                               \\
                                   & BoW + word2Vec (+ HP)              & 74.7                                                       & 71.1                                                         & 33.3                                                            \\ \hline
\end{tabular}
\end{table}

We then apply our method to generate linking in 6.00x. In Table~\ref{table:600_res} the performance of automated linking systems using various models and features is listed. Here we also investigate the linking between video sentences and slides for the homologous linking; as for heterologous task, in addition to the linking between video vignettes and textbook sections, linking between vignettes and discussion threads is also studied. Similar to the experiment in Stat2.1x, we adopt sentence level F1 score as the evaluation metric.

In this thesis, we utilize 6.00x to investigate generalizability of the proposed methods. Thus, in this experiment we only explore the system configurations which have shown improvement in the experiment of Stat2.1x, and examine whether the improvement can be generalized to various courses and materials. Specifically, we study three techniques which have been shown to be helpful above: 1) modeling contextual dependency with transition features, 2) adding visual features to detect scene changes in the alignment task, and 3) integrating lexical similarity features computed from BoW and word2vec embeddings for encoding complementary information. Here, we also start with systems using meta-data features\footnote{Position features are deployed in all the three linking tasks, i.e., video segments to slides, textbook, and discussions. The learner tagging features are also used in linking video segments and discussions.} and lexical features from BoW embedding, and add other features incrementally for the investigation.

In Table~\ref{table:600_res} we observe that the three techniques also yield improvement. Comparing rows 1 and 2, CRF outperforms logistic regression consistently in the three tasks (8.4\% in linking videos to slides, 3.3\% in linking videos to textbook, and 0.8\% in linking videos to discussions). This improvement shows that the benefit of modeling the contextual dependency can be generalized in these tasks. However, improvement in linking between videos and discussions is relatively small as compared to other tasks. We believe this is because the relations between videos and discussions are distinct from videos and others. Lecture videos, slides, and textbook are created by educators and aimed at transferring knowledge systematically. In contrast, most content in forums is created by learners for resolving specific confusions. The topic organization is very different between learner-created and educator-created materials. Hence, encoding contextual dependency, which is aimed at improving linking prediction by understanding topic organization in materials, is much less helpful in linking videos to discussions.

Comparing rows 2 and 3 we can find that the visual features also yield improvement in this programming course. Note that here we only explore the horizontal projection descriptor in homologous task, since in our experiment in Stat2.1x, HP yields better performance in homologous linking and none of the visual features is helpful in the heterologous task. As compared to the results observed in Table~\ref{table:stat21_res}, we find that performance enhancement from HP is smaller (71.7\% to 73.7\% here and 81.3\% to 85.8\% previously). We surmise that this is because of the differences in the video styles used in the two courses. We found in Stat2.1x that a large portion of lecture videos is simply shots of slides; however in 6.00x there is also a great deal of live coding demos and talking head sessions. The demos and talking head sessions introduce noise into our visual features and make detecting slide changes more challenging. From these results, we conclude that the benefit of our visual features can be generalized, but how much improvement the features can bring in depends on the styles of underlying lecture videos. 

We then further integrate the lexical features computed from BoW and word2vec embeddings. In the table we can observe that the combination of two embeddings also enhances the linking performance consistently (c.f., 73.7\% to 74.7\% in linking videos to slides, 69.3\% to 71.1\% in linking videos to textbook, and 32.1\% to 33.3\% in linking videos to discussions). These results imply that the complementarity of BoW and word2vec embeddings could be a general fact found in different courseware; thus the integration could yield generalized improvement over various courses.

From Table~\ref{table:stat21_res} and Table~\ref{table:600_res} we can observe that F1 scores range widely (the best result of each linking task ranges from 33.3\% to 86.7\%). In order to investigate the reason of this variation, we compare the best F1 scores to the annotator agreement (described in detail in Section 3.1.4) in each linking task in Table~\ref{table:kappa_f1}. The first two rows summarize the best performance of our automated linking systems in F1 scores, and the following two rows list the labeling consistency among annotators evaluated in terms of kappa scores. The kappa scores can be interpreted as a measurement of how difficult and ambiguous the underlying linking task is for humans. In this table we find that machine performance is highly correlated with the ambiguity of the linking task. This finding is reassuring, especially for the video-to-discussion task where only 33.3\% F1 score is achieved, since a significant portion of difference between the linking labeled by machines and humans might simply come from the ambiguity of tasks, which is much less harmful to learner experience than linking irrelevant learning objects.

\begin{table}[]
\centering
\caption{The comparison between the best performance of our automated linking system (evaluated with F1 scores and listed in the first two rows) and the annotator agreement (evaluated with kappa scores and listed in the third and fourth row) in each linking task.}
\label{table:kappa_f1}
\begin{tabular}{|c|l|c|cc|}
\hline
\multicolumn{2}{|c|}{\multirow{2}{*}{}}                                                                      & Homologous                                                 & \multicolumn{2}{c|}{Heterologous}                                                                                                \\ \cline{3-5} 
\multicolumn{2}{|c|}{}                                                                                       & \begin{tabular}[c]{@{}c@{}}Videos\\ to slides\end{tabular} & \begin{tabular}[c]{@{}c@{}}Videos \\ to textbook\end{tabular} & \begin{tabular}[c]{@{}c@{}}Videos \\ to discussions\end{tabular} \\ \hline
\multirow{2}{*}{\begin{tabular}[c]{@{}c@{}}Automatic linking\\ (F1 scores, \%)\end{tabular}}      & Stat2.1x & 86.7                                                       & 47.2                                                          & -                                                                \\
                                                                                                  & 600x     & 74.7                                                       & 71.1                                                          & 33.3                                                             \\ \hline
\multirow{2}{*}{\begin{tabular}[c]{@{}c@{}}Annotator agreement\\ (Kappa scores, \%)\end{tabular}} & Stat2.1x & 86.5                                                       & 59.9                                                          & -                                                                \\
                                                                                                  & 600x     & 81.0                                                       & 76.1                                                          & 43.4                                                             \\ \hline
\end{tabular}
\end{table}

In this section, we show that the CRF-based linking method can integrate information from various features and yield better performance than the conventional logistic regression method. By modeling the contextual dependency and combining the complementary BoW and word2vec embeddings, we observe consistent improvement in all the linking tasks we studied. As for the visual features, they are only helpful in homologous tasks. In these experiments, we show the extensibility of CRF to various features. We believe this characteristic is a great fit to our problem. Since both MOOCs and machine learning are booming in recent years, the styles of learning materials and algorithms for understanding content are ever-changing. With the extensibility our model can grow with new machine learning techniques, pedagogies, and content by adding corresponding features (e.g., our CRF can take posterior probabilities or classification results predicted by a neural network as features). Besides, the improvement across courses and materials to some extent demonstrates the generalizability of the proposed method, which is crucial for our framework to be widely applied in various conditions.

In the previous discussion we also examined the correlation between model performance and annotator agreement. We surmise that it is likely that the low F1 scores will not harm learner experience by much. In the next section, we attempt to provide evidence for this conjecture with a user study experiment.

\section{Evaluation: benefit in learning}
We then investigate the second issue of evaluation: can linking generated automatically still benefit learners? For this issue, we conduct a user study similar to what we did in Section 3.4. We study learners' performance when course materials are presented with various strategies: presenting different types of materials separately (i.e., presented in $null$ interface) or in linking by humans (i.e., $linking$) or machines (i.e., $auto~linking$). If similar performance improvement as described in Section 3.4 can be observed after replacing manually labeled linking with the one generated by machines\footnote{In this section, the best linking results achieved in Section 4.4 are used in the interface for conducting a user study.}, we can conclude that our automated linking system is beneficial to learners. In the following, we also study the effect of automated linking on learners by learning tasks: search and retention.

\subsection{How automated linking affects search}
\begin{figure}[t]
\includegraphics[width=9cm]{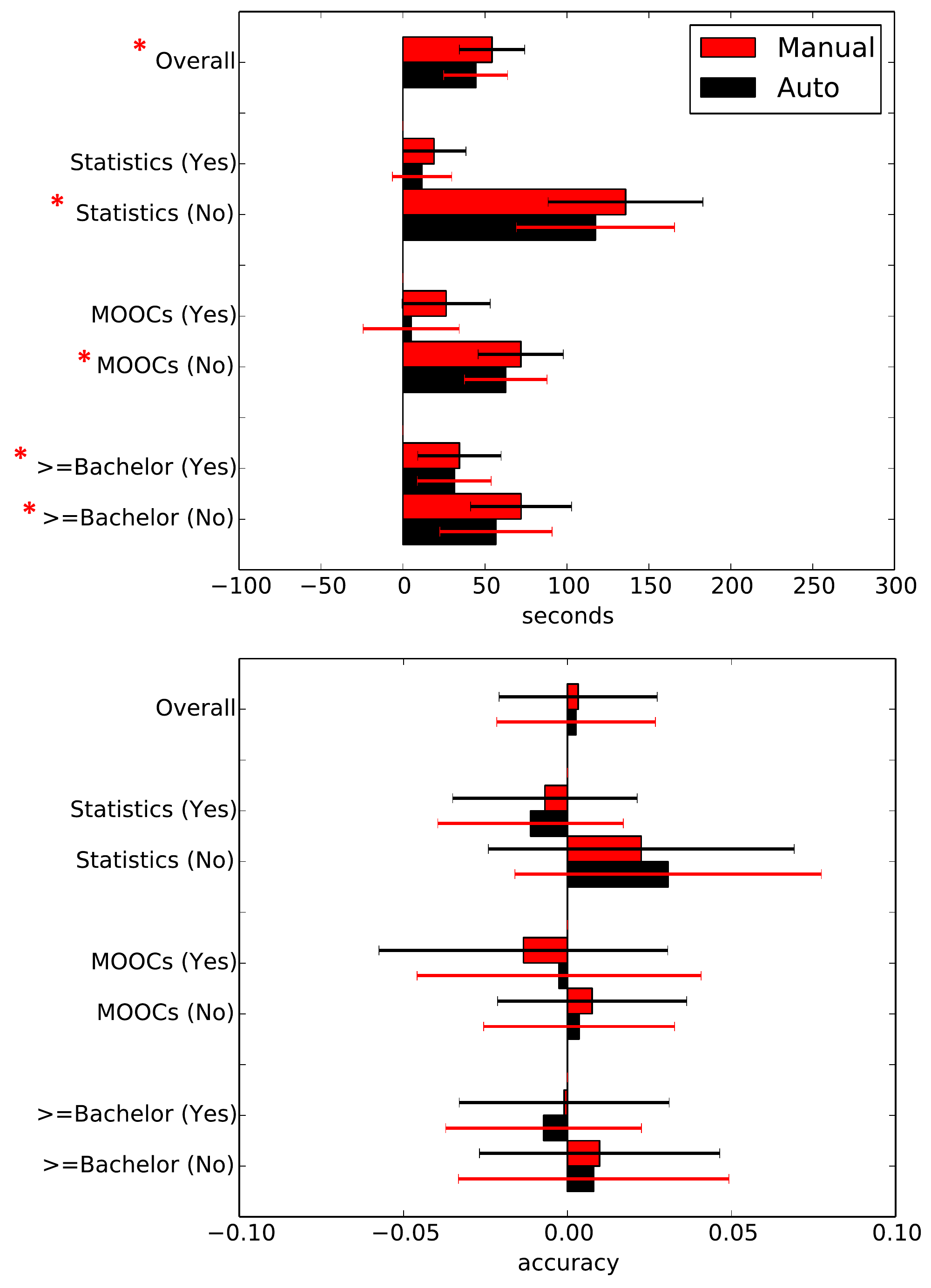}
\centering
\caption{The improvement in search time (upper panel) and accuracy (lower panel) when $linking$ (red bars) or $auto~linking$ (black bars) interface is used, with the baseline of deploying $null$. Learning performance improvement is measured in the study of Stat2.1x. The 95\% confidence intervals (shown as error bars) and significance test results (marked with red asterisk if the difference is statistically significant) are also provided.}
\label{fig:info_stat21x_diff}
\end{figure}

Here we investigate how the automated linking affects learners' performance in the information search scenario. Performance is also evaluated with two metrics: average searching time and average accuracy. In Table~\ref{table:info_stat21x} the performance in Stat2.1x is summarized. The results of the $null$ and $linking$ interfaces are identical to the ones discussed in Section 3.4.1. $Auto~linking$ corresponds to the condition where learners utilize an interface presenting automatic linking for accomplishing assigned tasks. Except for the deployed interfaces, other experimental procedures in the study of $auto~linking$ are identical\footnote{We also collected 1,000 HITs for the same 10 questions on Amazon mechanical turk. Since a between-subjects design is adopted, only online workers who did not participate in our experiment with $null$ and $linking$ interfaces are allowed in the study. The study of $auto~linking$ was conducted three months after the experiment of $null$ and $linking$ interfaces was complete. Besides, the same quality control mechanism is applied here.}. Moreover, to examine how automatic linking can benefit learners, we also focus on the performance difference when various interfaces were deployed. In Fig.~\ref{fig:info_stat21x_diff} the improvement from $linking$ (red bars) and $auto~linking$ (black bars) as compared to $null$ is visualized. The upper panel corresponds to the time reduction the $linking$ and $auto~linking$ can yield in different learner cohorts; the lower panel shows the accuracy increase from the two interfaces. Furthermore, the 95\% confidence interval of the difference is also provided.

\begin{table}[]
\centering
\caption{Learner performance in the information search scenario in the study of Stat2.1x. Performance is evaluated by the average searching time and average accuracy metrics, and measured within various cohorts using $null$, $linking$, or $auto~linking$ interfaces.}
\label{table:info_stat21x}
\begin{tabular}{|ll|ccc|ccc|}
\hline
 &  & \multicolumn{3}{c|}{Average searching time (seconds)} & \multicolumn{3}{c|}{Average accuracy (\%)} \\ \cline{3-8} 
 &  & \textit{null} & \textit{linking} & \textit{auto linking} & \textit{null} & \textit{linking} & \textit{auto linking} \\ \hline
\multicolumn{2}{|l|}{Overall} & 206 & 152 & 162 & 69.2 & 69.5 & 69.5 \\ \hline
\multicolumn{1}{|l|}{\multirow{2}{*}{Statistics}} & Yes & 166 & 147 & 154 & 71.1 & 70.5 & 70.0 \\ \cline{2-2}
\multicolumn{1}{|l|}{} & No & 295 & 160 & 178 & 64.9 & 67.1 & 68.0 \\ \hline
\multicolumn{1}{|l|}{\multirow{2}{*}{MOOCs}} & Yes & 166 & 139 & 160 & 72.0 & 70.6 & 71.7 \\ \cline{2-2}
\multicolumn{1}{|l|}{} & No & 225 & 154 & 163 & 68.2 & 68.9 & 68.5 \\ \hline
\multicolumn{1}{|l|}{\multirow{2}{*}{$\geq$Bachelor}} & Yes & 198 & 163 & 167 & 70.7 & 70.6 & 70.0 \\ \cline{2-2}
\multicolumn{1}{|l|}{} & No & 208 & 136 & 151 & 67.5 & 68.5 & 68.3 \\ \hline
\end{tabular}
\end{table}

Similar to the $linking$ interface, in Table~\ref{table:info_stat21x} and Fig.~\ref{fig:info_stat21x_diff} we can observe that an interface driven by machine generated linking also helps learners finding information with a shorter period of search time in general; as for the search accuracy, no statistically significant difference can be found. Among the studied learner cohorts, novice learners (subjects without prior knowledge in statistics, without prior exposure to MOOCs, and without a degree higher than bachelor's) as well as subjects with a degree higher than bachelor's show statistically significant improvement in search time reduction. On the other hand, comparing the improvement yielded by $linking$ and $auto~linking$, we find that in each cohort subjects using the $linking$ interface consistently take less amount of time in search.

\begin{figure}[t]
\includegraphics[width=9cm]{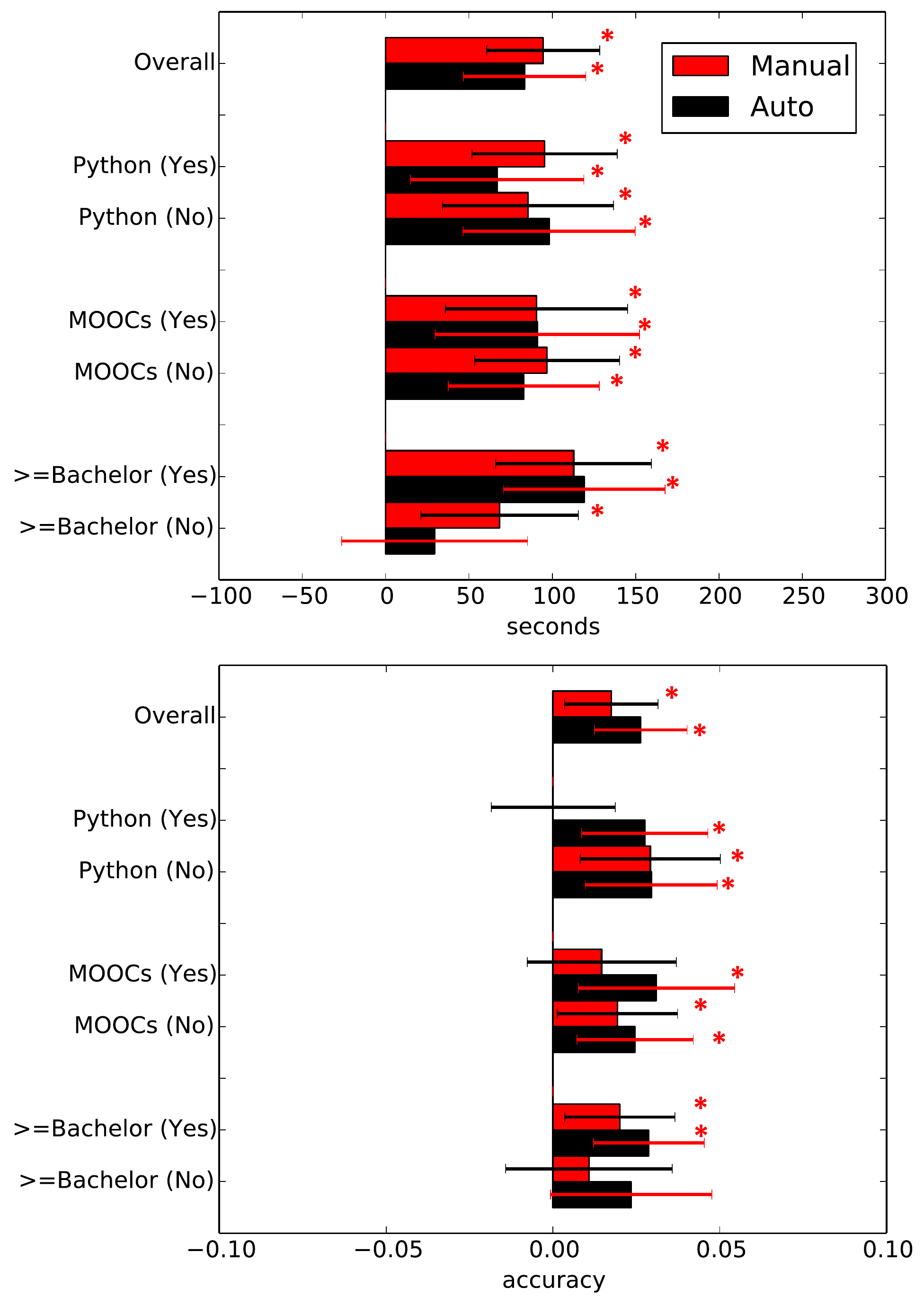}
\centering
\caption{The improvement in search time and accuracy when various interfaces were used in the study of 6.00x. The improvement from using $linking$ (red bars) and $auto~linking$ (black bars) as compared to $null$ is plotted.}
\label{fig:info_600x_diff}
\end{figure}

We also explore how automated linking affects search in 6.00x. The results are summarized in Table~\ref{table:info_600x}. In this table, in addition to the learning performance reported in Table~\ref{3-table:search_res_600}, the result of the user study deploying the $auto~linking$ interface is also listed\footnote{In this study, each experimental procedure except the deployed interface is the same as in Section 3.4.1. 1,000 HITs for the same 10 questions have been collected on Amazon mechanical turk. A between-subjects design is adopted. In this study we conduct the experiment with the three interfaces (i.e., $null$, $linking$, and $auto~linking$) simultaneously. Additionally, the same quality control mechanism is employed.}. Furthermore, we also visualize the performance difference when various interfaces were deployed in Fig.~\ref{fig:info_600x_diff}. The improvement (i.e., time reduction in the upper panel and accuracy increase in the lower) from $linking$ (red bars) and $auto~linking$ (black bars) as compared to $null$ is presented.

\begin{table}[]
\centering
\caption{Learner performance in the information search scenario in the study of 6.00x. Performance is evaluated by the average searching time and average accuracy metrics, and measured within various cohorts using $null$, $linking$, or $auto~linking$ interfaces.}
\label{table:info_600x}
\begin{tabular}{|l|l|ccc|ccc|}
\hline
\multicolumn{2}{|l|}{\multirow{2}{*}{}} & \multicolumn{3}{c|}{Average search time (seconds)} & \multicolumn{3}{c|}{Average accuracy} \\ \cline{3-8} 
\multicolumn{2}{|l|}{} & \textit{null} & \textit{linking} & \textit{auto linking} & null & linking & \textit{auto linking} \\ \hline
\multicolumn{2}{|l|}{Overall} & 443 & 349 & 360 & 87.7 & 89.5 & 90.4 \\ \hline
\multirow{2}{*}{Python} & Yes & 419 & 323 & 352 & 90.3 & 90.3 & 93.0 \\ \cline{2-2}
 & No & 463 & 378 & 365 & 85.6 & 88.6 & 88.6 \\ \hline
\multirow{2}{*}{MOOCs} & Yes & 427 & 336 & 336 & 88.0 & 89.4 & 91.1 \\ \cline{2-2}
 & No & 454 & 357 & 371 & 87.6 & 89.5 & 90.0 \\ \hline
\multirow{2}{*}{$\geq$Bachelor} & Yes & 472 & 359 & 353 & 89.5 & 91.5 & 92.3 \\ \cline{2-2}
 & No & 399 & 331 & 370 & 85.1 & 86.2 & 87.4 \\ \hline
\end{tabular}
\end{table}

In Table~\ref{table:info_600x} and Fig.~\ref{fig:info_600x_diff}, we find that the $auto~linking$ interface also allows learners to complete tasks with less time in most cohorts (except for subjects without a bachelor's degree), as compared to the $null$ interface. Search accuracy is also improved in the entire group of subjects, subjects with or without experience in Python, subjects with or without previous exposure to MOOCs, and subjects with a bachelor's degree or higher. Besides, comparing the improvements yielded by the $linking$ and $auto~linking$ interface, we can observe that they are highly correlated and statistically significant in mostly the same cohorts (except for the time reduction of subjects without a bachelor's degree, the accuracy of subjects with experience in Python, and the accuracy of subjects with previous exposure to MOOCs).

From the user study results discussed above, we can observe that in most cases the $auto~linking$ interface can still help learners in the search task, but generally yield slightly less improvement than the $linking$ interface (except for the accuracy of the study of 6.00x). The observation is interesting: our results in Section 4.4 show that some of the linking used in the $auto~linking$ interface is very different from the one labeled by humans (e.g., the linking between videos and discussions in 6.00x); however, based on our user study experiment, learners seem to be able to benefit from both interfaces using somewhat different linking annotations.

We believe the user study results support our previous conjecture that even though there are some discrepancies between linking labeled by humans and machines, many of the differences could come from the ambiguity of underlying linking tasks. Therefore, both humans and machines make reasonable linking decisions, and learners can benefit from both $linking$ and $auto~linking$ interfaces. However, from the results we also believe that machines cannot reach the same depth of understanding of the learning content as humans, and thus still make some linking errors which confuse learners. Hence, usually less improvement in the learning performance is measured when we replace human-labeled linking with the machine generated one. In Section 4.6, we will continue the discussion about the difference between linking labeled by humans and machines, as well as look into the difference patterns to explain why $auto~linking$ is still helpful, but next we will investigate our other learning scenario: concept retention. 

\subsection{How automated linking affects information memorization}
Here we explore how the automated linking affects learners' performance in the concept retention scenario. Performance is measured by the numbers of unique key-terms in the essays submitted by learners. Table~\ref{table:retention_stat21x} compares the performance when an $auto~linking$ interface was deployed to the results when $linking$ or $null$ was used (the ones reported in Table~\ref{3-table:retention_res_stat} in the experiment of Stat2.1x). The user study of the $auto~linking$ interface follows the same experimental procedures described in Section 3.4.2, e.g., collecting 1,000 HITs for the same 10 sampled topics on AMT, adopting a between subjects design, and applying a plagiarism check for quality control\footnote{This study was also conducted three months after the experiment of $null$ and $linking$ interfaces was complete.}. Moreover, the performance improvement (increased number of key-terms) from $linking$ (red bars) and $auto~linking$ (black bars) as compared to $null$ is visualized.

\begin{table}[]
\centering
\caption{Learner performance in the concept retention scenario in the study of Stat2.1x. Performance is evaluated by the number of unique key-terms in submitted essays and measured within various cohorts using $null$, $linking$, or $auto~linking$ interfaces.}
\label{table:retention_stat21x}
\begin{tabular}{|ll|ccc|}
\hline
 &  & \multicolumn{3}{c|}{Number of unique key-terms} \\ \cline{3-5} 
 &  & \textit{null} & \textit{linking} & \textit{auto linking} \\ \hline
\multicolumn{2}{|l|}{Overall} & 4.39 & 4.91 & 4.83 \\ \hline
\multicolumn{1}{|l|}{\multirow{2}{*}{Statistics}} & Yes & 4.71 & 5.11 & 5.02 \\ \cline{2-2}
\multicolumn{1}{|l|}{} & No & 3.98 & 4.60 & 4.50 \\ \hline
\multicolumn{1}{|l|}{\multirow{2}{*}{MOOCs}} & Yes & 4.83 & 5.14 & 5.07 \\ \cline{2-2}
\multicolumn{1}{|l|}{} & No & 4.27 & 4.77 & 4.75 \\ \hline
\multicolumn{1}{|l|}{\multirow{2}{*}{$\geq$Bachelor}} & Yes & 4.73 & 5.23 & 5.04 \\ \cline{2-2}
\multicolumn{1}{|l|}{} & No & 3.98 & 4.60 & 4.46 \\ \hline
\end{tabular}
\end{table}

\begin{figure}[t]
\includegraphics[width=9cm]{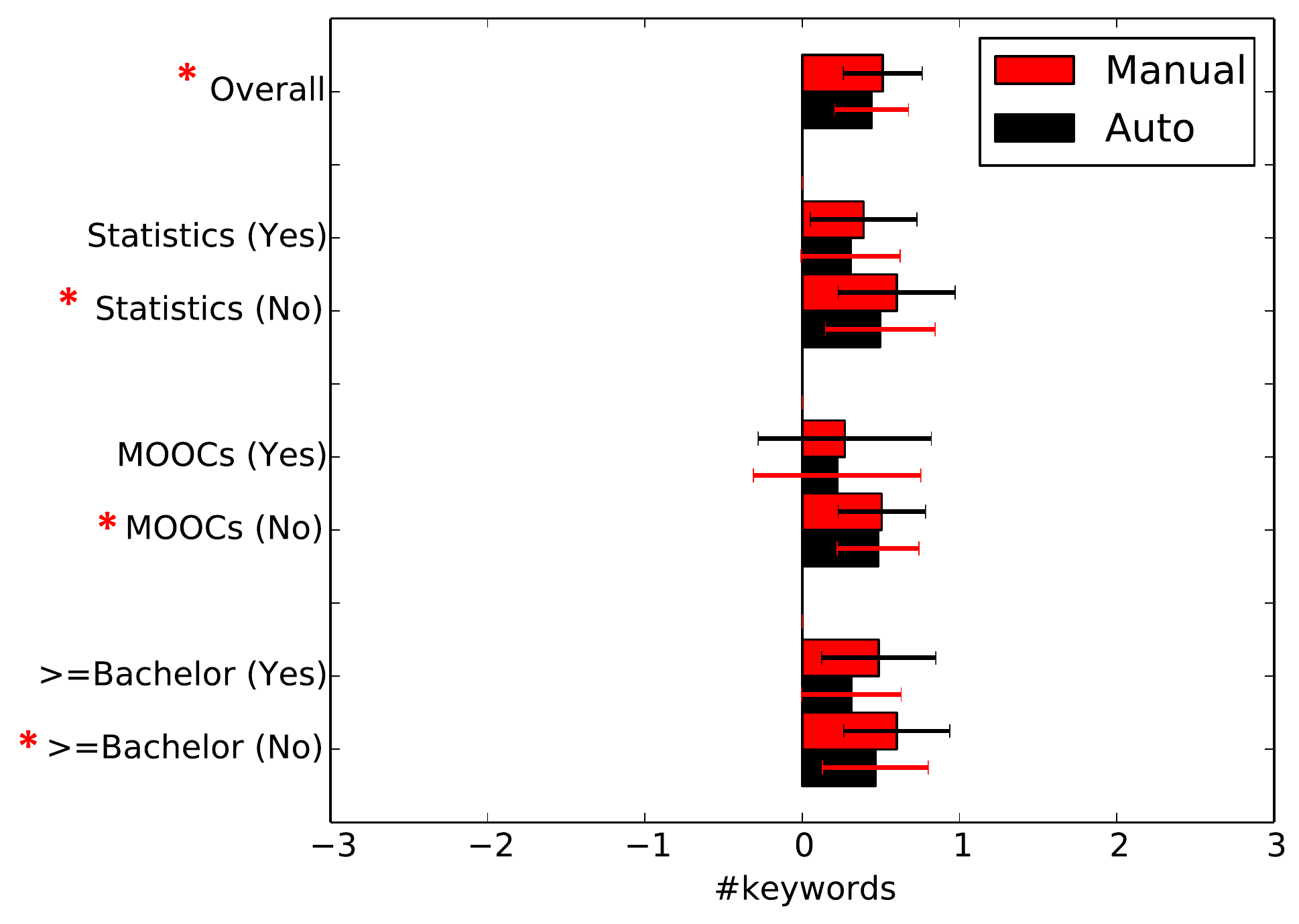}
\centering
\caption{The improvement in the number of unique key-terms contained by submitted essays when a $linking$ (red bars) or $auto~linking$ (black bars) interface is used, with the baseline of deploying $null$. Learning performance is measured in the study of Stat2.1x. The 95\% confidence intervals and significance test results are also provided.}
\label{fig:retention_stat21x_diff}
\end{figure}
Similar to the results in the information search scenario, automatic linking also yields performance improvement in the retention task. As compared to subjects assigned with the $null$ interface, learners who use $auto~linking$ mentioned more key-terms in their essays. The improvement is statistically significant in the entire group of experimental subjects as well as among the novice learners. Moreover, if we compare the improvement yielded by $linking$ and $auto~linking$, we can also find that the latter one has consistently a smaller increase in the number of key-terms.

\begin{figure}[t]
\includegraphics[width=9cm]{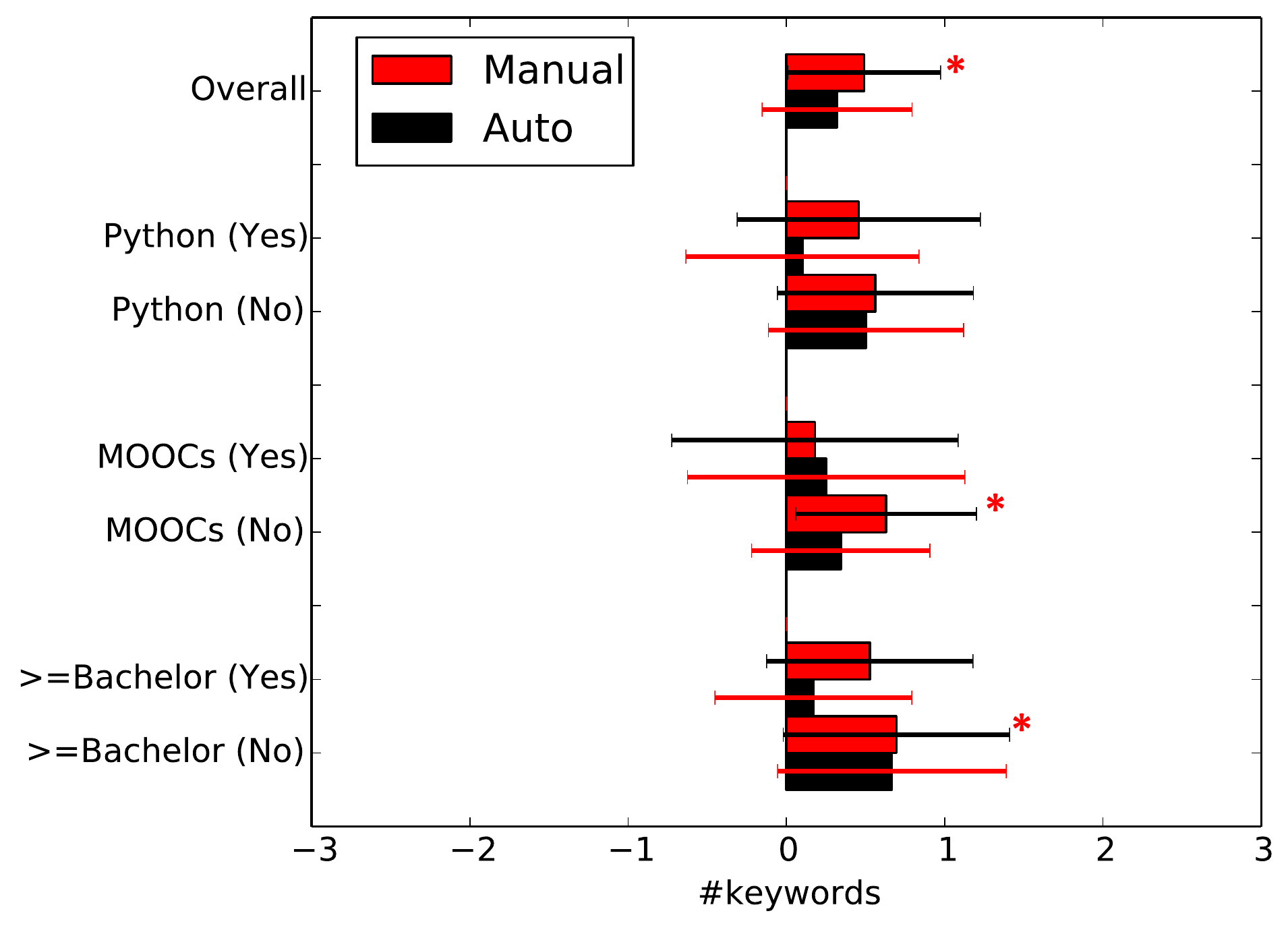}
\centering
\caption{The improvement in the number of key-terms contained by submitted essays when various interfaces were deployed in the study of 6.00x. The improvement from using $linking$ (red bars) and $auto~linking$ (black bars) as compared to $null$ is visualized.}
\label{fig:retention_600x_diff}
\end{figure}

The effect of automated linking on the retention task in 6.00x is also summarized in Table~\ref{table:retention_600x}. Columns of $null$ and $linking$ correspond to the results discussed in Table~\ref{3-table:retention_res_600}. The learning performance observed in user study where the $auto~linking$ interface is deployed is listed in the column $auto~linking$\footnote{In the study, the same experiment setup is adopted, such as 1,000 HITs for the same 10 topics and a between subjects design. Besides, the experiment with the three interfaces (i.e., $null$, $linking$, and $auto~linking$) is conducted at the same time.}. Additionally, the performance difference observed when various interfaces were used is visualized in Fig.~\ref{fig:retention_600x_diff}. The increase in the number of key-terms from using $linking$ (red bars) and $auto~linking$ (black bars) as compared to $null$ is plotted.

\begin{table}[]
\centering
\caption{Learner performance in the concept retention scenario in the study of 6.00x. Performance is evaluated by the number of unique key-terms in submitted essays and measured within various cohorts using $null$, $linking$, and $auto~linking$ interfaces.}
\label{table:retention_600x}
\begin{tabular}{|l|l|ccc|}
\hline
\multicolumn{2}{|l|}{\multirow{2}{*}{}} & \multicolumn{3}{c|}{Number of unique key-terms} \\ \cline{3-5} 
\multicolumn{2}{|l|}{} & \textit{null} & \textit{linking} & \textit{auto linking} \\ \hline
\multicolumn{2}{|l|}{Overall} & 8.07 & 8.56 & 8.39 \\ \hline
\multirow{2}{*}{Python} & Yes & 8.64 & 9.09 & 8.74 \\ \cline{2-2}
 & No & 7.64 & 8.20 & 8.14 \\ \hline
\multirow{2}{*}{MOOCs} & Yes & 8.37 & 8.55 & 8.63 \\ \cline{2-2}
 & No & 7.93 & 8.56 & 8.28 \\ \hline
\multirow{2}{*}{$\geq$Bachelor} & Yes & 8.60 & 9.13 & 8.77 \\ \cline{2-2}
 & No & 7.21 & 7.91 & 7.88 \\ \hline
\end{tabular}
\end{table}

From Fig.~\ref{fig:retention_600x_diff} and Table~\ref{table:retention_600x}, we can find automatic and manual linking perform similarly. Both of the $linking$ and $auto~linking$ interfaces allow learners in each cohort to mention more key-terms in their summary of assigned topics. However, at 95\% confidence interval, the improvement yielded from the $auto~linking$ interface is not statistically significant in the seven cohorts studied here.

Similar to what we found in the information search scenario, the user study result discussed here also shows that, disregarding the difference between the manual and automatic linking, learners can benefit from both. The results provide additional evidence for our conjecture that many disagreements between the manual and automatic linking come from the task ambiguity, and our CRF model still makes reasonable decisions in linking learning objects. Hence, only a small degradation in learning performance was found when manual linking was replaced by an automated one.

\clearpage

\section{Difference pattern analysis}
In the previous section, we provide evidence that the automated linking can still benefit learning, and that the low similarity between manual and automated linking only causes a small degradation in observed learning performance. In this section, we attempt to provide explanations for why automatic linking is also helpful in learning although less, by looking into the difference pattern between manual and automatic linking.

For the analysis, we choose the task of linking between video vignettes and discussion threads, which is the task of the lowest F1 score. We compare the linking predicted by our best automated system to human annotation. In the comparison, we first sampled 50 discussion threads from those threads which are linked to different vignettes in human and machine labeling. After skimming through the sampled threads, we summarize four difference patterns:

\begin{enumerate}
  \item Pattern 1: only annotators linked some vignettes to the thread.
  \item Pattern 2: only machines linked some vignettes to the thread.
  \item Pattern 3: both machines and annotators linked some but not the same vignettes to the thread, and the non-overlapping vignettes belong to the same lecture video.
  \item Pattern 4: both machines and annotators linked some but not the same vignettes to the thread, and the non-overlapping vignettes belong to various lecture videos.
\end{enumerate}
We categorize the sampled 50 threads into the four patterns in Table~\ref{table:diff_pattern}. Here we find that pattern 1 dominates in numbers. With the categorization, in the following we analyze each pattern and how it can affect learning experience in order to explain the user study results.

\begin{table}[]
\centering
\caption{The number of threads categorized into the four difference patterns.}
\label{table:diff_pattern}
\begin{tabular}{|l|cccc|}
\hline
 & Pattern 1 & Pattern 2 & Pattern 3 & Pattern 4 \\ \hline
Number of threads & 31 & 6 & 8 & 5 \\ \hline
\end{tabular}
\end{table}

Pattern 1 includes 62\% of the sampled threads. However, we believe that this pattern has the least negative effect on learning experience among the four. Because of the way we present learning content and visualize linking, in this pattern, the $auto~linking$ interface simply regresses to the $null$; since the thread is not linked to any video vignette by machines, this thread is presented under a separate discussion tab. Although the regression increases the difficulty to access this thread, the user experience in interacting with the rest of the learning materials is almost the same. Thus, learners can still access desired information from the rest of the materials to accomplish their tasks as they did in the interface driven by manual linking.

\begin{figure}[t]
\includegraphics[width=14cm]{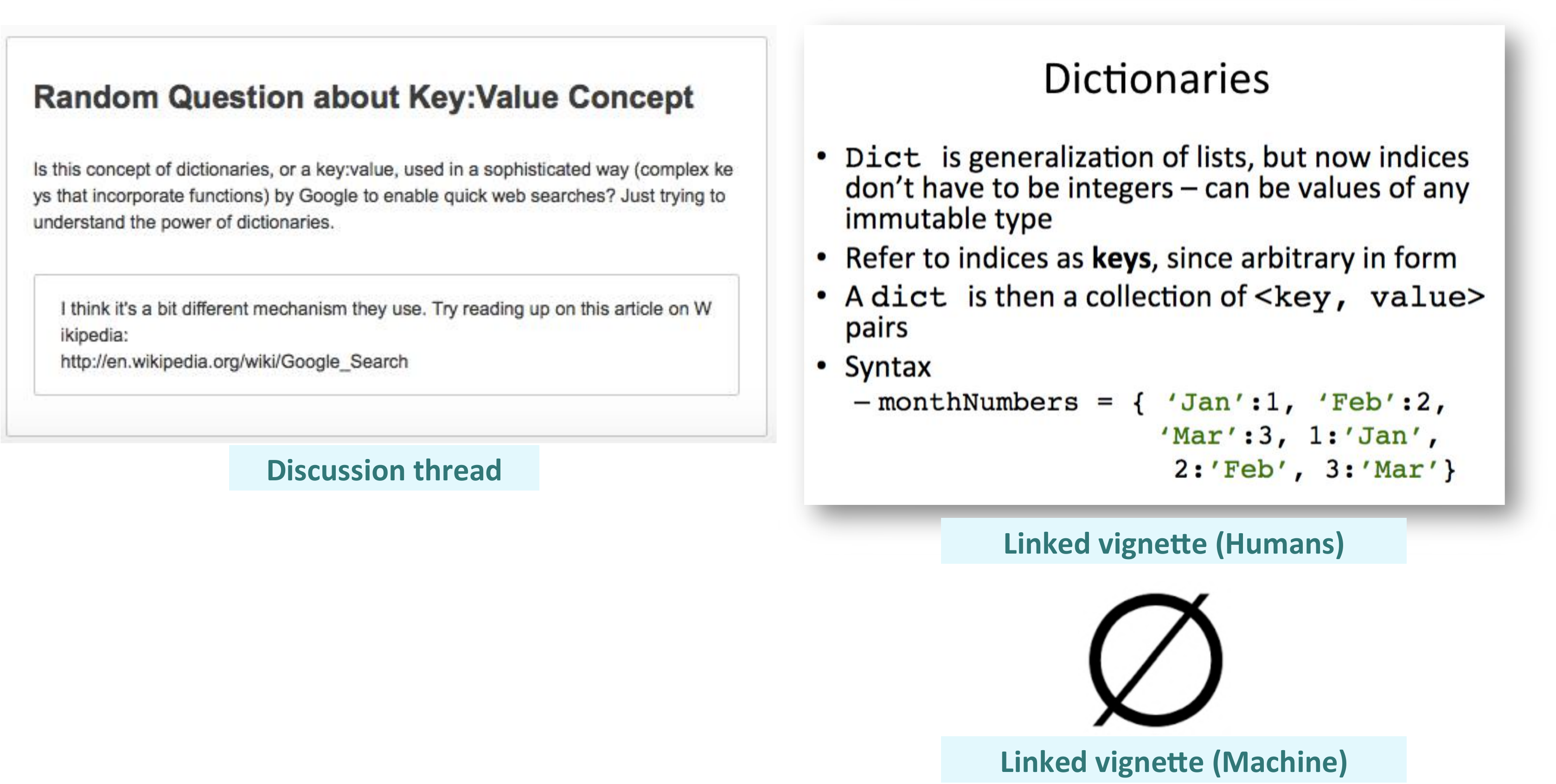}
\centering
\caption{An example of difference pattern 1. The left panel shows a sampled discussion thread; the right panel presents the vignette linked to the thread by human annotators (upper right) and our CRF algorithm (lower right). The condition where none of the vignettes is linked is represented by $\emptyset$.}
\label{fig:pattern_1}
\end{figure}

To illustrate this difference pattern and give a concrete example, we consider one specific thread from our set of 31. We present the content of this thread and its linked vignette in Fig.~\ref{fig:pattern_1}. In the left panel of the figure, we see that the discussion is about how dictionaries enable quick web searches in Google. In the right panel we observe that our TAs related this discussion to a vignette\footnote{Note that since a video vignette is defined as the video chunk aligned to one slide page, the content in a vignette can be well summarized by the aligned slide. For simplicity, here we show the aligned slide to represent the vignette.} introducing the basic idea of a dictionary, while the machines linked nothing to this post. Thus, in our $linking$ interface, when learners are surveying this vignette of dictionary introduction, this discussion about web search will be rendered under the video, while no discussion will be presented in the $auto~linking$ interface. However, the discussion only adds fun facts and additional information to the vignette. Without the post, the concept of a dictionary can still be properly learned from the vignette. This example implies that the effect of difference pattern 1 on learning can be negligible.

In addition, we examine the relation between this difference pattern and task ambiguity. We found that in 19 out of the 31 threads one of the three annotators\footnote{Note that the manual annotation is obtained by taking majority voting over the labeling of three TAs.} agrees with the machines (i.e., linking nothing to the thread). From this result we believe that many disagreements between machines and annotators come from the conservativeness of the machines: our automated algorithm is inclined not to link videos and threads when the linking is ambiguous. This fact also supports our belief that this difference pattern has little negative effect on learning, since learners can focus on the case when the relation between learning objects is strong.

Six out of the 50 sampled threads are categorized into the second pattern. This pattern deteriorates learning experience and makes learners confused. In this case, the machines relate some vignettes which have been decided irrelevant by annotators to discussion threads. Under these vignettes the discussion threads are shown in our $auto~linking$ interface. Learners have to spend their cognitive capacity to understand these unrelated threads and might feel confused about why these threads are presented. Therefore with this type of difference pattern, learning performance is lowered.

\begin{figure}[t]
\includegraphics[width=14cm]{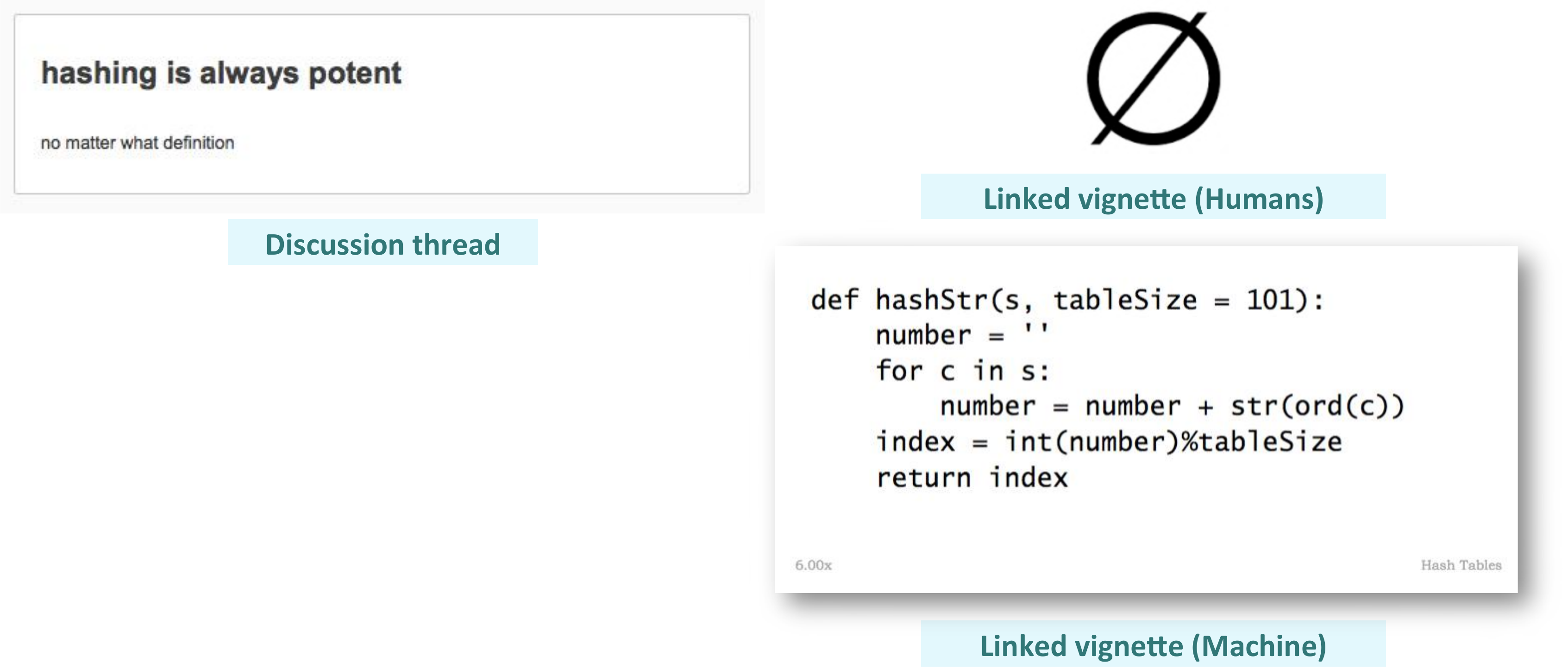}
\centering
\caption{An example of difference pattern 2. The left panel shows a sampled discussion thread; the right panel presents the vignette linked to the thread by human annotators (upper right) and our algorithm (lower right).}
\label{fig:pattern_2}
\end{figure}

We also give an example of difference pattern 2. In Fig.~\ref{fig:pattern_2}, we consider one specific thread from the set of six on the left; the content of the linked vignette labeled by humans and machines is presented on the right. The thread is meaningless discussion, but the machines link it to a vignette describing a hash function. It is obvious that this type of difference may distract or confuse learners, and thus deteriorate their learning experience.

We also examine the question of whether all disagreements result from algorithmic errors or whether some of them result from task ambiguity. We found that in the case of difference pattern 2 one of the three annotators agrees with the machine's linking decision in 2 out of the 6 threads. Hence, we believe that not all differences in this category are the result of machine mistakes, and only a portion of the differences causes a decline in the learners' performance.

As for the third pattern, 8 out of the 50 sampled threads are classified to this category. Similar to the first pattern, we also believe this difference pattern has little negative effect on learners with our interface design. In this pattern, although the machines and humans link different vignettes to a thread, these vignettes belong to the same lecture video. With our interface design, this thread is presented under the same video but only aligned to different parts of the video scrubber. Therefore, this difference pattern does not change the user experience by much\footnote{In 4 out of the 8 threads, one of the three annotators link the same video vignettes as machine does. This fact also supports our claim that our automated system links reasonable vignettes and thus it is very likely the disagreement does not affect learners negatively.}.

\begin{figure}[t]
\includegraphics[width=14cm]{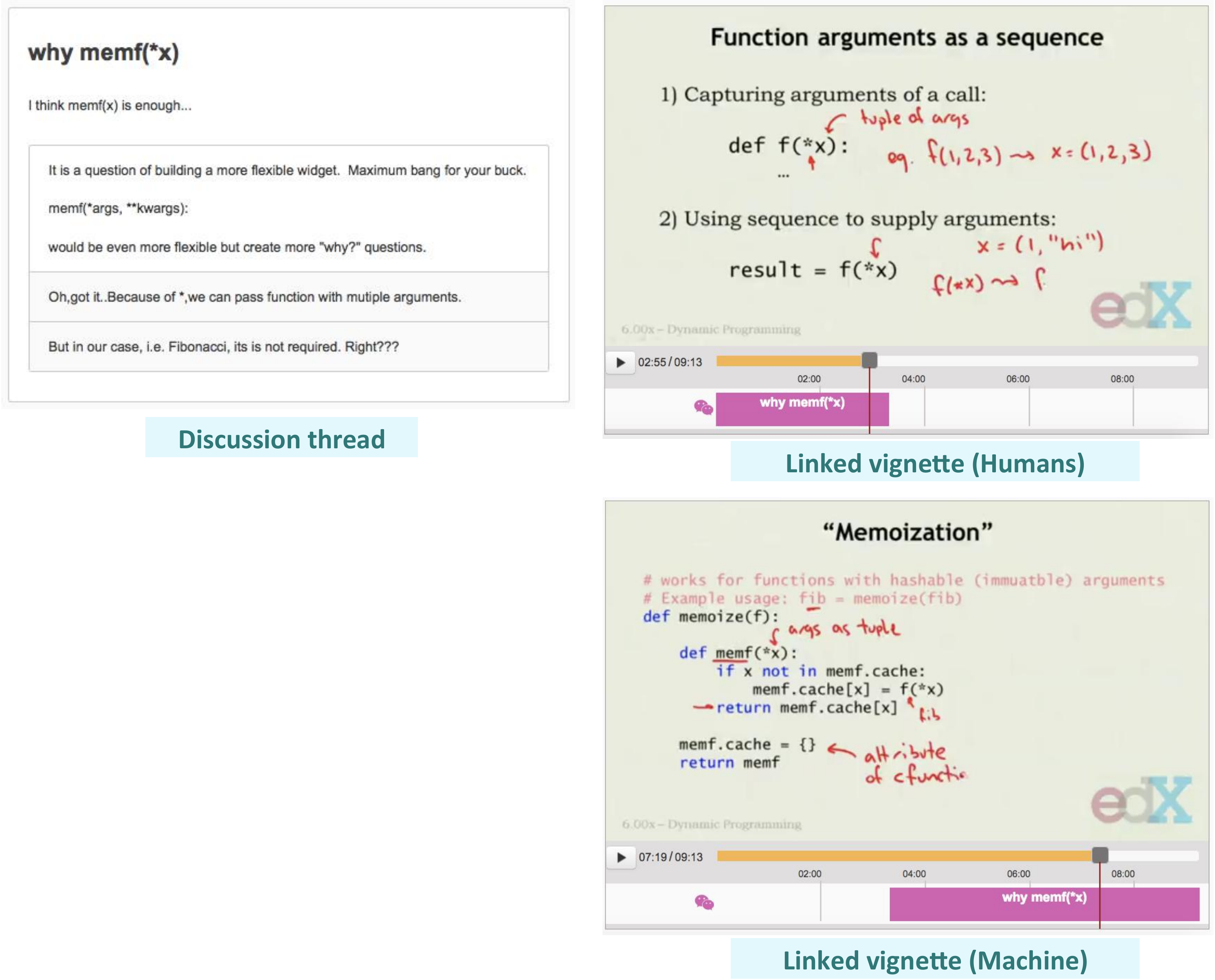}
\centering
\caption{An example of difference pattern 3. In this example, the thread (left panel) is linked by humans and machine to two vignettes (right panel) in the same lecture video. The two vignettes are closely related to each other, and the difference in presenting these two ways of linking is minor.}
\label{fig:pattern_3}
\end{figure}

An example of difference pattern 3 is shown in Fig.~\ref{fig:pattern_3}. As we can see, the thread is relevant to both vignettes linked by humans and machine. When the two various ways of linking are presented in our interface, qualitatively the difference is minor, which supports our previous claim.

A total of 10\% (5 out of 50) of the sampled threads are categorized as the last pattern. As we discussed about pattern 2, this difference category also makes the learner confused and deteriorates learning experience, since in $auto~linking$ the discussion threads are linked to lecture videos which are totally different from the ones labeled by the humans, and here our automated linking algorithm agrees with any one of the three annotators in only 1 out of the 5 threads. Thus, we conclude that this difference pattern is less probably caused by task ambiguity, and it is very likely that the videos only linked in our automated system are irrelevant to the threads.

\begin{figure}[t]
\includegraphics[width=14cm]{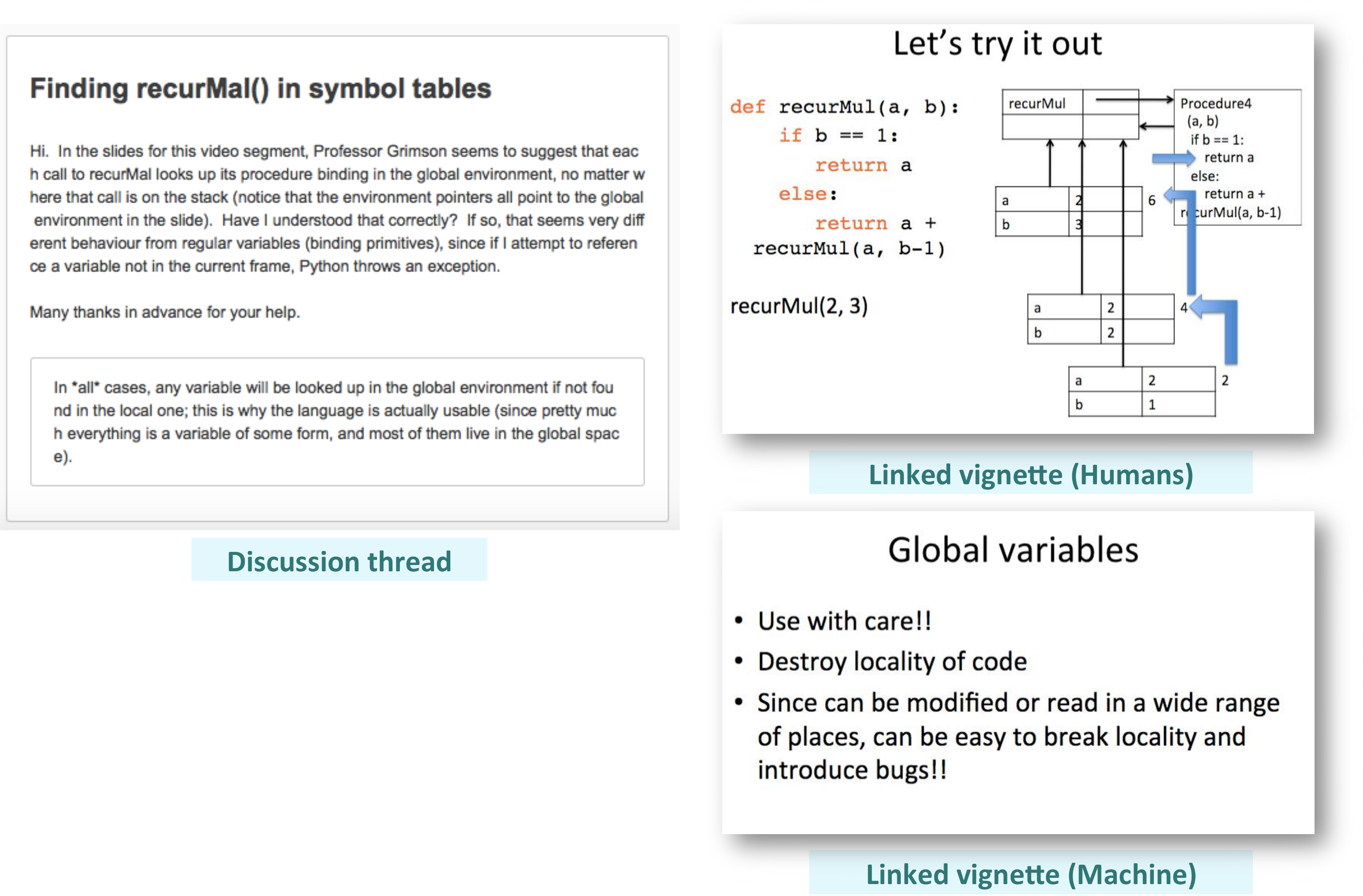}
\centering
\caption{An example of difference pattern 4. Here, humans and machine link two vignettes (right panel) from various lecture videos to the same discussion the thread (left panel).}
\label{fig:pattern_4}
\end{figure}

Fig.~\ref{fig:pattern_4} shows an example of difference pattern 4. In the posts here, learners discussed how the Python interpreter utilizes symbol tables to keep track of variable bindings in recursion, which is explained exactly in the vignette linked by our TAs. In contrast, our CRF algorithm links this thread to a less relevant vignette belonging to another lecture video describing global variables. Learners may be distracted by the discussion of the symbol tables when viewing this video.

From the analysis we can find that, although the similarity between automated and manual linking is low in some tasks, many of the differences resulting from the task ambiguity and the linking labeled by both machine and humans are reasonable, or the differences can be properly presented in our interface. Such differences could be recovered or ignored easily by learners and thus have little effect on user experience. Therefore, we also observe considerable improvement in learners' performance when the $auto~linking$ interface is deployed. Note that this analysis was done in the task of linking video vignettes and discussion threads. We believe that the conclusion can be generalized to the linking of other materials, since similar automated algorithms were utilized.

\section{Comparison with the \textit{edx} interface}

In studies above, we have shown that, when we present learning content to learners, if we are able to visualize the relations among content, learners can achieve better performance in completing assigned learning tasks. Furthermore, we demonstrated that we can obtain the relations from human annotators or the proposed CRF linking algorithm. These studies were conducted by comparing the $linking$ and \textit{auto linking} interfaces to the $null$ interface, which is a baseline that implements the conventional strategy for delivering learning materials online. In this section, we investigate whether our linking framework can provide added value to the interface currently deployed in MOOC platforms (here we choose the edX website as our baseline).

\subsection{The \textit{edx} interface}
\begin{figure}[t]
\includegraphics[width=14cm]{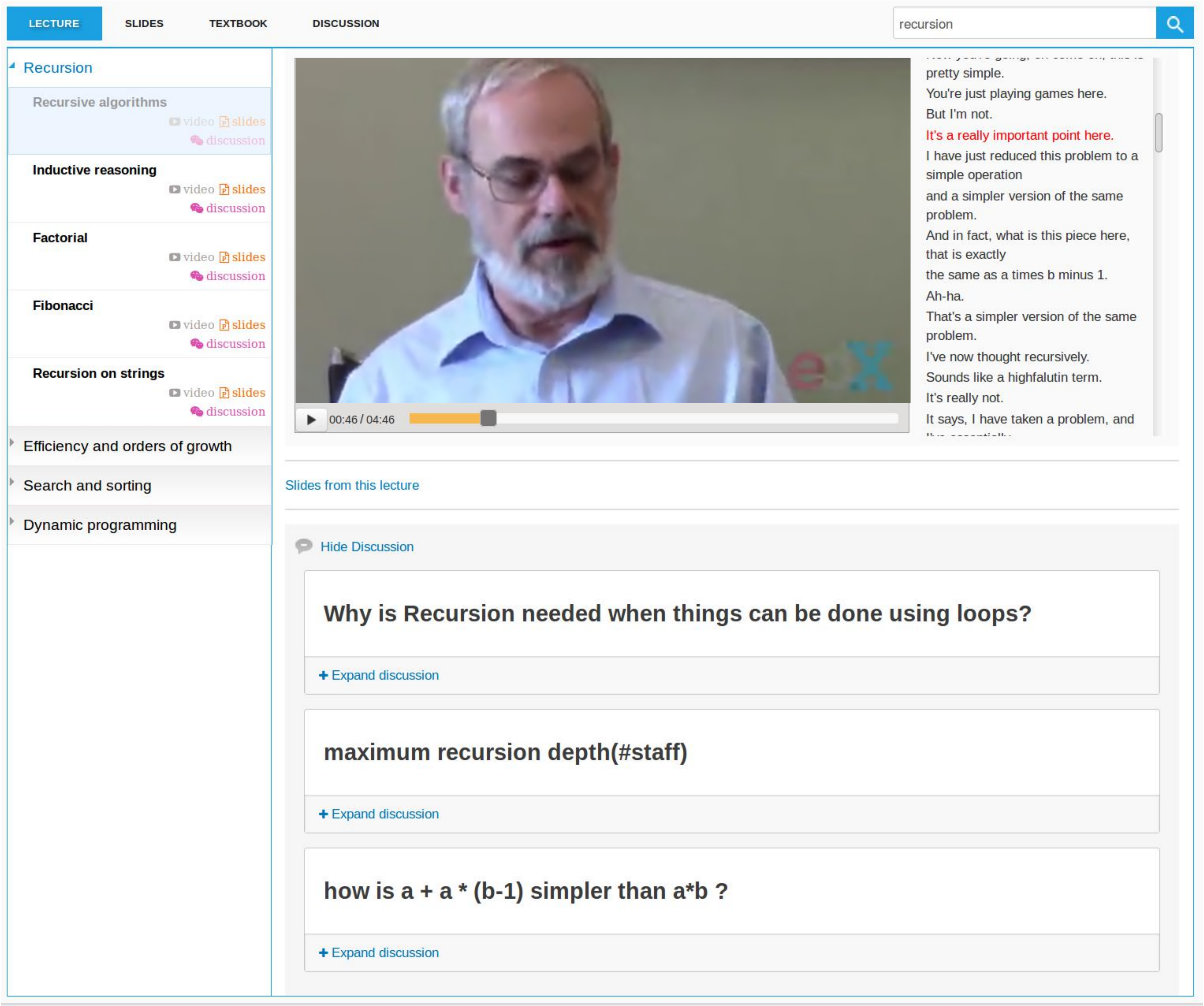}
\centering
\caption{The implemented $edx$ interface that reproduces the design and layout of the edX website to offer learners a user experience similar to that of a real MOOC, except that we also give the additional search tool for accessing course materials. This interface is used to investigate how much added value our linking framework can provide for the state-of-the-art MOOC platforms.}
\label{fig:edx_interface}
\end{figure}

For the user study being deployed in the consistent condition, rather than using the edX website directly, we also implement $our$ $edx$ interface to conduct the study on AMT. A screenshot of this interface is presented in Fig.~\ref{fig:edx_interface}. Here, we reproduce the design of the interface and the layout of content from the edX website in order to offer learners a user experience that is identical to the one they used when engaged in a state-of-the-art MOOC platform. The only difference between our $edx$ interface and the real platform is that here we additionally provide the search mechanism for accessing course materials; this is also done in order to create a consistent comparison to our $linking$, $auto~linking$ and $null$ interfaces.

In this interface, instructors can upload a deck of lecture slides beneath the paired lecture video. The associated slides are presented as a link under the video player for learners to refer to. In the edX website, to motivate learners engaging in discussions and to organize the enormous forum posting, learners are allowed to post under a lecture video to specify the relation between their discussions and the lecture; these posts are directly rendered under the video for future learners. This functionality is also implemented in our $edx$ interface.

From the design of the $edx$ interface we can find that this interface can be interpreted as another realization of educational content linking on different information levels: lecture slides are linked on the lecture level rather than the page level as they are in the $linking$ interface; the relation between discussions and lecture videos is inferred from learners' choice rather than the content these materials contain; the textbook is still presented separately. Thus, as compared to the $null$ interface we build, the $edx$ interface serves as a baseline in the comparative study for a different purpose. We implement the $edx$ interface in order to investigate how much added value our linking framework can provide for the state-of-the-art MOOC platforms. In contrast, we utilize the $null$ interface for exploring the fundamental research question asked in this thesis: can linking help learning?.

\subsection{The user study, results, and discussions}
To examine the value of the proposed linking framework can add to the state-of-the-art MOOC platform, we conduct another user study. In the study, again we published 1,000 HITs on AMT for each of the two learning scenarios, and learners in this study had to use the $edx$ interface to accomplish tasks. By comparing the learning performance measured here to the results from the $null$, $linking$, and $auto~linking$ interfaces, we can investigate whether our educational content linking framework can potentially improve the current MOOC design. Note that in the study here, except for the deployed interface for completing tasks, the remaining experimental setup is identical to the other user studies discussed in this thesis (e.g., the same 10 sampled questions and topics, the same number of rewards, the same quality control mechanism, and the between-subjects design). Furthermore, for simplicity in this chapter we only investigate 6.00x for our study, and the user study was conducted together with the other three (i.e., $null$, $linking$, and $auto~linking$) interfaces.

Table~\ref{table:info_res} summarizes learner performances (evaluated by the average searching time and accuracy) in the information search scenario. Columns 1 to 3 and 5 to 7 correspond to the performance measured when the $null$, $linking$, and $auto~linking$ interfaces are utilized. These results are identical to the ones reported in Table~\ref{table:info_600x}. The user study result when the $edx$ interface was deployed is listed in columns 4 and 8. Here, we employ the same dividing criteria for stratifying learners' background (i.e., prior knowledge, experience in MOOCs, and highest degree). 

\begin{table}[]
\centering
\caption{Learner performance in the information search scenario in the study of 6.00x. In addition to the results reported in Table~\ref{table:info_600x}, performance (evaluated by the average searching time and average accuracy) measured when the $edx$ interface was used is listed.}
\label{table:info_res}
\begin{tabular}{|l|l|cccc|cccc|}
\hline
\multicolumn{2}{|l|}{\multirow{2}{*}{}} & \multicolumn{4}{c|}{Average search time (seconds)} & \multicolumn{4}{c|}{Average accuracy} \\ \cline{3-10} 
\multicolumn{2}{|l|}{} & \textit{null} & \textit{linking} & \textit{auto linking} & \textit{edx} & null & linking & \textit{auto linking} & \textit{edx} \\ \hline
\multicolumn{2}{|l|}{Overall} & 443 & 349 & 360 & 401 & 87.7 & 89.5 & 90.4 & 88.6 \\ \hline
\multirow{2}{*}{Python} & Yes & 419 & 323 & 352 & 352 & 90.3 & 90.3 & 93.0 & 89.5 \\ \cline{2-2}
 & No & 463 & 378 & 365 & 443 & 85.6 & 88.6 & 88.6 & 87.9 \\ \hline
\multirow{2}{*}{MOOCs} & Yes & 427 & 336 & 336 & 386 & 88.0 & 89.4 & 91.1 & 89.3 \\ \cline{2-2}
 & No & 454 & 357 & 371 & 409 & 87.6 & 89.5 & 90.0 & 88.2 \\ \hline
\multirow{2}{*}{$\geq$Bachelor} & Yes & 472 & 359 & 353 & 393 & 89.5 & 91.5 & 92.3 & 90.8 \\ \cline{2-2}
 & No & 399 & 331 & 370 & 411 & 85.1 & 86.2 & 87.4 & 85.8 \\ \hline
\end{tabular}
\end{table}

\begin{figure}[t]
\includegraphics[width=9cm]{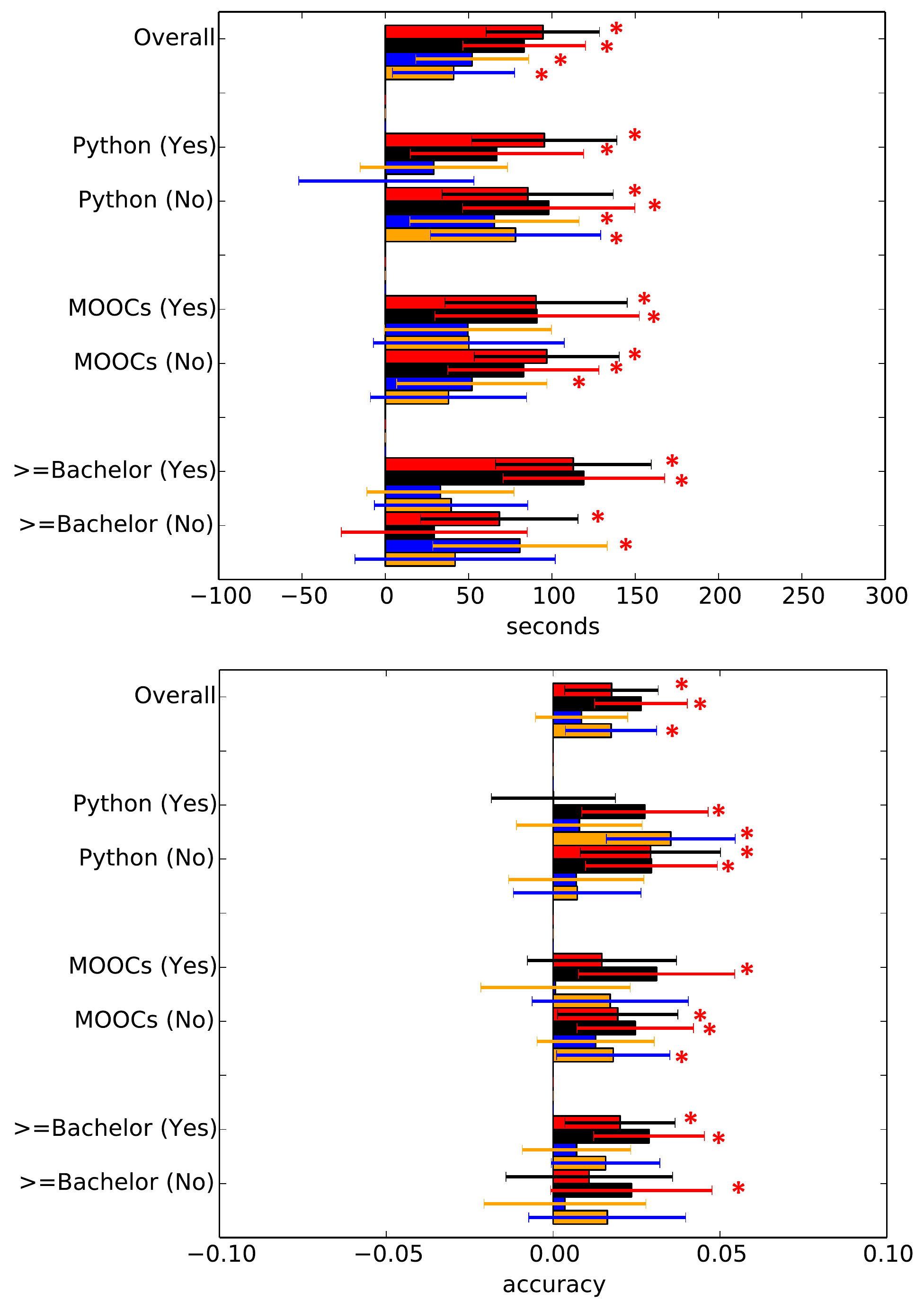}
\centering
\caption{The improvement in search time and accuracy when various interfaces were deployed. The improvement from using $linking$ (red bars) and $auto~linking$ (black bars) as compared to $null$, as well as the improvement from $linking$ (blue bars) and $auto~linking$ (orange bars) as compared to $edx$ is plotted.}
\label{fig:info_diff}
\end{figure}

Since our goal is to examine how much added value our educational content linking framework can bring us, we again visualize the differences of performance yielded by different pairs of interfaces in Fig.~\ref{fig:info_diff}. In this figure, the improvement (i.e., time reduction in the upper panel and accuracy increase in the lower) from $linking$ (red bars) and $auto~linking$ (black bars) as compared to $null$, as well as the improvement from $linking$ (blue bars) and $auto~linking$ (orange bars) as compared to $edx$ is presented. By looking at the blue and orange bars, we can understand the potential of our linking framework in improving the content delivery in current MOOC platforms. The red and black bars are plotted here for comparison to our previous results, and they are identical to the ones in Fig.~\ref{fig:info_600x_diff}.

\begin{figure}[t]
\includegraphics[width=9cm]{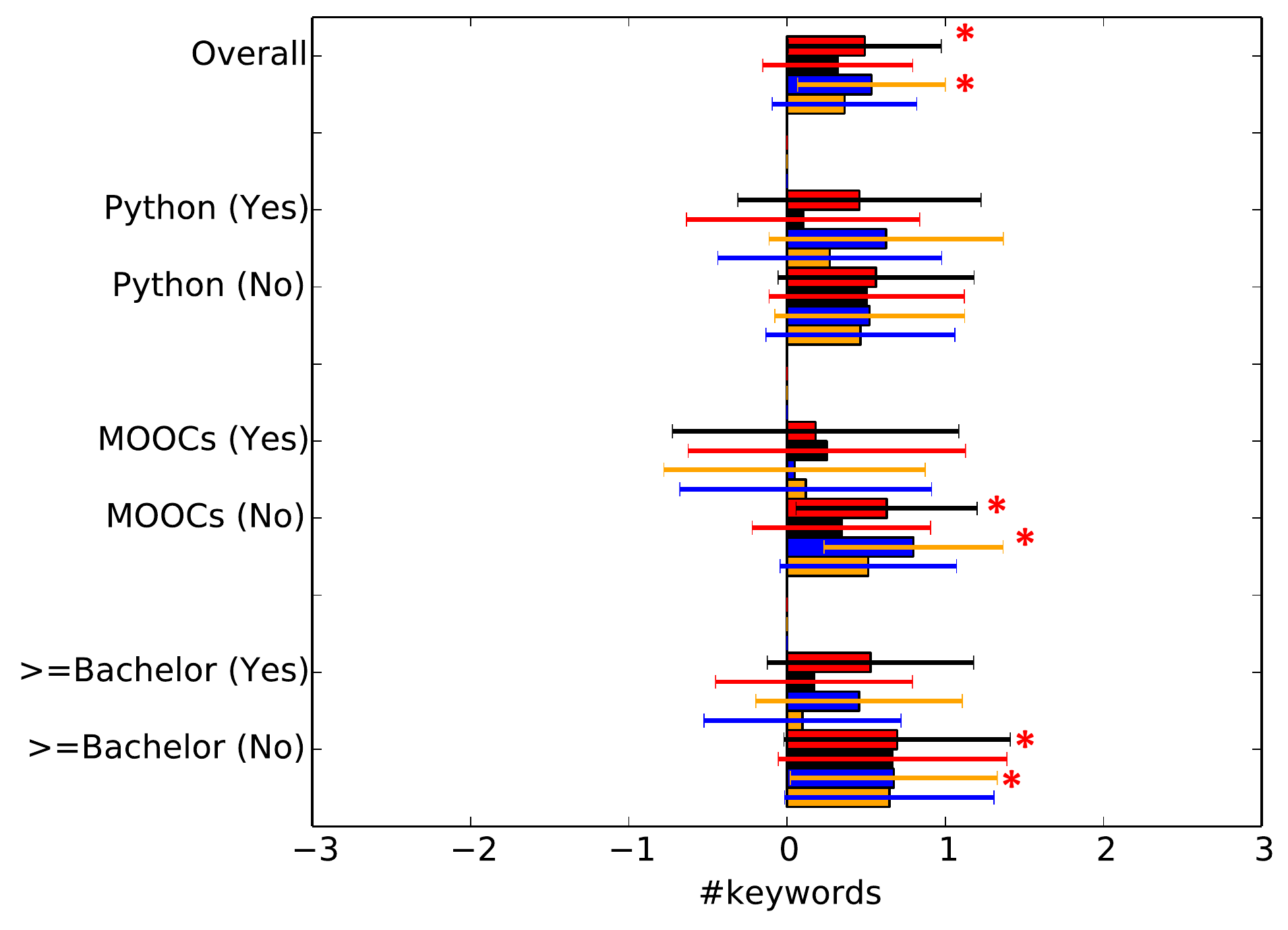}
\centering
\caption{The improvement in the number of unique key-terms contained by submitted essays when various interfaces were deployed. The bars are pictured in the same way as in Fig.~\ref{fig:info_diff}.}
\label{fig:retention_diff}
\end{figure}

We report the observed performance in the concept retention scenario in Table~\ref{table:retention_res}. In addition to the results discussed in Table~\ref{table:retention_600x}, which are listed in columns 1 to 3 here, the performance measured for learners who use the $edx$ interface for their tasks is presented in column 4. Additionally, to visualize the added value our linking framework may yield, we also plot the differences in the number of key-terms when various interfaces were deployed in Fig.~\ref{fig:retention_diff}.

\begin{table}[]
\centering
\caption{Learner performance in the concept retention scenario in the study of 6.00x. In addition to the results reported in Table~\ref{table:retention_600x}, the number of unique key-terms in submitted essays measured when the $edx$ interface was used is listed.}
\label{table:retention_res}
\begin{tabular}{|l|l|cccc|}
\hline
\multicolumn{2}{|l|}{\multirow{2}{*}{}} & \multicolumn{4}{c|}{Number of unique key-terms} \\ \cline{3-6} 
\multicolumn{2}{|l|}{} & \textit{null} & \textit{linking} & \textit{auto linking} & \textit{edx} \\ \hline
\multicolumn{2}{|l|}{Overall} & 8.07 & 8.56 & 8.39 & 8.03 \\ \hline
\multirow{2}{*}{Python} & Yes & 8.64 & 9.09 & 8.74 & 8.47 \\ \cline{2-2}
 & No & 7.64 & 8.20 & 8.14 & 7.68 \\ \hline
\multirow{2}{*}{MOOCs} & Yes & 8.37 & 8.55 & 8.63 & 8.51 \\ \cline{2-2}
 & No & 7.93 & 8.56 & 8.28 & 7.76 \\ \hline
\multirow{2}{*}{$\geq$Bachelor} & Yes & 8.60 & 9.13 & 8.77 & 8.67 \\ \cline{2-2}
 & No & 7.21 & 7.91 & 7.88 & 7.23 \\ \hline
\end{tabular}
\end{table}

From the $linking-edx$ and $auto-linking-edx$ bars (i.e., the blue and orange bars) in Fig.~\ref{fig:info_diff}, we find that our linking interfaces (driven by either manual or automated linking) allow learners to search for content using less time but with similar accuracy. Comparing each $linking-edx$ and $auto-linking-edx$ bar to the corresponding $linking-null$ (i.e., red) and $auto-linking-null$ (i.e., black) one, less time reduction is observed in general, and the reduction is significant in fewer groups of subjects. For the $linking-edx$ and $auto-linking-edx$ bars of subjects with experience in Python, subjects with previous exposure to MOOCs, and subjects with a bachelor's or higher degree, as well as the $auto-linking-edx$ bar of subjects without previous exposure to MOOCs, the differences are not statistically significant. However, their counterparts in the $linking-null$ and $auto-linking-null$ bars are significant. Results here suggest the added value our $linking$ and $auto~linking$ interface may provide for the current MOOC platforms. Furthermore, in the search accuracy, only the $auto-linking-edx$ bars of the entire group of subjects, subjects with experience in Python, and subjects without previous exposure to MOOCs show statistically significant improvement. As compared to the $linking-null$ and $auto-linking-null$ bars, statistically significant difference is observed in much fewer cohorts. These observations imply that the $edx$ interface is a better baseline (in terms of yielding better learning performance) than the $null$ one. The reason is self-evident: the $edx$ interface implements some ideas of linking, which have been shown to be able to make course materials more accessible and improve learning.

Our linking interfaces can also yield more key-terms consistently over each cohort as compared to the interface reproducing the current MOOC design (Fig.~\ref{fig:retention_diff}). Furthermore, in this learning scenario, the $null$ and $edx$ interfaces seem to perform similarly, and therefore the difference between the $linking-edx$ and $auto-linking-edx$ bars to the $linking-null$ and $auto-linking-null$ ones in the same cohorts is much less obvious than in the information search scenario. We surmise that this difference results from the nature of the two scenarios. By comparing the $null$ and $edx$ interfaces, one of the most significant changes is that in the $edx$ interface relevant discussions are stacked beneath the lecture video. Discussions are usually initiated because of confusions about specific problems or concepts in the learning content. These discussion threads contain useful information to solve questions asked in the search scenario. However, information in these posts is fractional and thus it is challenging to learn about a topic systematically from these posts. Therefore, the improved navigation over discussions is much more helpful in our search tasks.

Our results show that our educational content linking framework can potentially allow learners to find desired learning content faster than the current MOOC interface design with comparable accuracy; with our proposed interfaces, subjects can also retain more information after learning with the assigned topic using the same amount of time. Since the $edx$ interface also partially implements our linking idea, we surmise that the improvement results from several differences in the design of interfaces. First, the page-level alignment between slides and lecture videos can better visualize the structure and emphasize the sub-goals of videos. Second, the relation between discussions and videos tagged by learners can be noisy and distract other learners; considering the topical relevance between content contained in the two types of materials can control the quality of tagging. Third, some discussions might be related to only parts of the videos. This information is not available in the current MOOC platforms. Fourth, it might be worthwhile to visualize the relations between lecture videos and external resources, such as a recommended textbook. We believe that, by integrating these features properly into the current platform design, learning experience in MOOCs can be enhanced significantly.

\section{Conclusions}
This chapter investigates our second research question: can linking be generated at scale? For this question, we formulate the linking annotation as a sequential tagging problem, and propose an automatic content linking algorithm based on CRFs. In the algorithm, to infer the linking a variety of features are utilized: lexical similarity, transition, visual, and meta-data features. By analyzing the difference pattern between automated and manual linking, we find that many differences result from the task ambiguity and have little negative effect on learners. Hence, in our user research we observe similar improvement in learning performance when the \textit{auto linking} interface replaces the $linking$ one. We conclude that our linking framework can be realized at scale with an automated algorithm.

Furthermore, we also compare our $linking$ and \textit{auto linking} interfaces to the reproduction of the edX website, and explore how these interfaces support learners in completing assigned tasks. In the user study we also demonstrate that, as compared to the current design in MOOC platforms, our educational content linking framework can still help learners find information faster and retain more concepts. We believe that the linking framework we propose enriches the current MOOC interface design. We envision engaging learners with more accessible course materials and better learning experience powered by content linking.

\chapter{Conclusions}

This dissertation introduced educational content linking: a framework for organizing course materials to make content more accessible for learners. To conclude, in this chapter we will summarize the main contributions of this thesis and discuss possible directions for future work.

\section{Summary and contributions}
This thesis contributes to the research community by proposing the framework of educational content linking. This framework provides better navigation over learning materials and improves learning experience. Around the framework we conduct two lines of studies to answer two research questions: 1), if we are able to link learning materials with humans, would the linking help learners?, and 2), can the courseware be linked at scale with machine learning techniques?

With the exploration of the first question, this thesis makes three main contributions: the linking annotation, the interface design, and the evaluation.\begin{itemize}
  \item \textbf{The linking annotation}. We provide a definition of linking and a formulation of labeling relations among learning materials as an alignment and binary classification problem. We design the workflow of annotating linking with researchers or with collaboration between course staff and online workers. Linking is an abstract concept and this contribution makes it a reality.
  \item \textbf{The interface design}. We design an interface to deliver learning content with the visualization of linking among the materials. The interface can provide pedagogical benefits through improved content navigation.
  \item \textbf{The evaluation}. We conduct a large-scale user study with online workers and two learning scenarios to investigate specific learning mechanisms: search and retention. We argue that this study can measure the benefits of pedagogical intervention reliably with reasonable cost, when the underlying learning goals in the study are clear to participants. The study result shows that the proposed linking framework can indeed improve learning outcomes in the investigated search and retention scenarios.
\end{itemize}

In the second part of this thesis, we investigate the second research question, and make two main contributions: the automated linking algorithm and the comparison to a currently deployed MOOC interface.
\begin{itemize}
  \item \textbf{The automated linking algorithm}. We propose an automated linking algorithm based on CRFs and multimodal features. In our large-scale user study we demonstrate that, although there are some differences between the manually and automatically generated linking, most differences can be properly presented in our interface or easily ignored by learners, and thus the interface powered by automated linking can still lead to better learning performance than the unlinked interface. This result suggests that the proposed linking framework can be realized at scale with an automatic algorithm based on machine learning techniques.
  \item \textbf{The comparison to a currently deployed MOOC interface}. In addition to the conventional unlinked content delivery, we explore the added value our linking framework can potentially provide to a currently used MOOC interface. The user research result suggests that the framework proposed here can possibly enrich the design of the studied MOOC interface, engage learners with more accessible learning content, and improve learning outcomes.
\end{itemize}

\section{Future work}
This dissertation shows the potential of educational content linking in engaging learners with better learning experience. This conceptual idea can be further verified, refined and applied to various circumstances to improve learning.

\subsection{Learning platforms of the future}
Our user research results demonstrate several possible ways in which linking can offer learners a better experience. We envision that the future learning platforms can engage learners with more accessible learning content. While in this thesis we investigate linking in two learning tasks, search and retention, there are many other aspects of learning that can potentially benefit from our proposed framework. For instance, in solving the problem sets or performing online lab experiments, with better organized learning materials, learners are more likely to receive proper assistance from the content when they are confused. Furthermore, our studies were conducted with online workers. Research implementing the proposed experimental pipeline in a MOOC environment is valuable to clarify the mechanism by which linking helps actual learners.

Instead of reproducing our entire implementation of linking, separate components in our pipeline also inspire directions of design for future platforms. For example, the automated linking algorithm can be applied to filter noisy posts and improve the quality of discussions; instructors can utilize the algorithm to discover relevant learning content to enrich or reorganize their lectures. Our presentation of lecture videos provides design implication for better video interaction with visualized structure and subgoals. The design of our $linking$ interface suggests a neat way to offer recommended readings. These components lead to new avenues to improved MOOC platforms.

\subsection{Towards a variety of course subjects and material types}
This thesis focuses on two STEM courses: statistics and the Python programming language. However, as we mentioned in Chapter 1, there have been over 4,000 MOOCs on the Web with subject fields ranging from science and engineering to humanities and law. In addition to the topics, these MOOCs span different applied pedagogies, course designs, and methods for content organization or delivery. \textit{"Whether our linking framework can be applied to other subject fields"}, \textit{"which fields, pedagogies, designs, and content organization can benefit from linking"}, and \textit{"how various conditions interact with the idea of linking"} are several interesting questions that should be explored. We believe a wider deployment of this framework in various MOOCs can elucidate these questions.

With a wider deployment, the involved types of course materials can also vary. For example, some MOOCs may put strong emphasis on problem sets, while others may stress on online labs. For the proposed framework to be of more general use, we also have to answer questions such as \textit{"how to extend our implementation, from linking annotation to visualization, to accommodate these variations"} and \textit{"whether and how the conclusions made in this dissertation are affected by using various materials"}. Our initial foray of adding forums to our implementation can be good illustration showing how to investigate these questions.

\subsection{More sophisticated algorithm for linking at scale}
In Chapter 4, we have demonstrated that many differences between manual and automated linking have little negative effect on learners. However, there is still a considerable portion of disagreement which may confuse learners. We believe that a more sophisticated machine learning model can achieve deeper and more comprehensive understanding of the course materials, and thus generate linking which is more similar to the human annotated one. 

One promising model is the attention-based neural network. This model was first proposed for machine translation \cite{139}; for each word in the sentence of the target language, this neural network learns a weight for each word in the source sentence. The weight represents how relevant the source word is in predicting the target word. Applying this model to our linking problem, the target tokens to be predicted are the linking configuration (e.g., the linked slide index or whether two objects are linked), and the source tokens are two sequences of learning objects to be linked. This neural network can place attention on informative learning content when deciding linking configuration; such a mechanism is similar to how humans generate the linking.

More informative features are also capable of yielding better linking results. In addition to the information extracted from learning content, such as the lexical and visual features utilized in our method, user behavior is another excellent resource for predicting linking. In this thesis, we demonstrate the usage of learner-generated tags about discussion posts in the linking algorithm. Aside from these tags, click log and browsing history are very likely to help linking inference. Kim shows that the aggregation of learners' video interaction can reveal the underlying video structure and provide implication for video authoring and interface design \cite{22}. We believe that, by grouping the browsing history and summarizing patterns of clickstream, we can understand the relevance among learning materials and extract informative features for linking prediction.

Beyond the machine learning techniques, crowdsourcing (or learnersourcing) is an alternative for linking at scale. Li and Mitros proposed a learning module where learners can collaboratively recommend additional learning objects and manage the recommended materials for future learners \cite{138}. We envision a linking system which allows learners and machines to author, edit, and manage the linking of course materials in a collaborative way.

Furthermore, portability is another issue that should be investigated towards a scalable linking system. In this thesis, we adopt a 5-fold cross validation technique to obtain training and testing sets from the same MOOC. This experimental setup makes people wonder why we need the automated linking, since the manual labeling of the entire MOOC is available. Hence, more realistic conditions should be explored; for instance, the automated algorithm is trained and tested on the same course subjects but different offerings, or even on different MOOCs.

\clearpage
\newpage

\begin{singlespace}
\bibliography{main}
\bibliographystyle{plain}
\end{singlespace}

\end{document}